# Software Design Document, Testing, Deployment and Configuration Management of the UUIS--a Team 2 COMP5541-W10 Project Approach

**Omer Shahid Ahmad**
**Faisal Alrashdi**
**Jason (Jun-Duo) Chen**
**Najah Ilham**
**Jianhai Lu**
**Yiwei Sun**
**Tong Wang**
**Yongxin Zhu**

# Table of Contents [1]

## 1. Overview of System Architecture

The UUIS consists of six functional modules supplemented by a presentation layer. The presentation layer is in charge of parsing through the data to be displayed and outputting the relevant information as described in the requirement specification, while the logic layer will be divided into six modules. The database layer consists of a 20-table database.

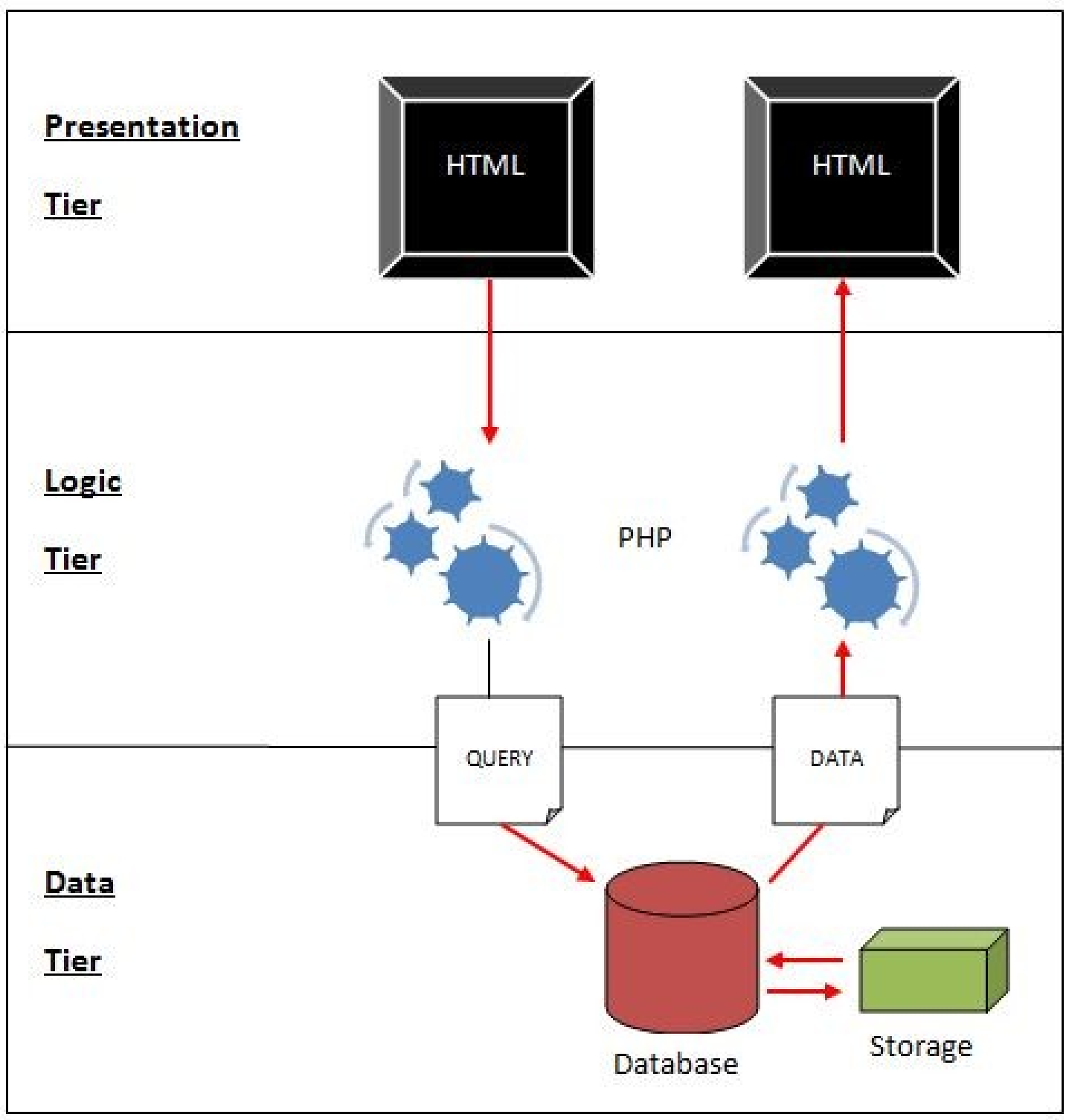

Figure 1.1[7]:Three-tier architecture diagram

### 1.1. Presentation Layer

An Apache 2.2 server will parse through code written in PHP 5 to generate browser-readable output. In addition, it is responsible for ensuring that large datasets are properly truncated (as requested by the user, where applicable), and that controls for navigating through truncated datasets are provided. When truncation occurs, the presentation layer will display controls to display the data (1) immediately preceding the current records, (2) immediately following the current records, (3) at a given position by pagination, (4) the first set of records and/or (5) the last set of records.

The output from the UUIS will also contain Javascript code in order to perform client-side presentation of information, as well as allow preliminary validation of data entered by the user before it reaches the logic layer. All operations are hidden from the external viewer with a wrapper class.

**1.2 Logic Layer**

The user-accessible functions are distributed across six functional modules. One such module groups together the miscellaneous functions accessible by all users (login, logout, submit requests, search, etc.). All university management operations are grouped into another functional module. A third module relates to asset-related operations, while a fourth interfaces with review options (audits and reports). There is a separate module for managing database errors, and the remaining module concerns privileged operations related to requests.

The functions accessible to all users are of necessity those where no background check of the user's privilege is necessary, but merely require validation of the user's identity. The functions are: (1) login, (2) logout, (3) change password, (4) view personal information, (5) submit requests, (6) view request status (for requests submitted by the user), (7) cancel a request (for requests submitted by the user) and (8) search.

Those related to University Management require differing levels of permissions. Provided that such users exist, all users are able to change the privileges of users of compatible affiliation but lesser permission level than themselves. L1 users (users with permission level 1, only higher in rank than L0 users) are allowed to enter new users into the database from a CSV (comma-separated values) file if the new records are of compatible affiliation (same department) and to manually initiate a database back-up. L2 users are able to create a new department and add a location to the database in addition to the functions available to L1 users (though the compatibility extends to all departments within the L2 user's faculty). L3 users are able to perform the same functions as L2 and L1 users on a university-wide basis, as well as create new faculties.

Asset management includes the ability to view assets, add assets either in single item or in large quantities from CSV files, update asset information either for a single asset or for large numbers of assets and group assets. Each of these operations may be performed by L1 users for asset(s) within their department, by L2 users for asset(s) within their faculties or by L3 users for any asset(s) in the university.

There are two primary functions related to the review module: generating reports or audit data. As with asset management, the review functions are limited by the user's privileges, and the selection of options varies accordingly. Reports may be generated to compare data across various fields (such as the number of seats available vs. the maximum capacity of a room, or the number of desktop computers vs. the number of computer screens, or even simply a list of the users within a department). Auditing relates to reviewing the transactions in the database that have occurred, such as tracing the user who has added an item, or who updated which field for an asset.

A user with sufficient privileges (L1, L2 or L3 users with system administrator privileges) is able to manage error reports. Options for doing so are: listing errors (possibly after inputting constraints in the form of a search), printing error messages and viewing/annotating specific error messages.

The sixth module concerns request management. All users are able to submit requests as well as view and cancel requests they have submitted themselves. All other request-related operations are handled by the "request management" module. Those operations include: viewing pending requests submitted by other users, approval/formalization of pending requests and rejection of requests. In all cases, users may only access requests made by other users with compatible affiliation but lower permission level than themselves.

### 1.3 Database Layer

The database was designed in order to accommodate various situations. For instance, the number of properties for a given item will vary depending on the category of item, and each property may be set to be required or not. As a concrete example, a software title may have a license key which may be installed on a specified number of devices. The database will be able to record the license key, the number of devices on which the software title may be installed, and the number of devices on which it has already been installed.

In order to facilitate access to the database, we have installed PEAR into the server and encapsulated the abstract DB class such that it may be accessed as required by the six functional modules. This module provides a standardized interface throughout the application with which to interact with the database.

## 2. System Architecture [3,4]

In this section, we will present the details of the system architecture. For each module, we will first describe the functions with a use-case diagram and then the class diagram for all the relevant classes of the module. Each of the functions will then be shown as use-case and activity diagrams.

### 2.1. Common Functions

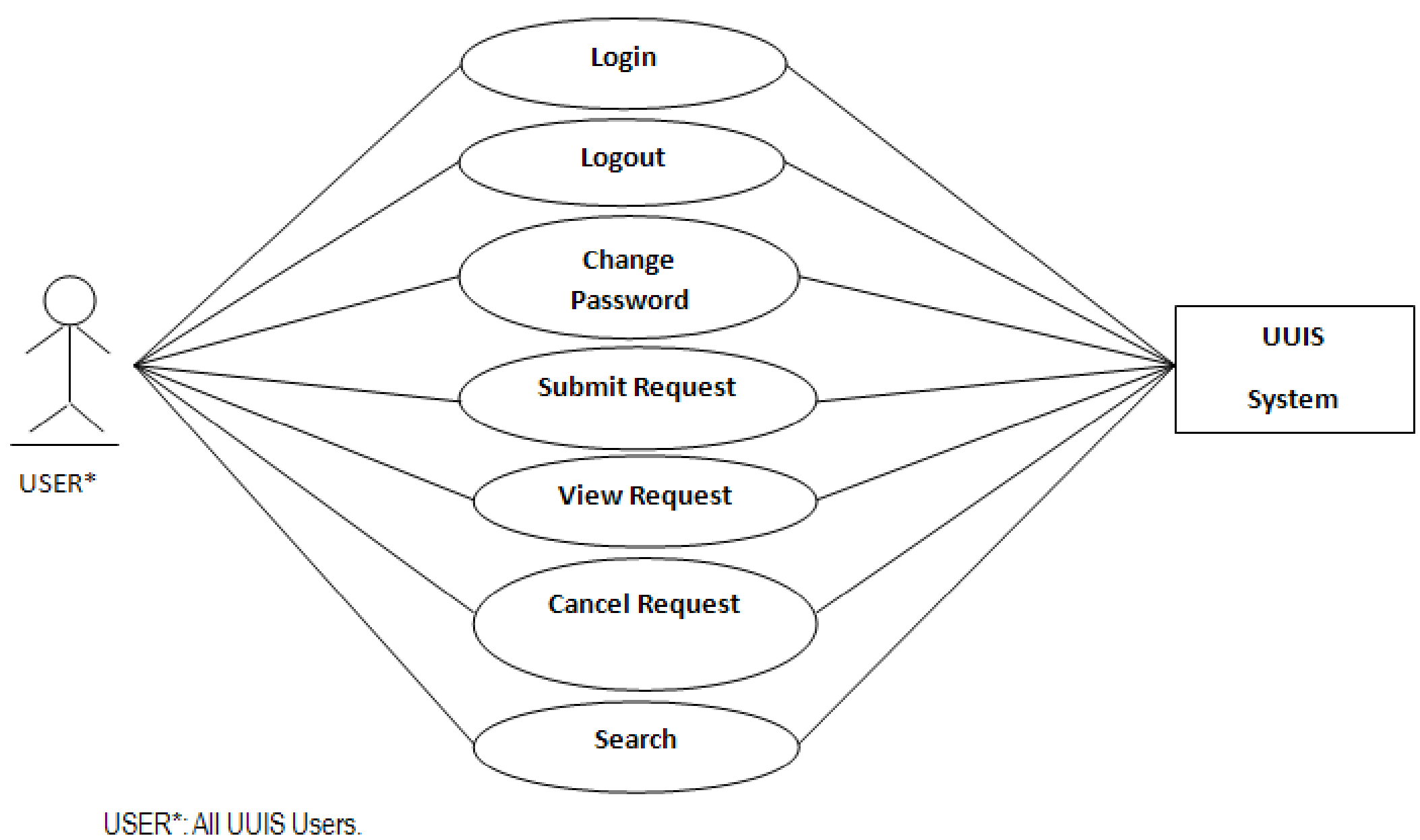

Figure 2.1.0.1 Overview of Functions Common To All UUIS Users. This use-case diagram lists all the functions that have been grouped into a module containing the basic functionality available to all users, regardless of user permissions.

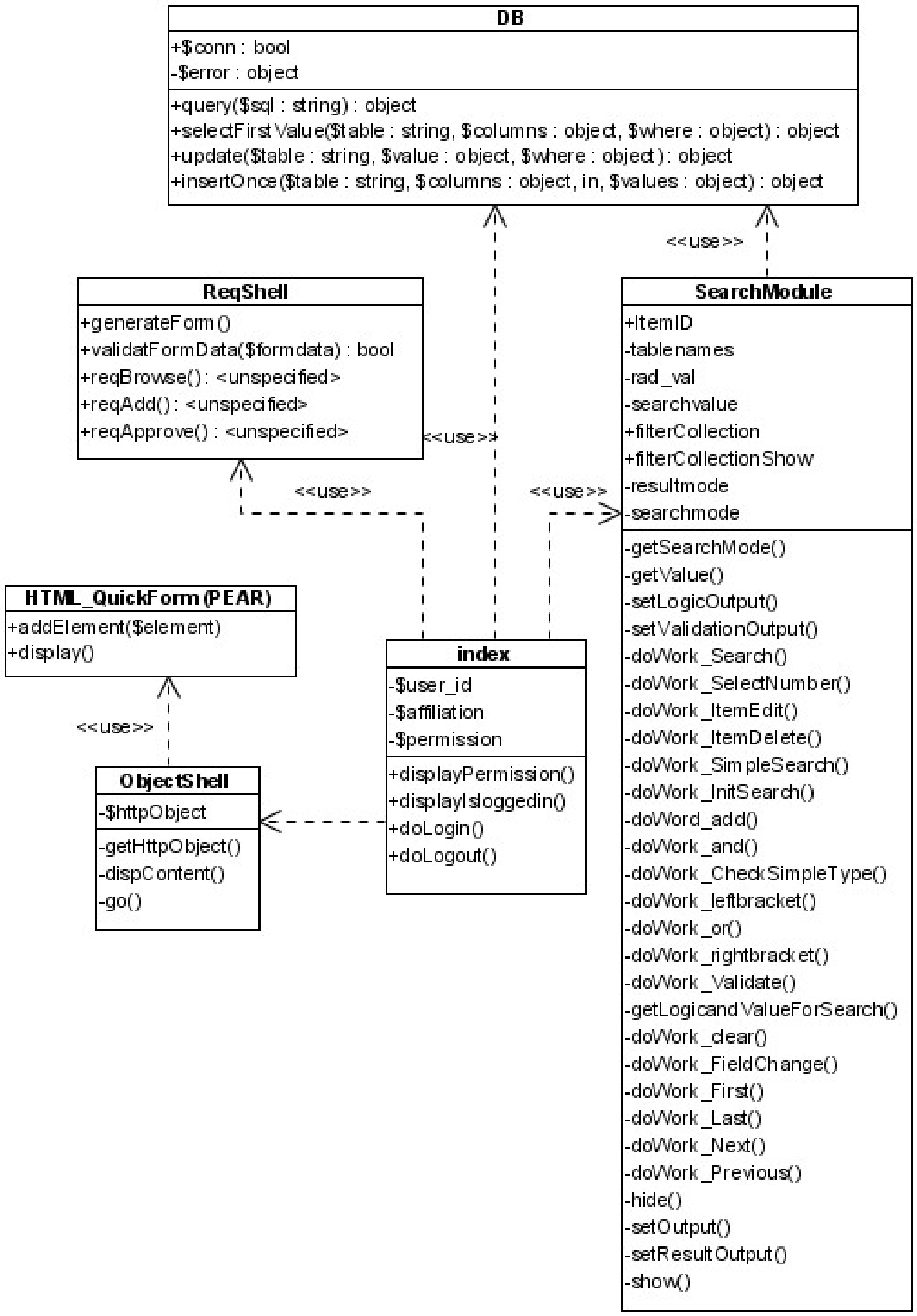

Figure 2.1.0.2 Classes in the Common Module. The above diagram depicts the classes in the module and their relationships to each other. Please note that the SearchModule and ReqShell are designed to communicate with the appropriate components of the "Search" and the "Request Management" modules (sections 2.3 and 2.6 below), respectively.

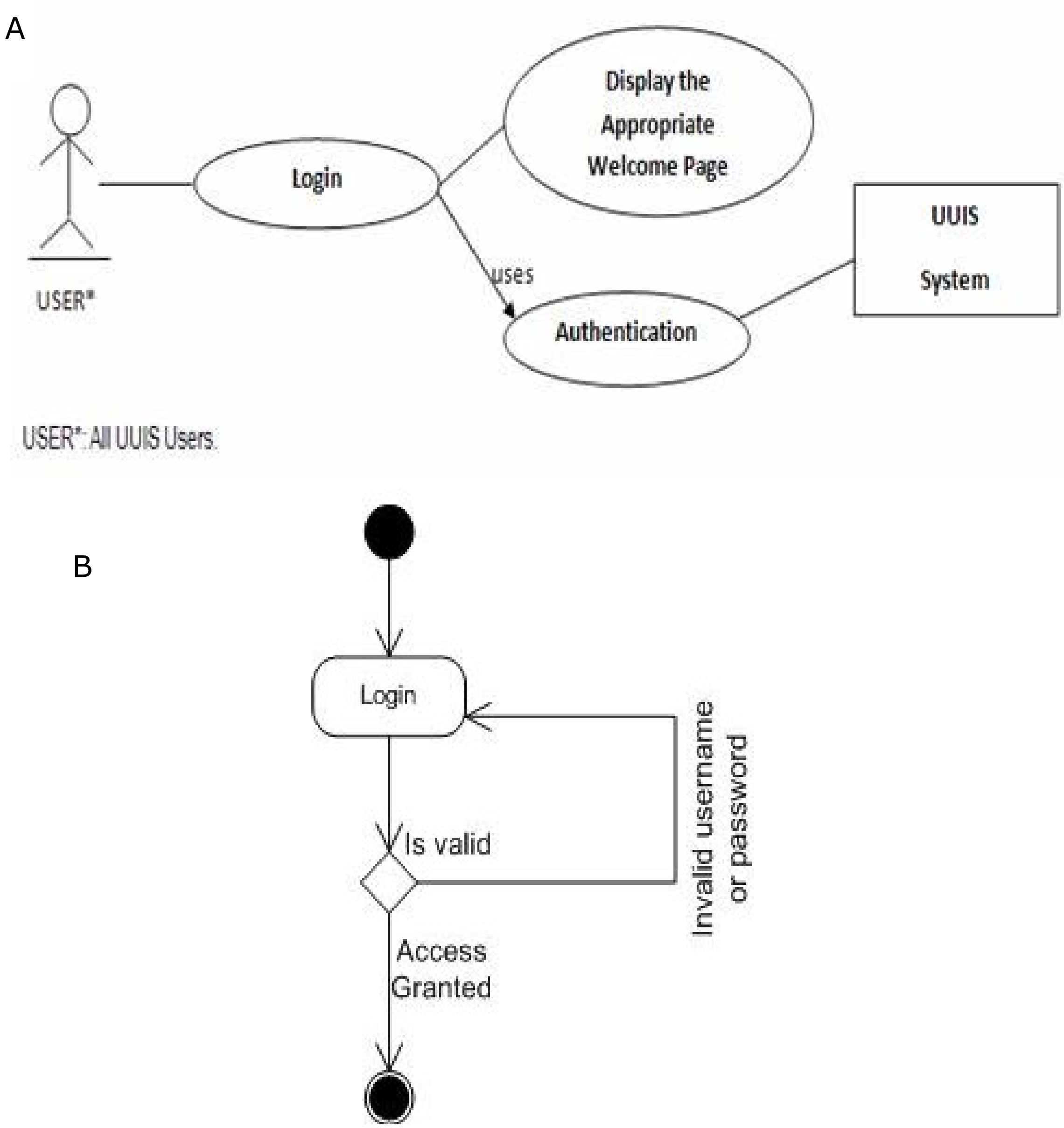

Figure 2.1.1 Login. A view of the function is shown with a use-case diagram (panel A) and an activity diagram (panel B).

A

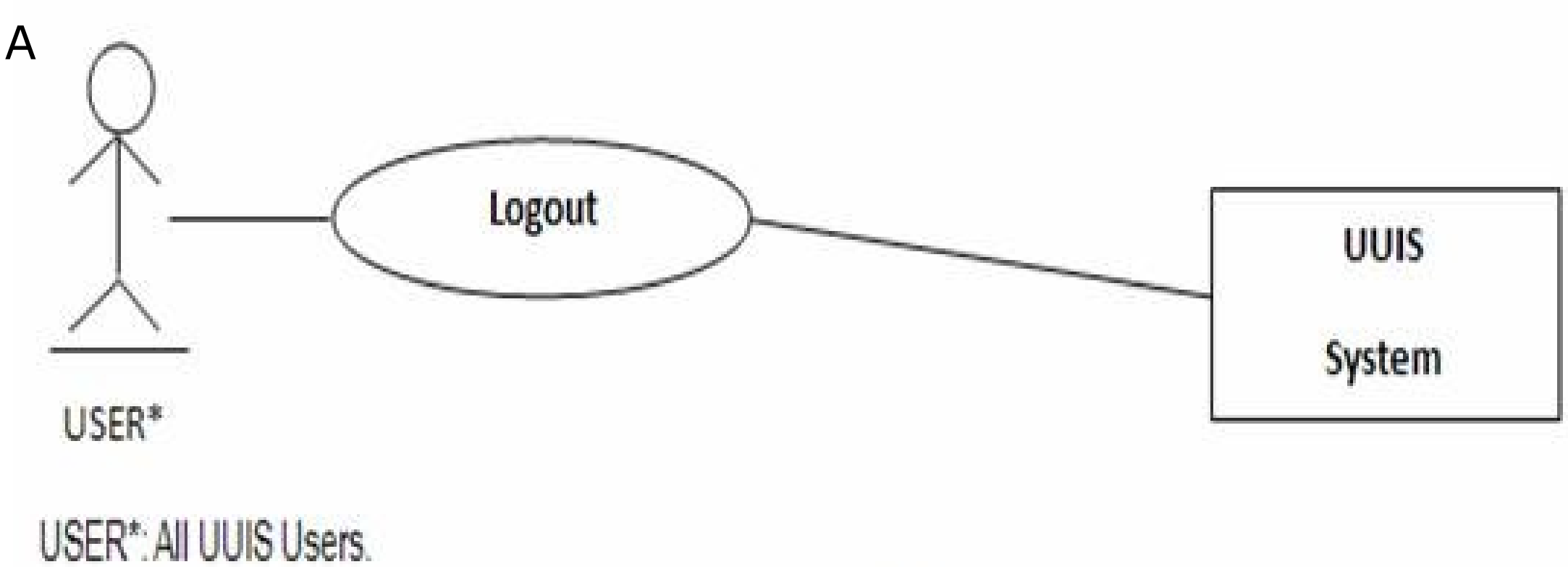

B

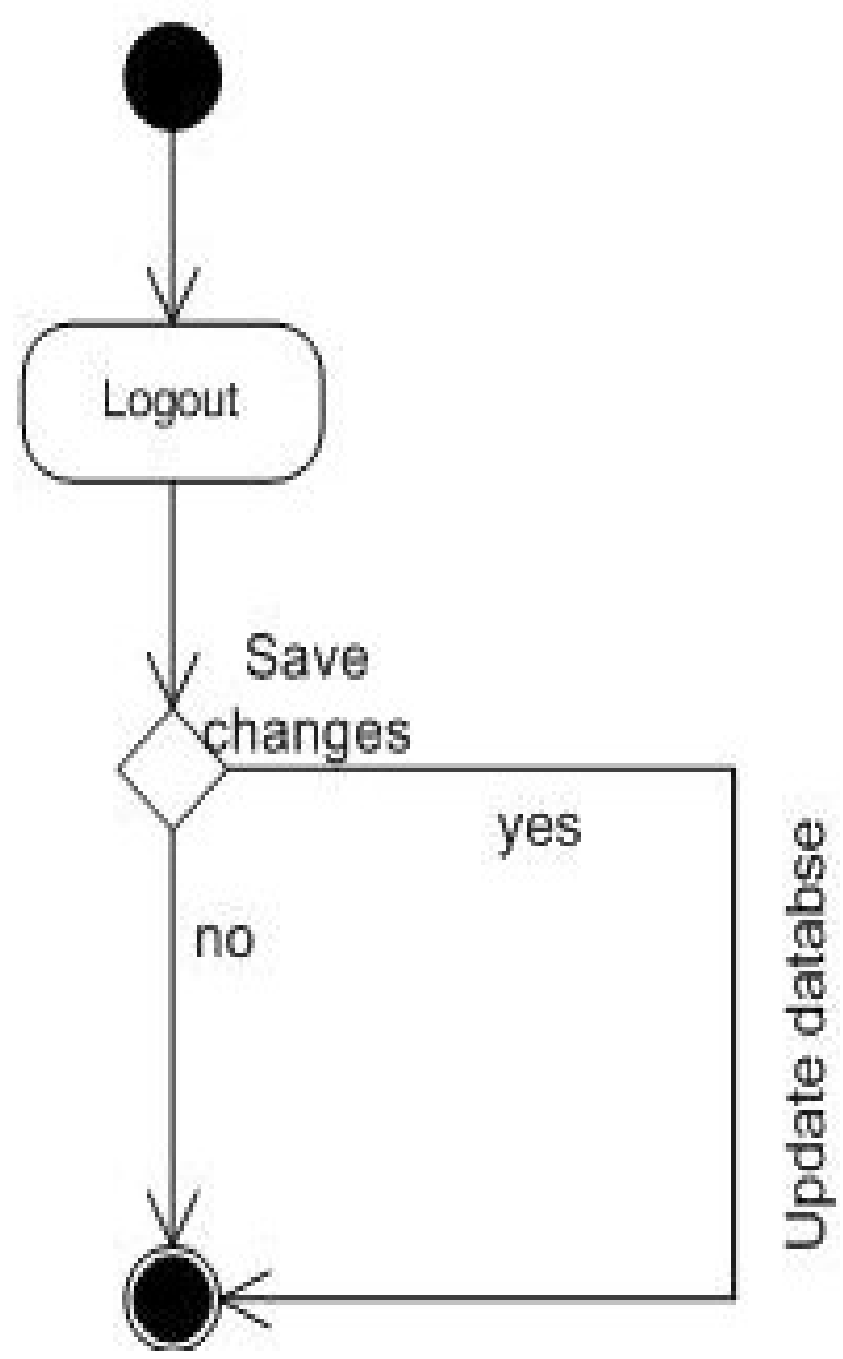

Figure 2.1.2 Logout. A view of the function is shown with a use-case diagram (panel A) and an activity diagram (panel B).

A

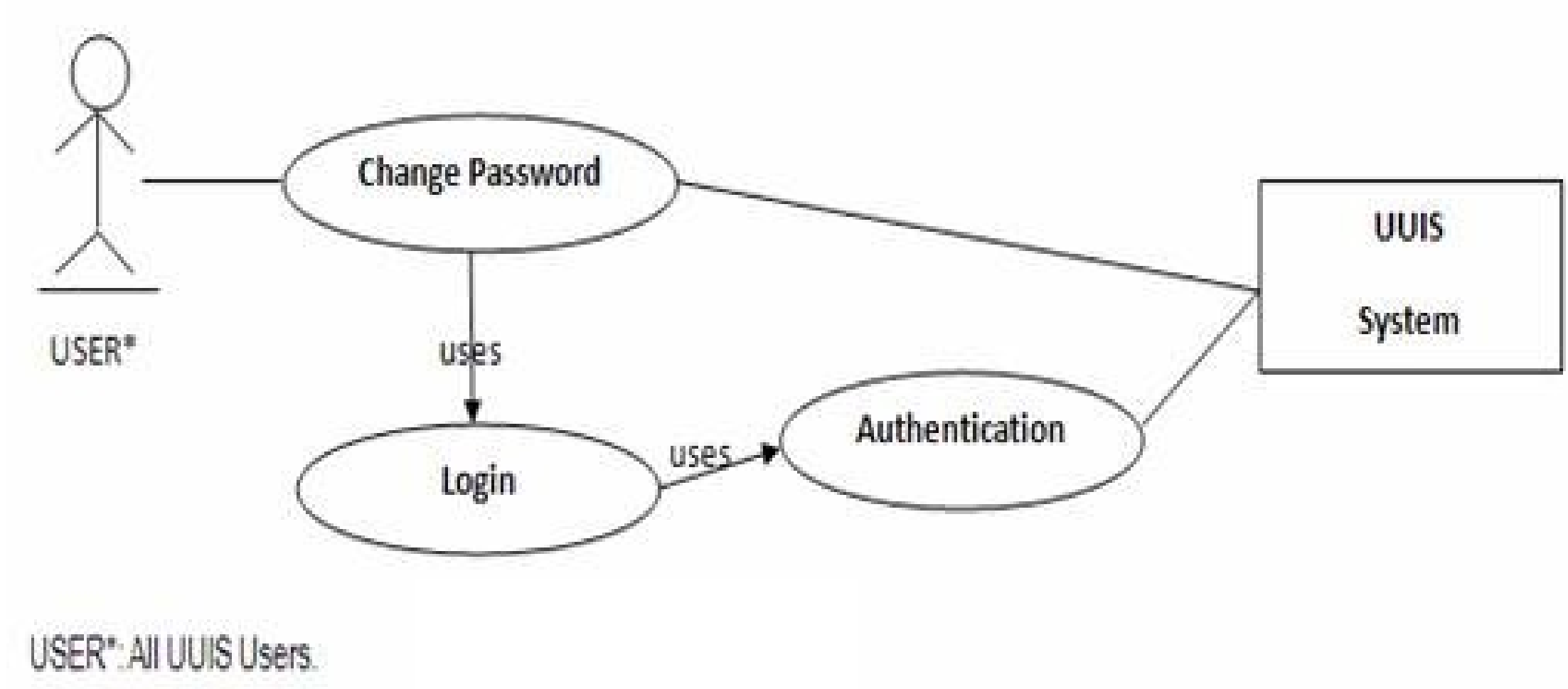

B

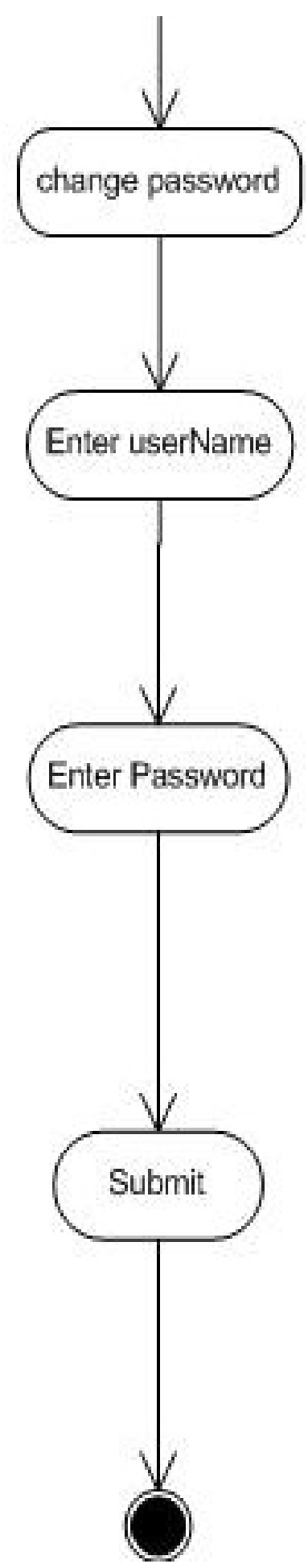

Figure 2.1.3 Change password. A view of the function is shown with a use-case diagram (panel A) and an activity diagram (panel B).

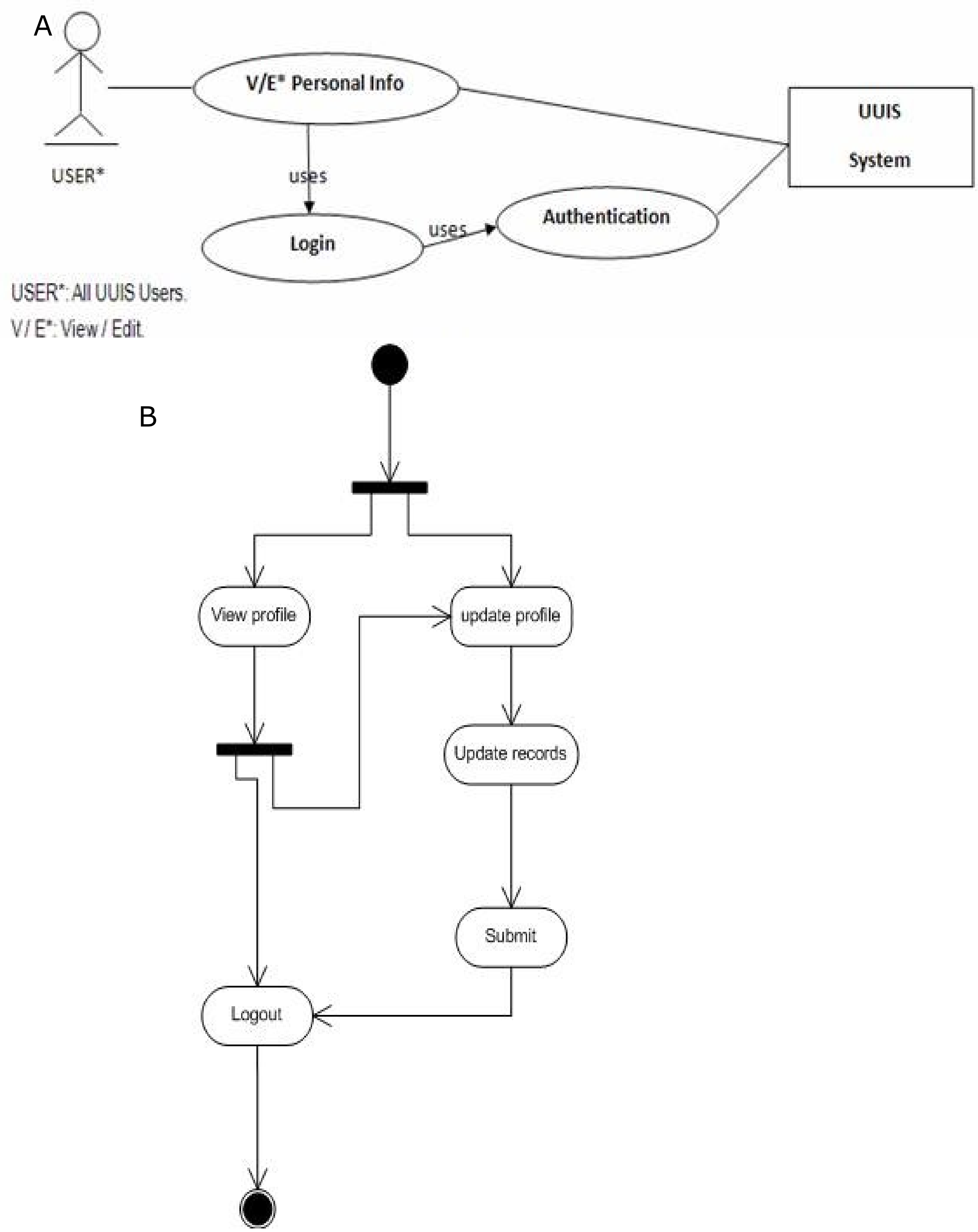

Figure 2.1.4 View personal information. A view of the function is shown with a use-case diagram (panel A) and an activity diagram (panel B).

A

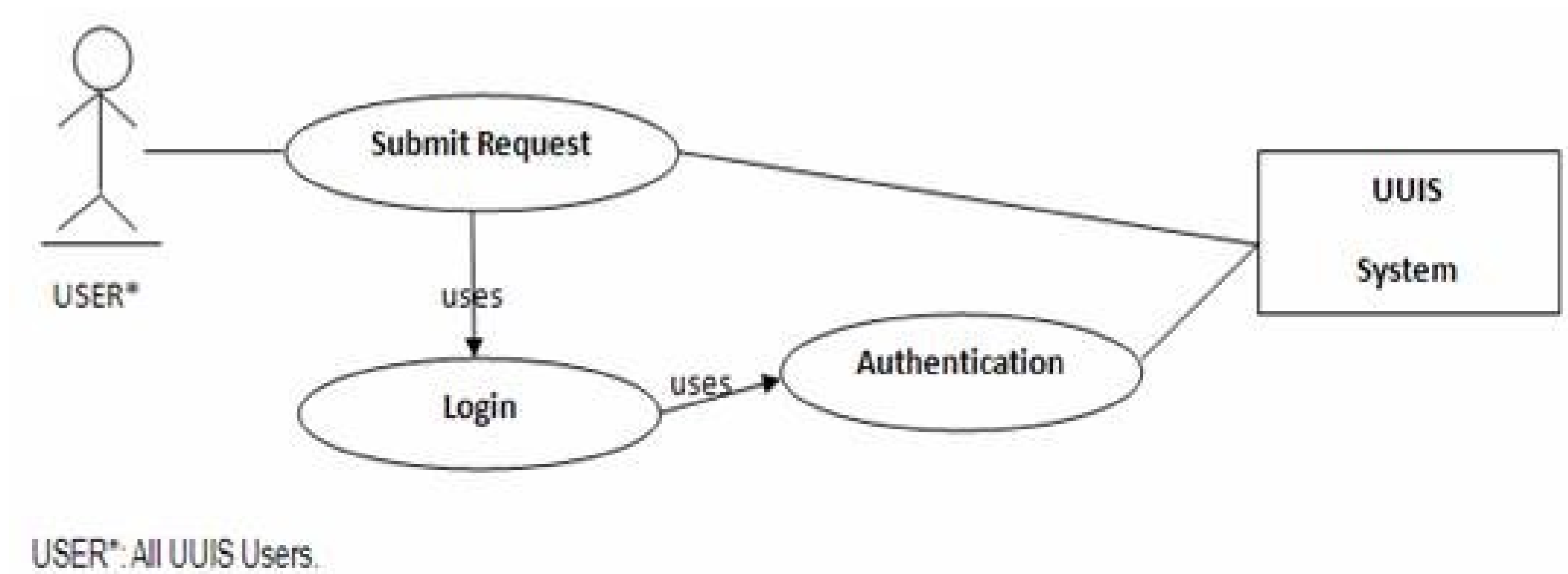

B

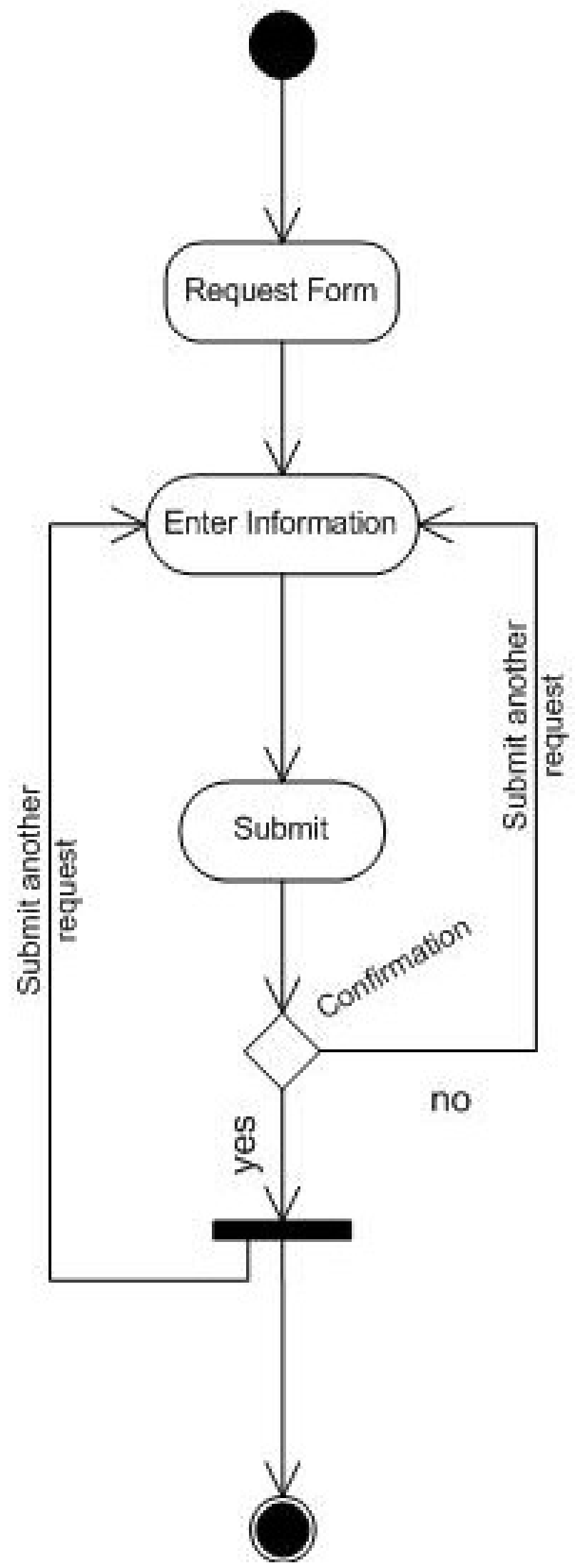

Figure 2.1.5 Submit requests. A view of the function is shown with a use-case diagram (panel A) and an activity diagram (panel B).

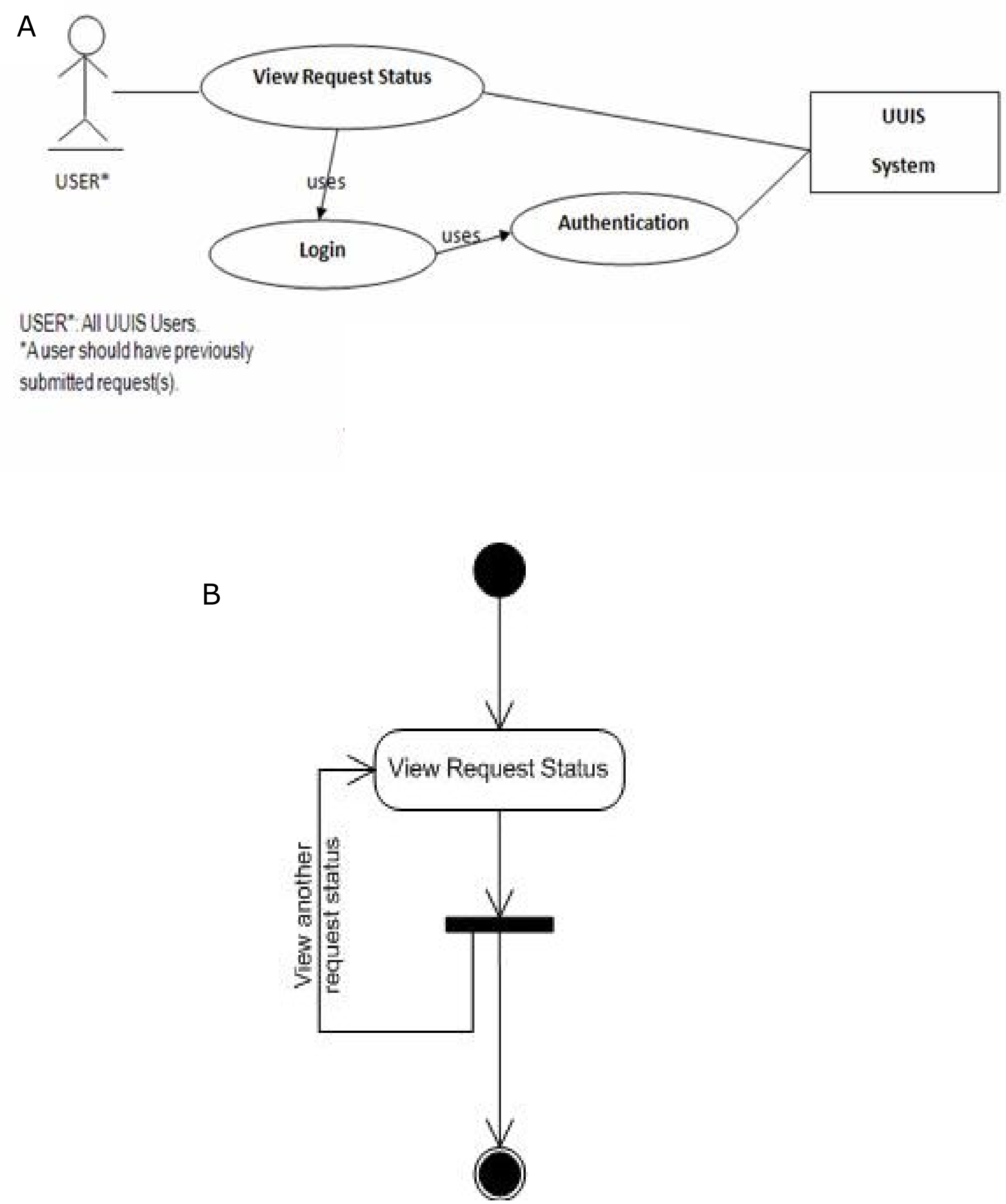

Figure 2.1.6 View request status. A view of the function is shown with a use-case diagram (panel A) and an activity diagram (panel B).

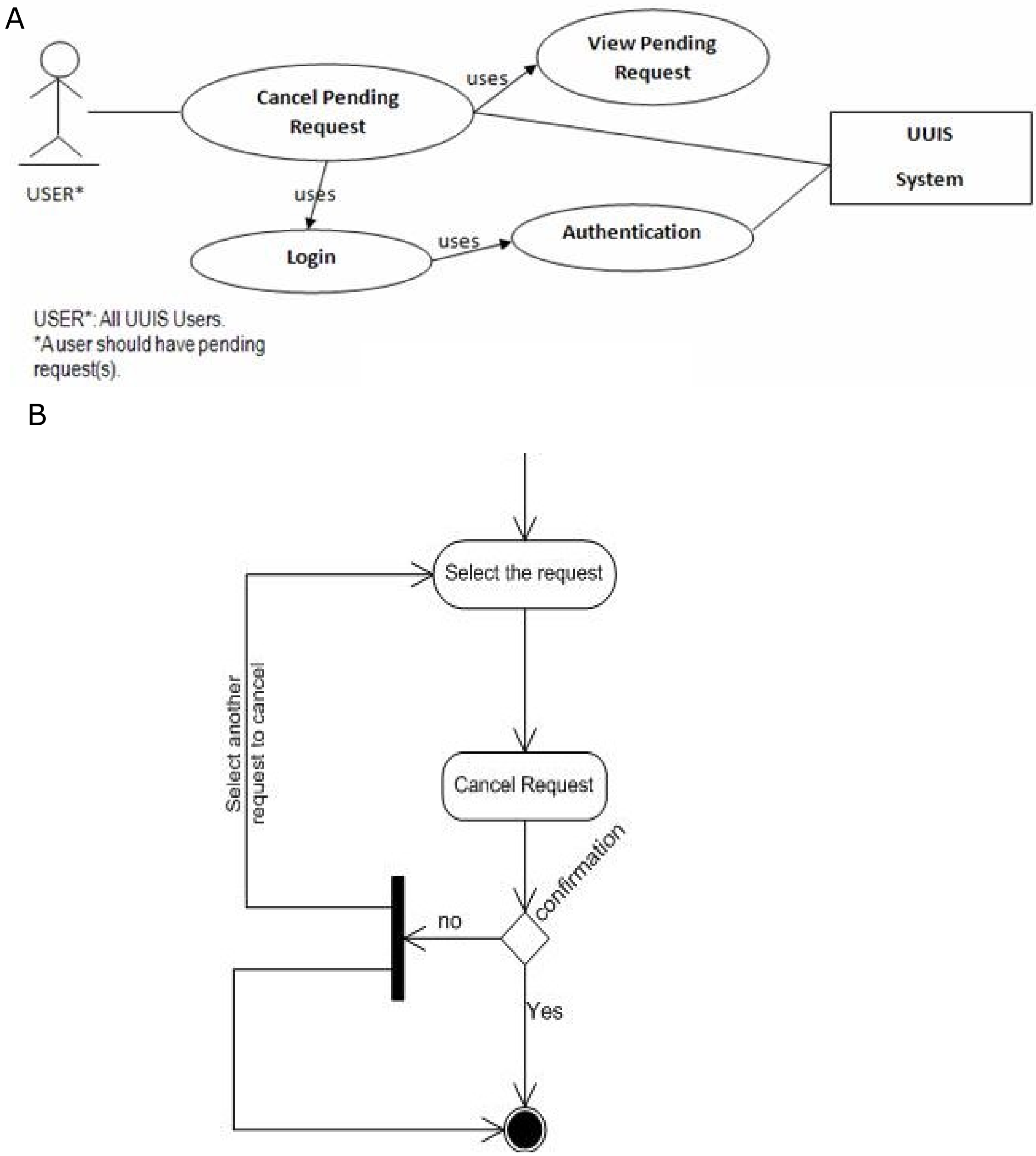

Figure 2.1.7 Cancel a request. A view of the function is shown with a use-case diagram (panel A) and an activity diagram (panel B).

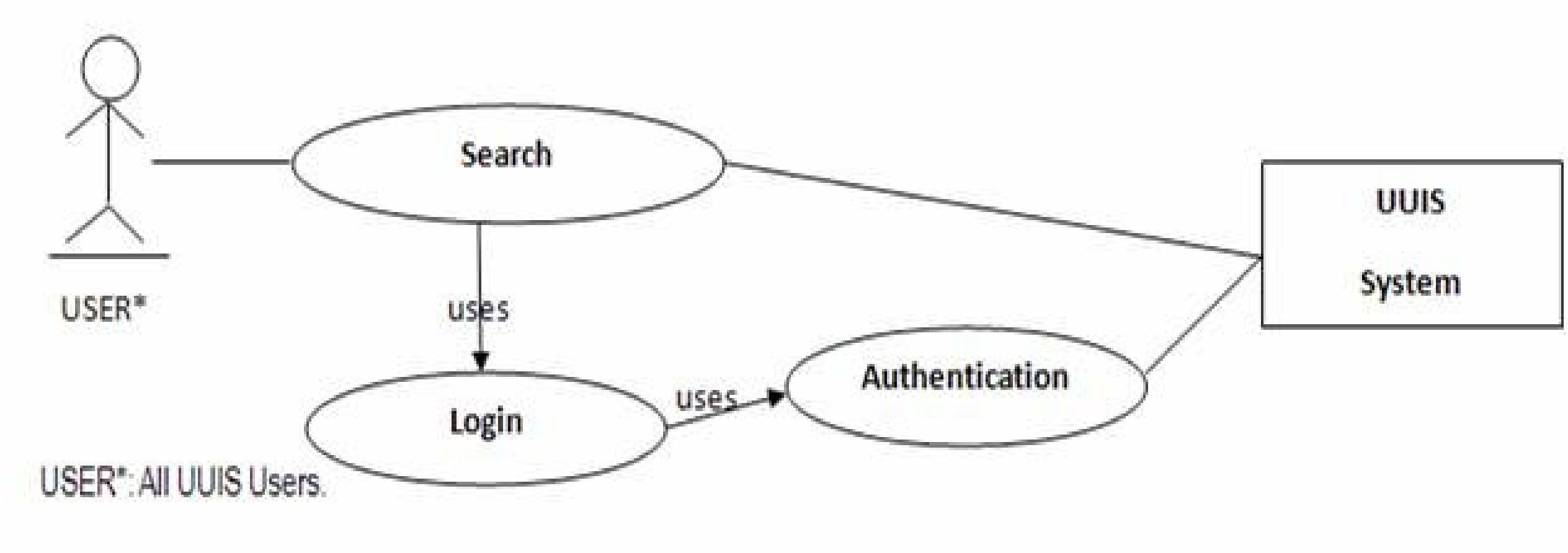

Figure 2.1.8 Search. This use-case diagram highlights the interactions between the search module and the authentication method. This is particularly relevant as the search options offered to the user are filtered according to the user's privileges, and the query itself is verified server-side to ensure that the request does not exceed the user's permission level. For a class diagram of the search module, please refer to Figure 2.3.0.2.

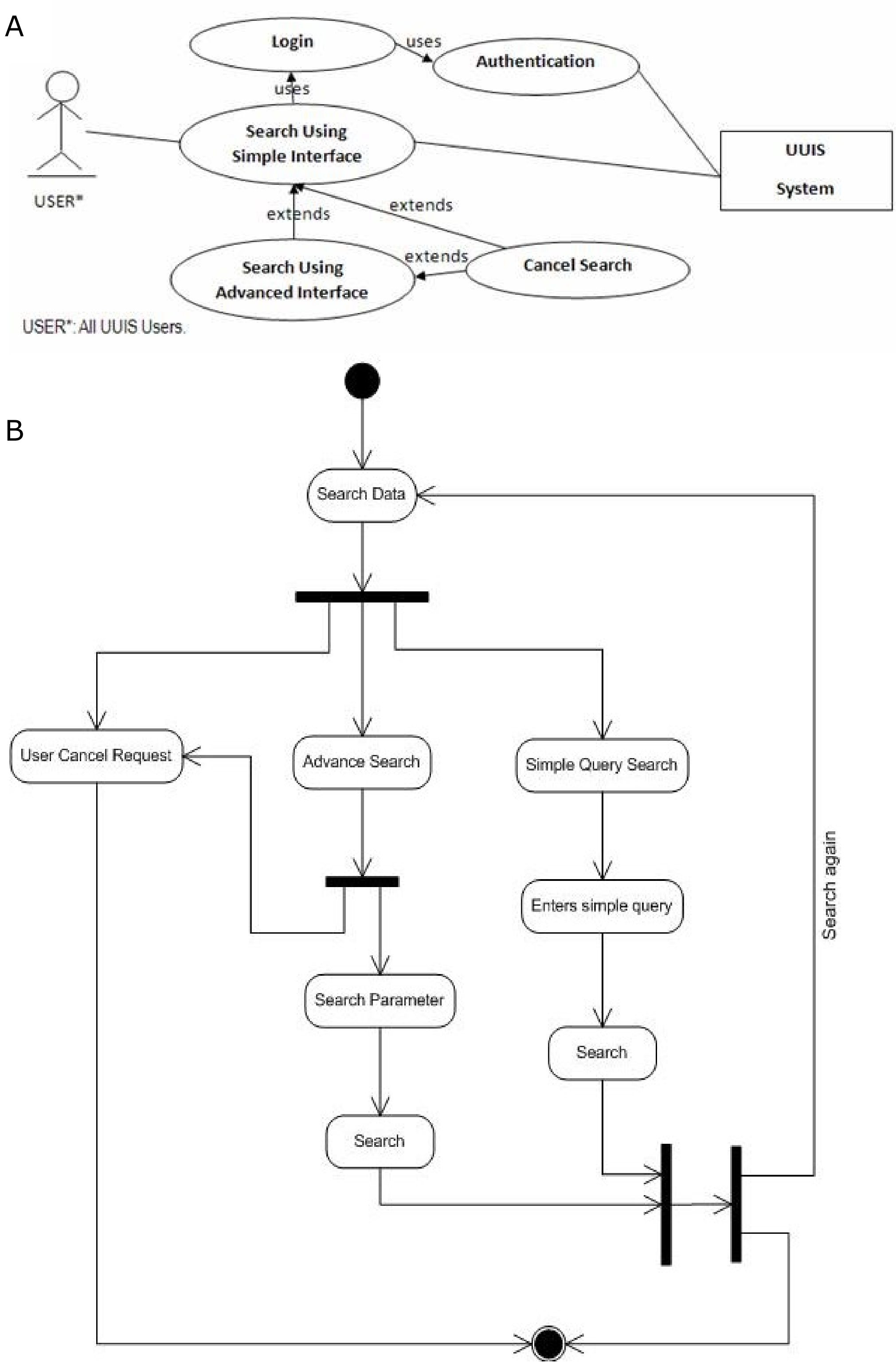

Figure 2.1.8.1. Search for data. A view of the function is shown with a use-case diagram (panel A) and an activity diagram (panel B).

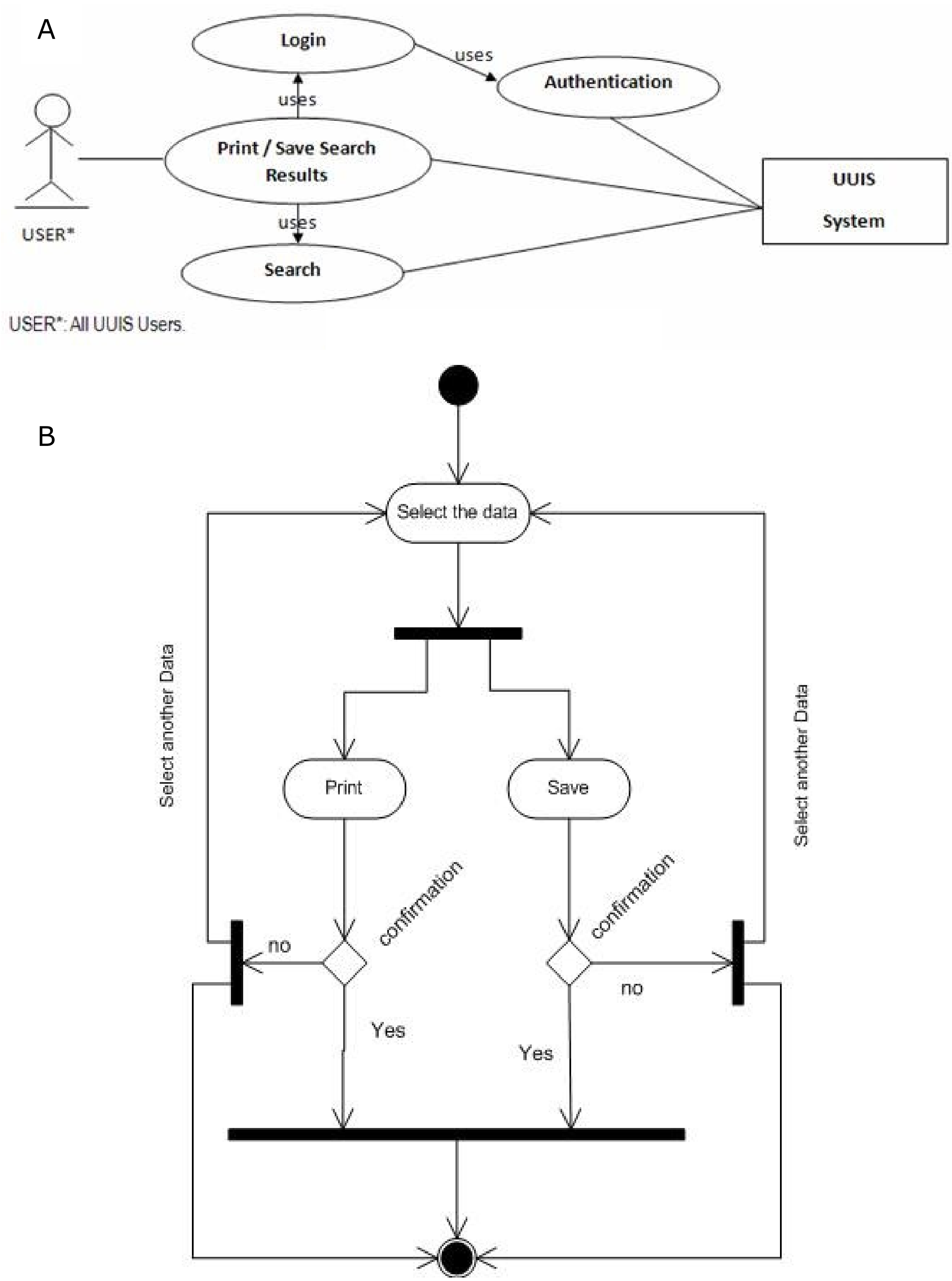

Figure 2.1.8.2. Print/save search results. A view of the function is shown with a use-case diagram (panel A) and an activity diagram (panel B).

## 2.2. University Management

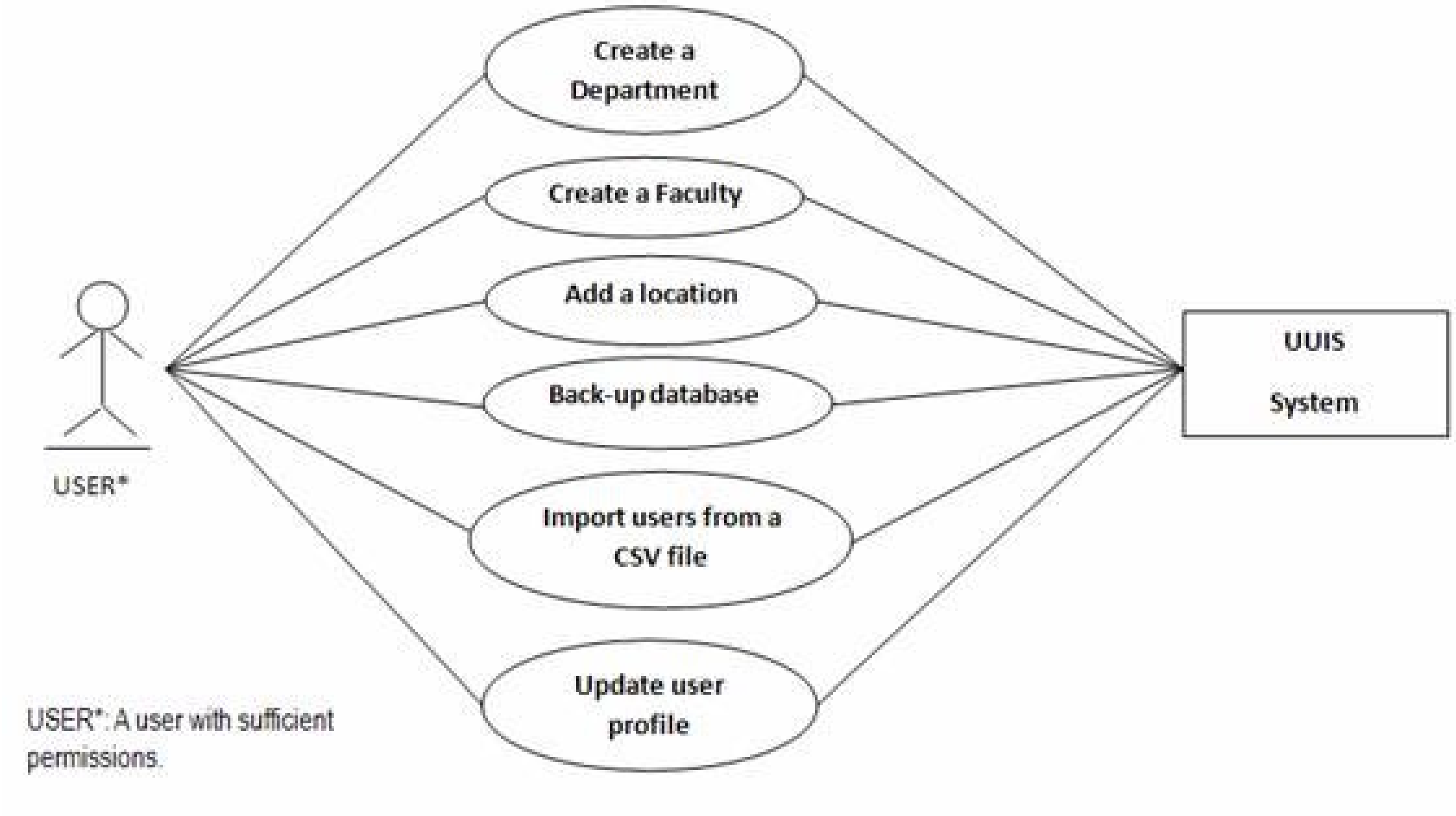

Figure 2.2.0.1 University Management. This use-case diagram provides a view of the functionality provided by the "university management" module.

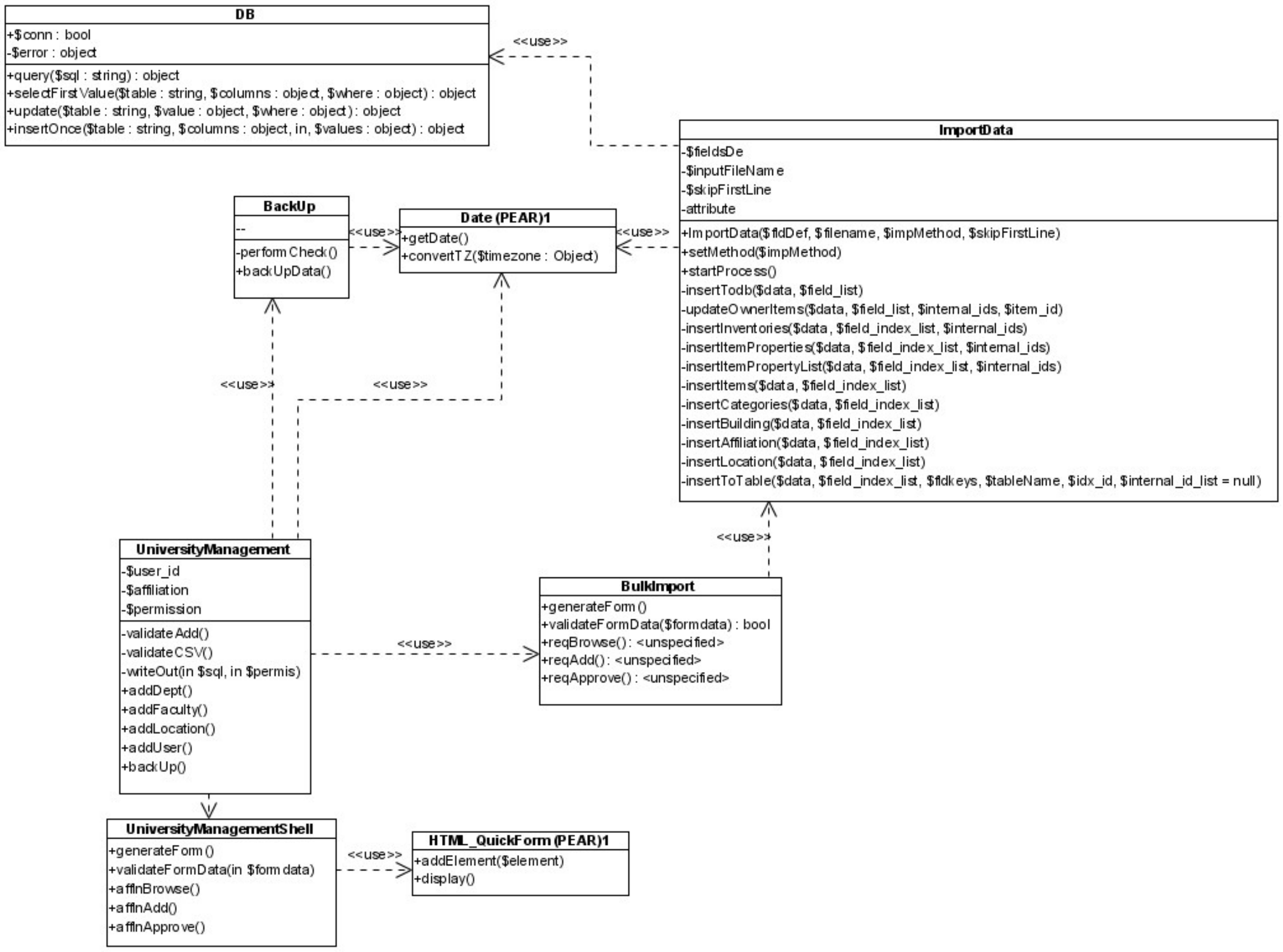

Figure 2.2.0.2 Classes in University Management. The above diagram illustrates the classes in the module and the relations between the classes.

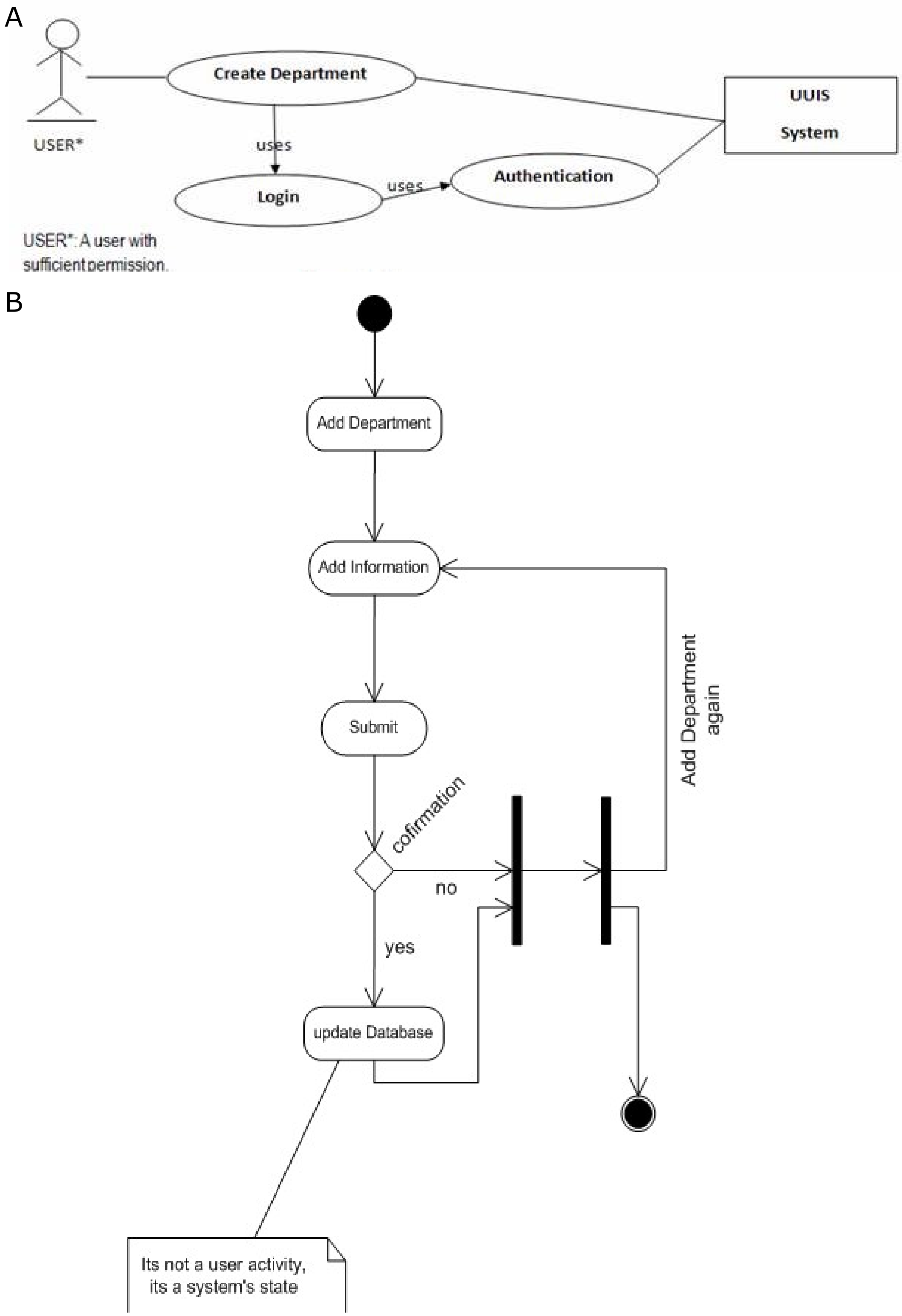

Figure 2.2.1 Create a Department. A view of the function is shown with a use-case diagram (panel A) and an activity diagram (panel B).

A

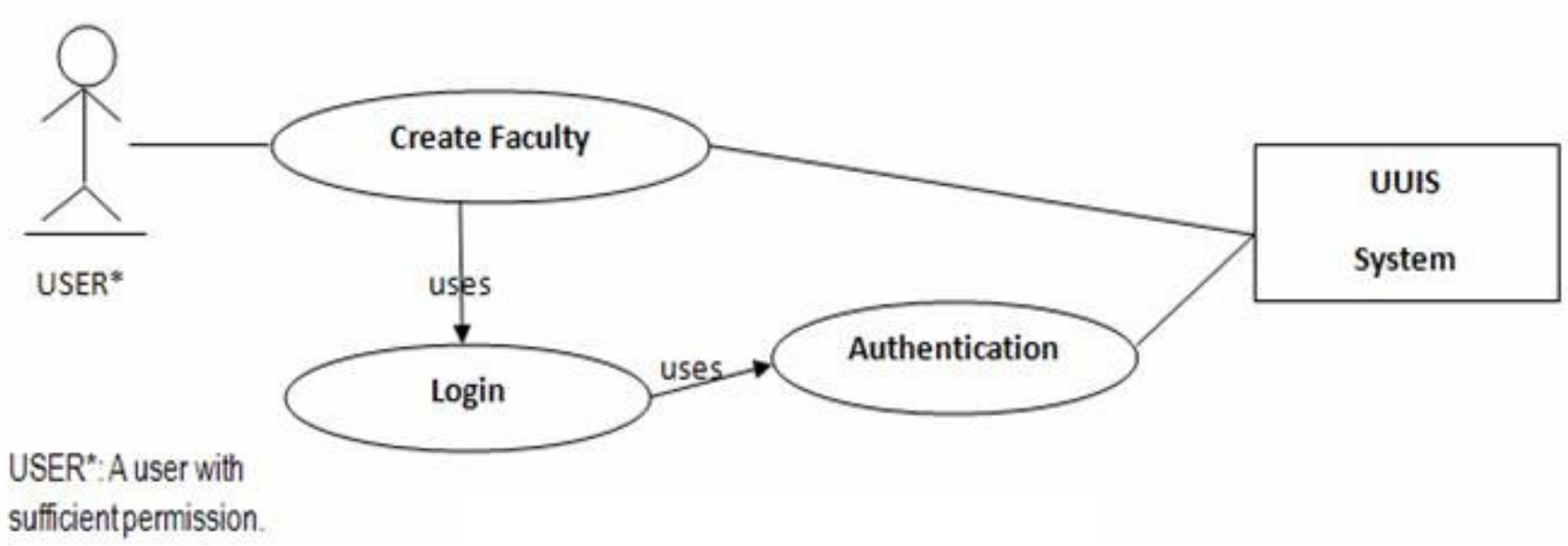

B

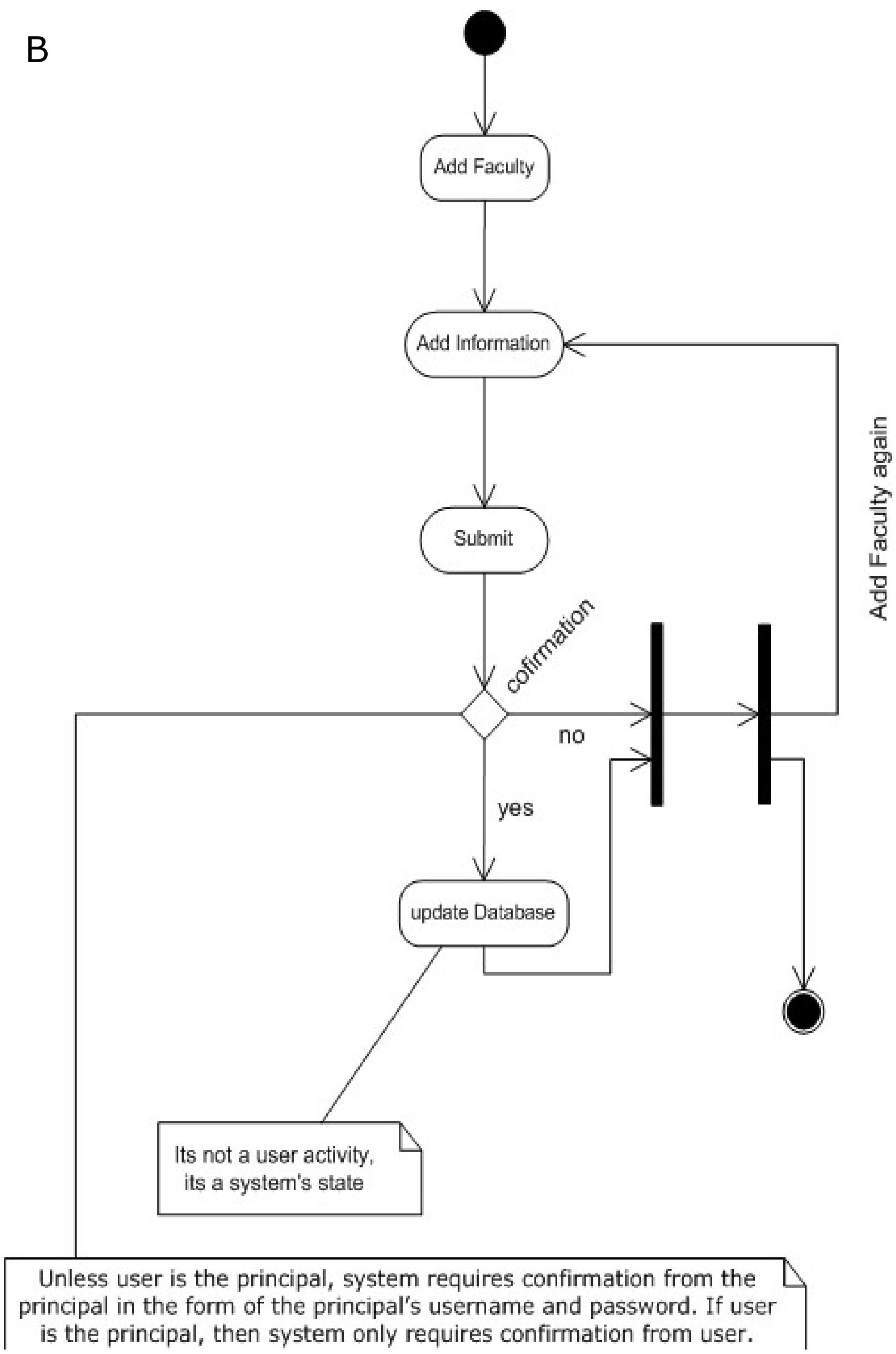

Figure 2.2.2 Create a Faculty. A view of the function is shown with a use-case diagram (panel A) and an activity diagram (panel B).

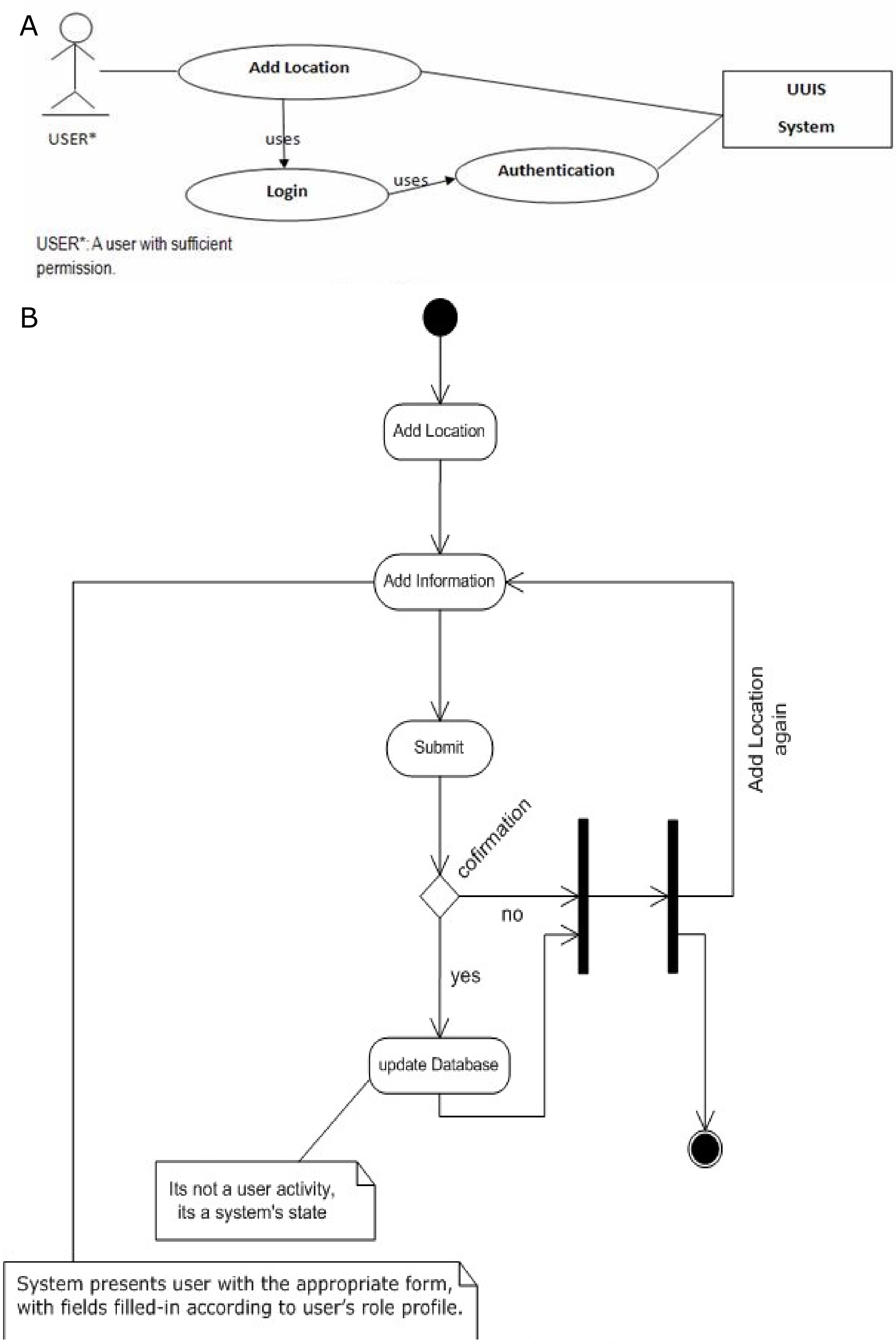

Figure 2.2.3 Add a Location. A view of the function is shown with a use-case diagram (panel A) and an activity diagram (panel B).

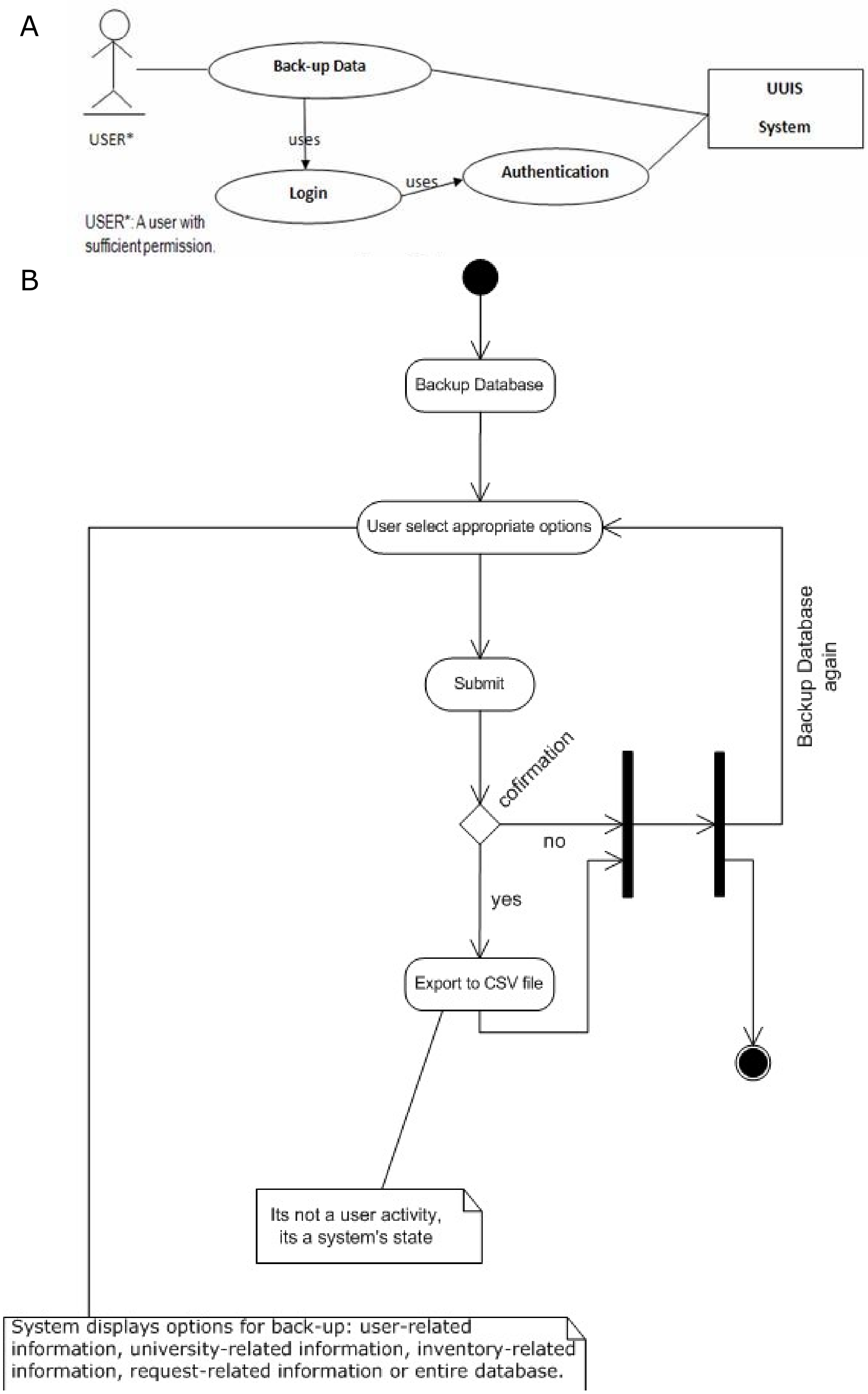

Figure 2.2.4 Back-Up Database. A view of the function is shown with a use-case diagram (panel A) and an activity diagram (panel B).

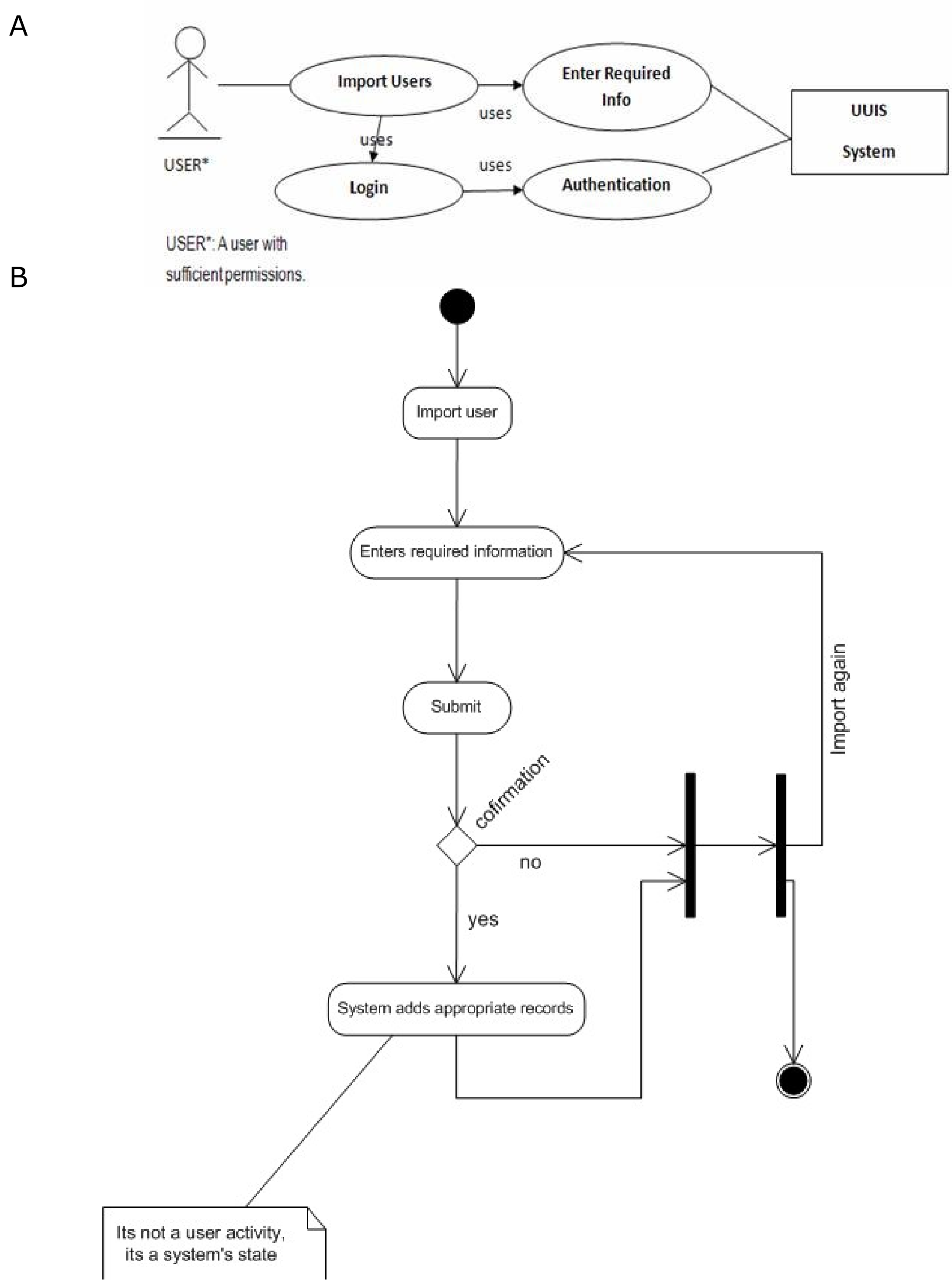

Figure 2.2.5 Bulk Import Users from a CSV File. A view of the function is shown with a use-case diagram (panel A) and an activity diagram (panel B).

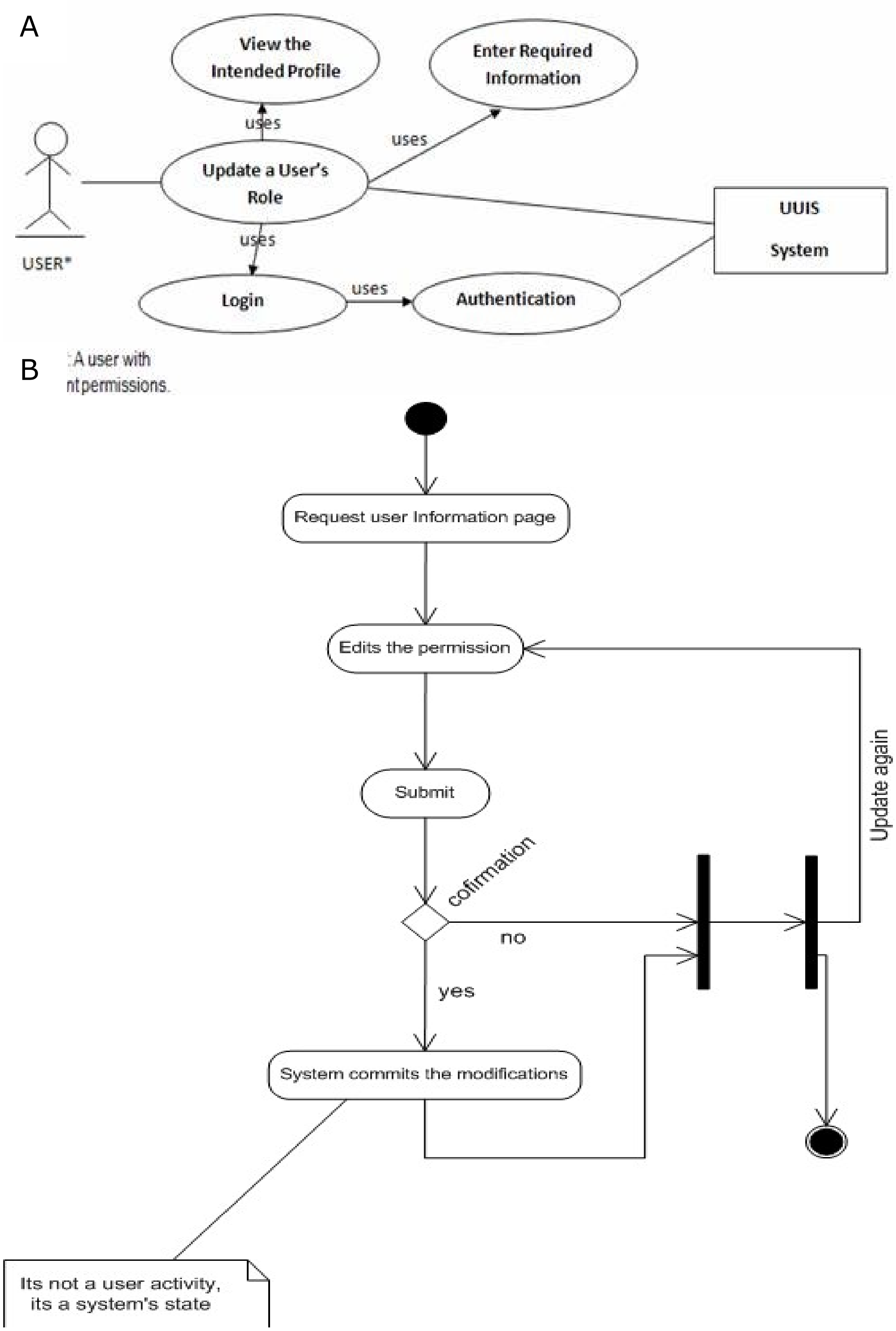

Figure 2.2.6 Update User Profile. A view of the function is shown with a use-case diagram (panel A) and an activity diagram (panel B).

## 2.3. Asset Management

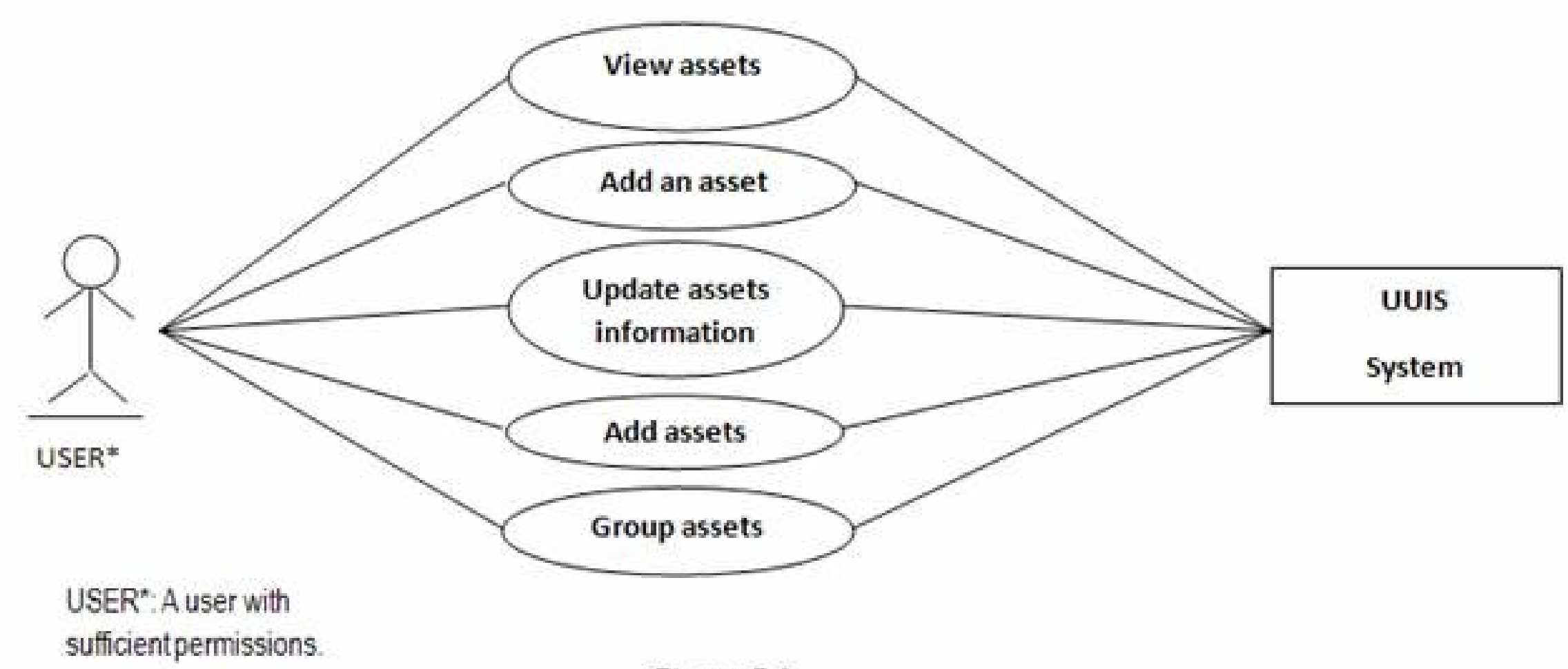

Figure 5.3
Use Cases to Manage Assets

Figure 2.3.0.1 Manage Assets. This use-case lists the functions related to asset management. Since all of the functions are accessible through the "search" module as mentioned in section 2.1, we have implemented them by performing a default search to retrieve all assets available according to the user's permission level.

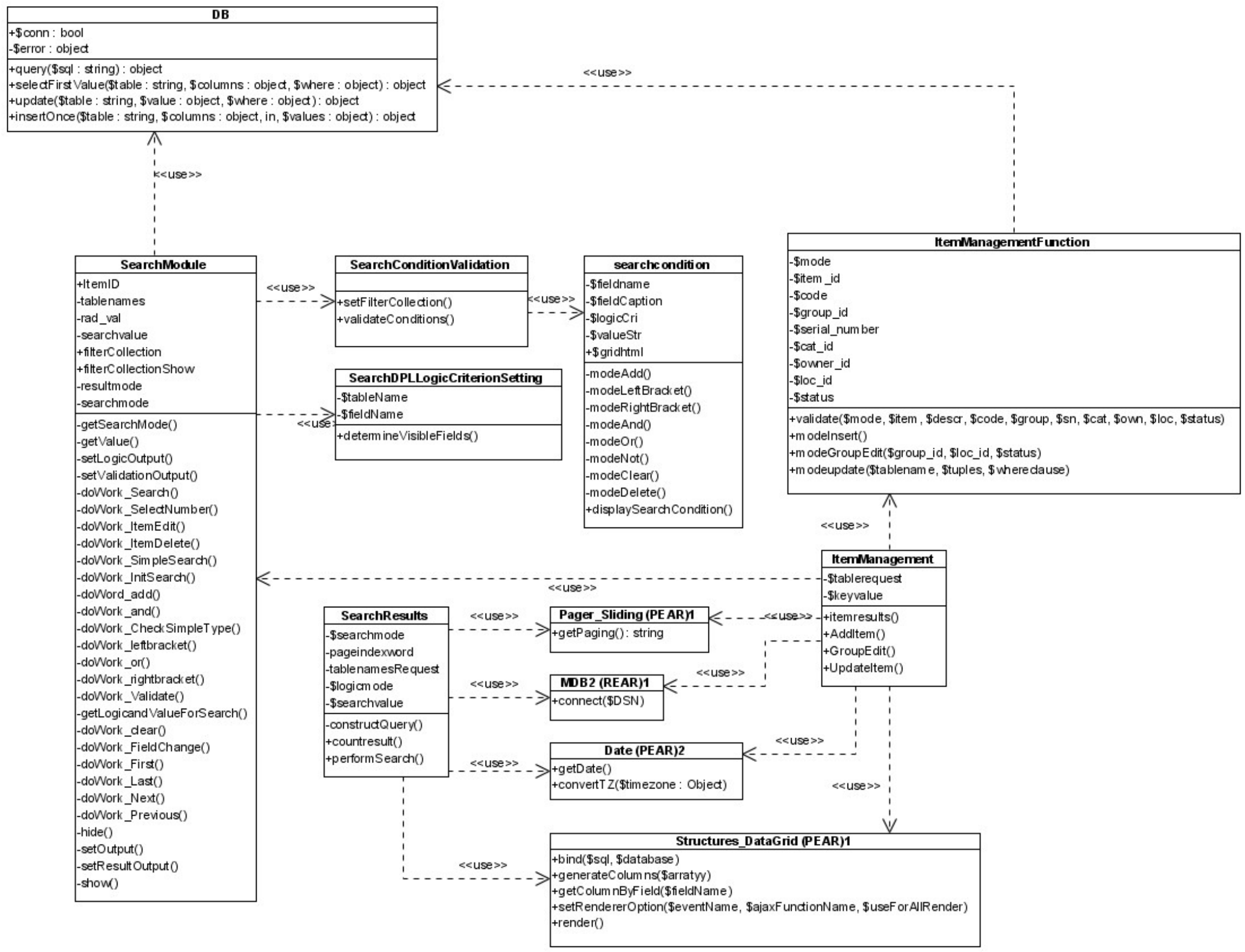

Figure 2.3.0.2 Classes in “Manage Assets” Module. The above class diagram illustrates the classes present in the “Manage Assets” module and their relationships to each other.

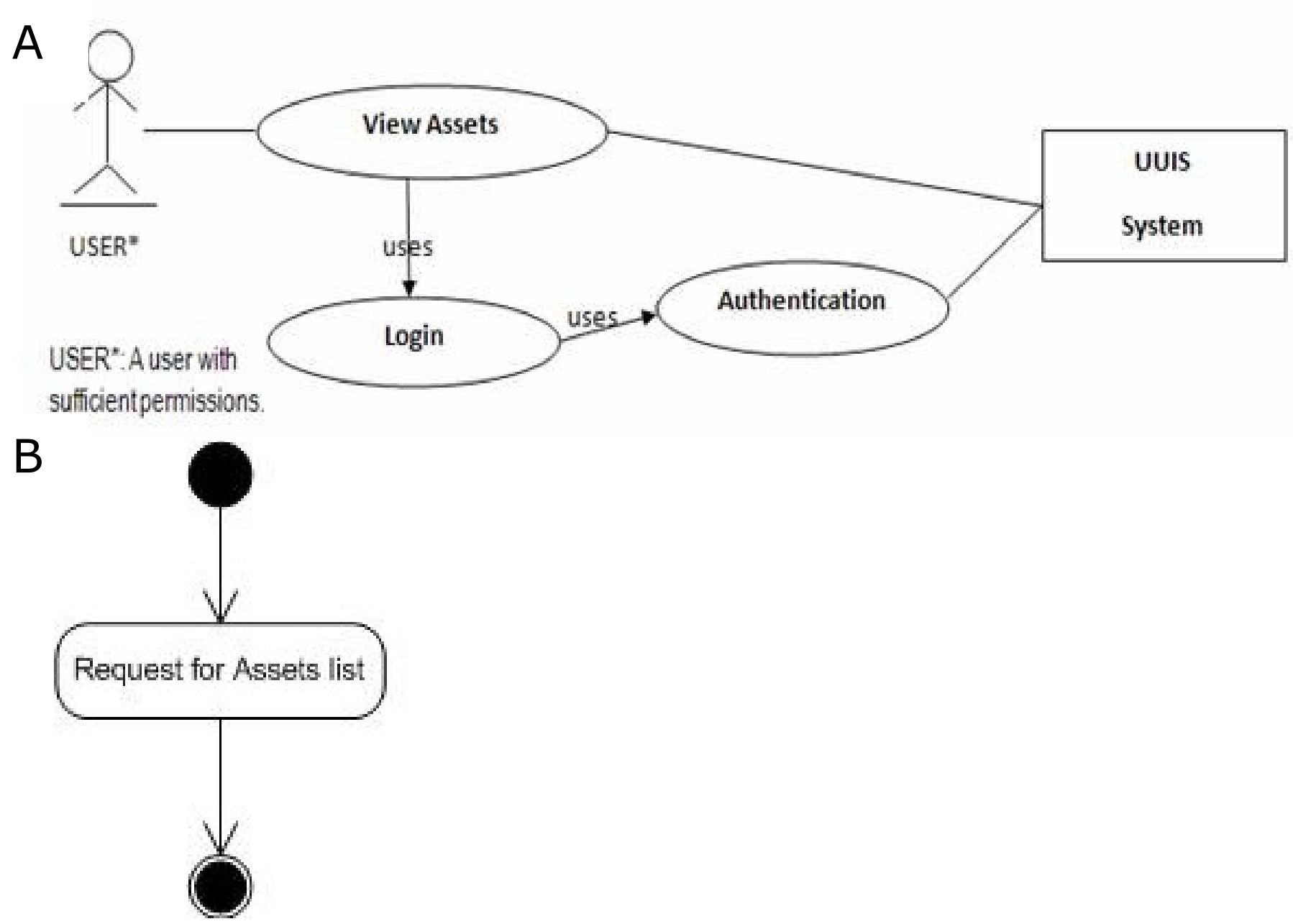

Figure 2.3.1. View Assets. A view of the function is shown with a use-case diagram (panel A) and an activity diagram (panel B).

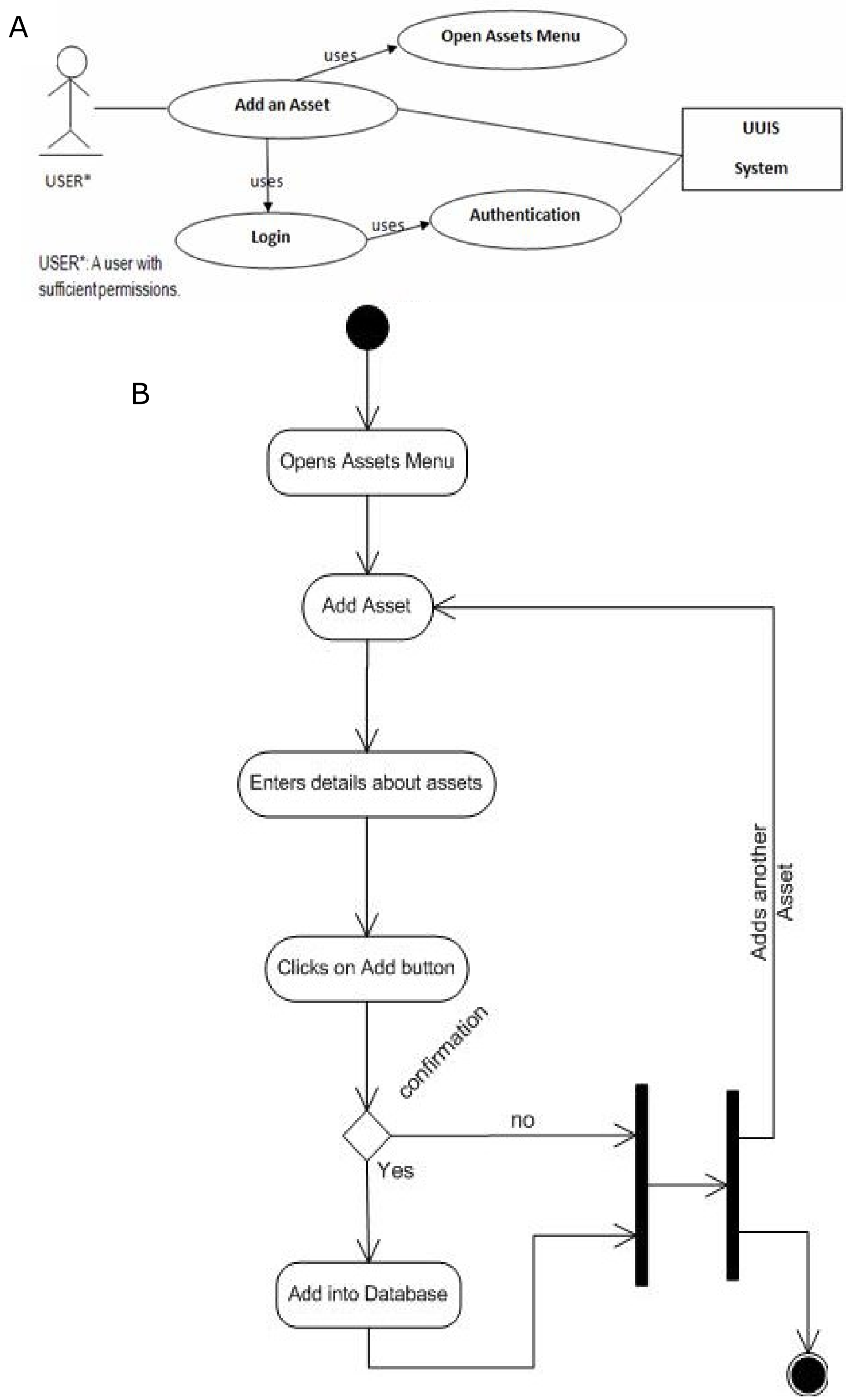

Figure 2.3.2 Add Asset. A view of the function is shown with a use-case diagram (panel A) and an activity diagram (panel B).

A

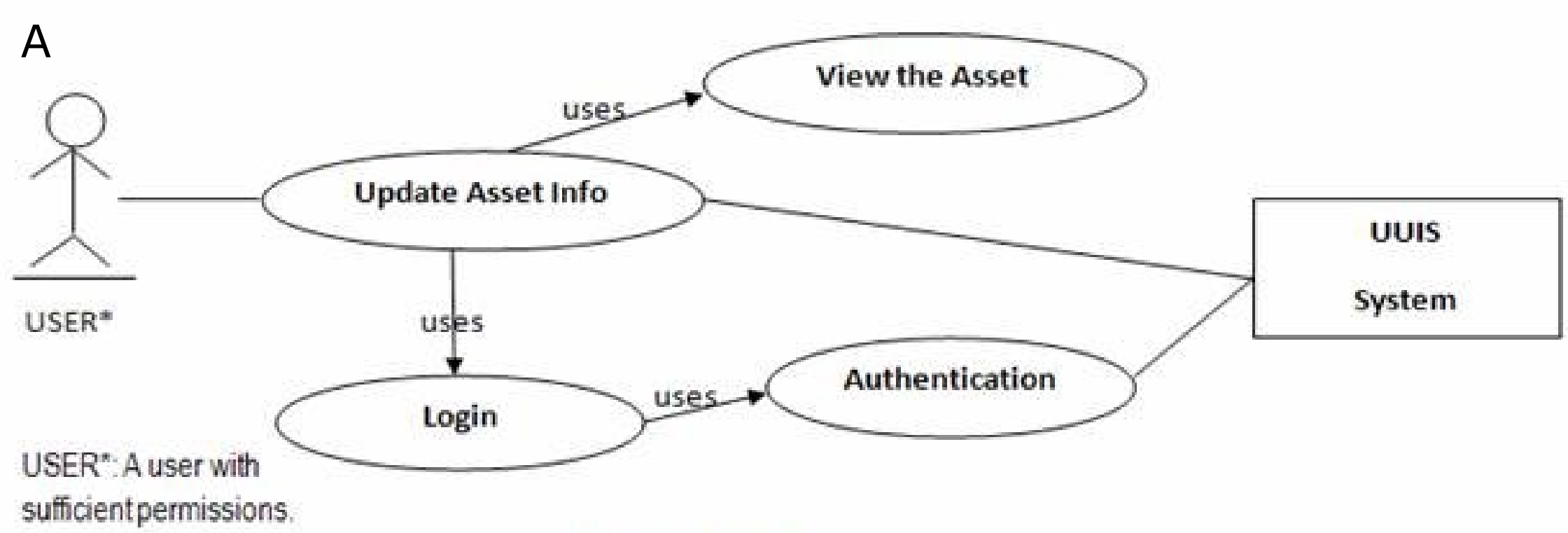

B

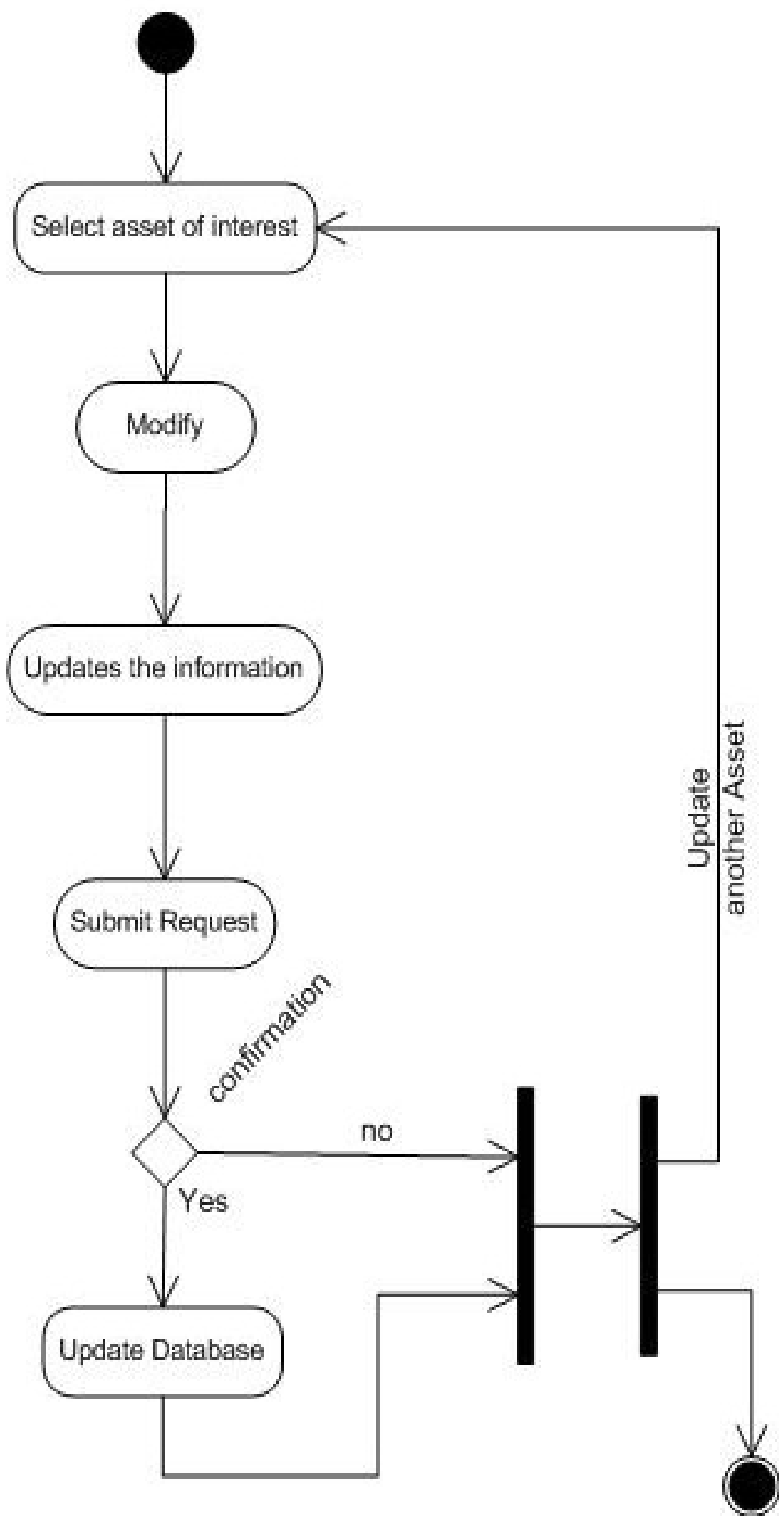

Figure 2.3.3 Update Asset(s) Information. A view of the function is shown with a use-case diagram (panel A) and an activity diagram (panel B).

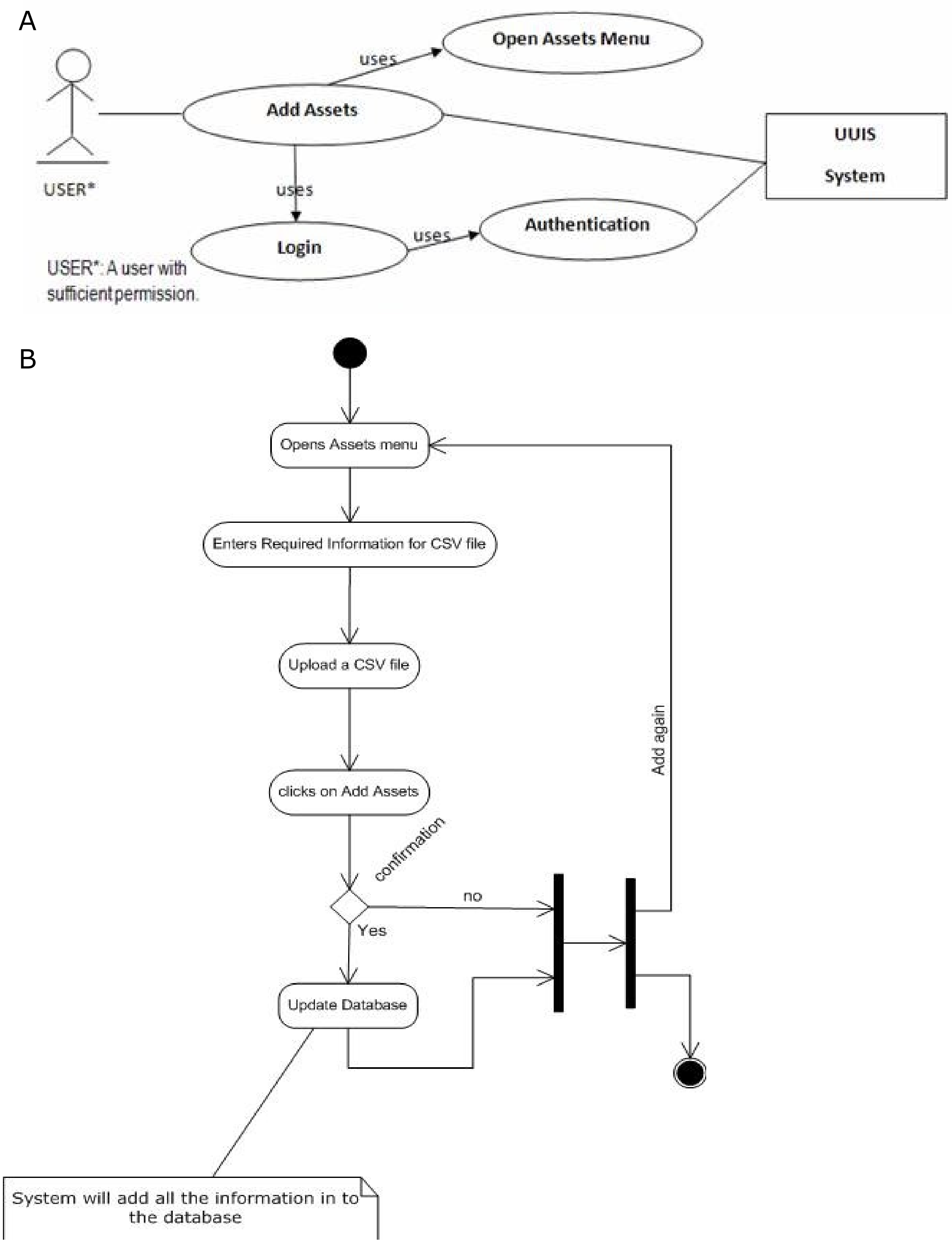

Figure 2.3.4 Bulk Add Assets. A view of the function is shown with a use-case diagram (panel A) and an activity diagram (panel B).

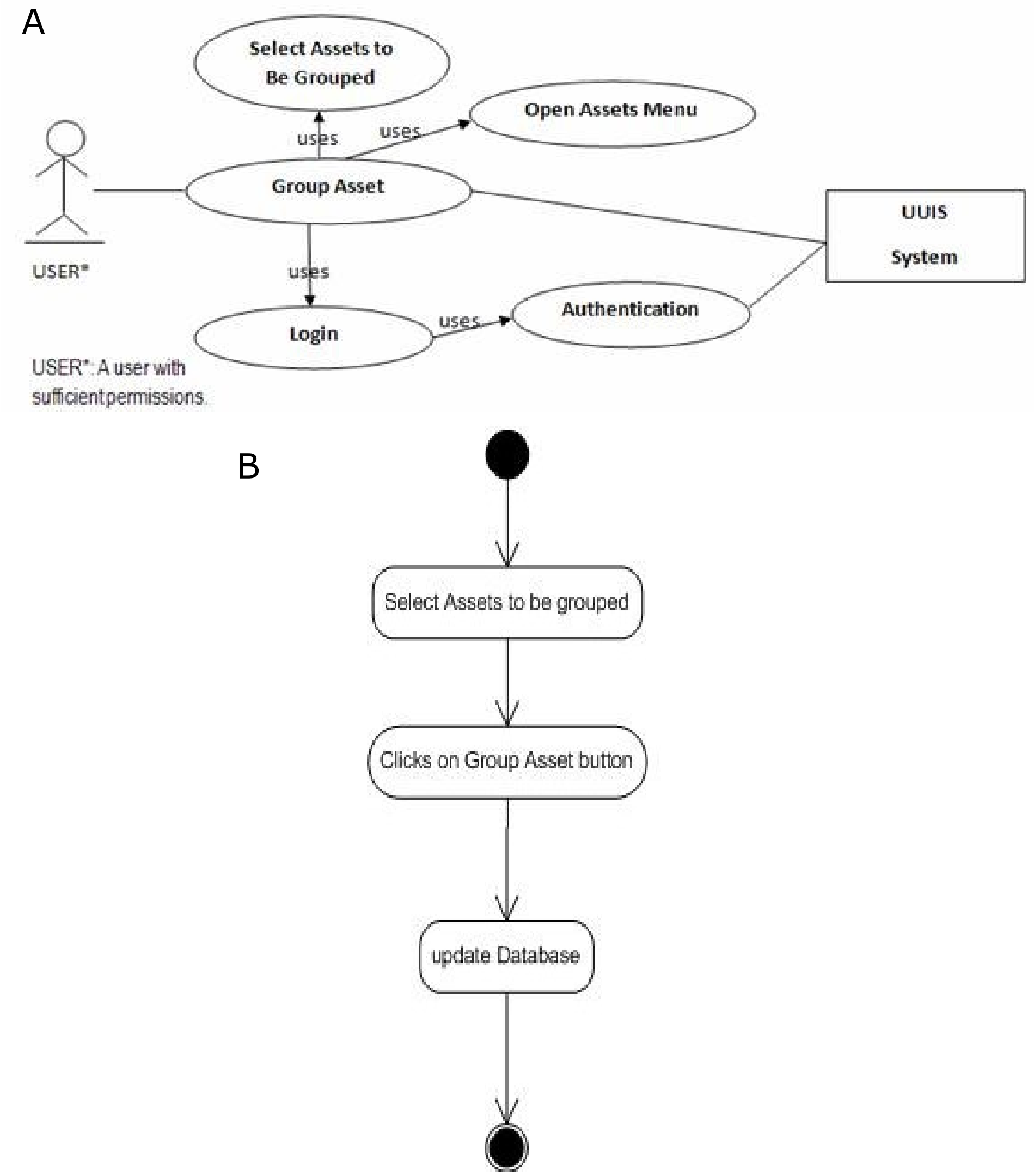

Figure 2.3.5. Group assets
A view of the function is shown with a use-case diagram (panel A) and an activity diagram (panel B).

## 2.4. Review Options

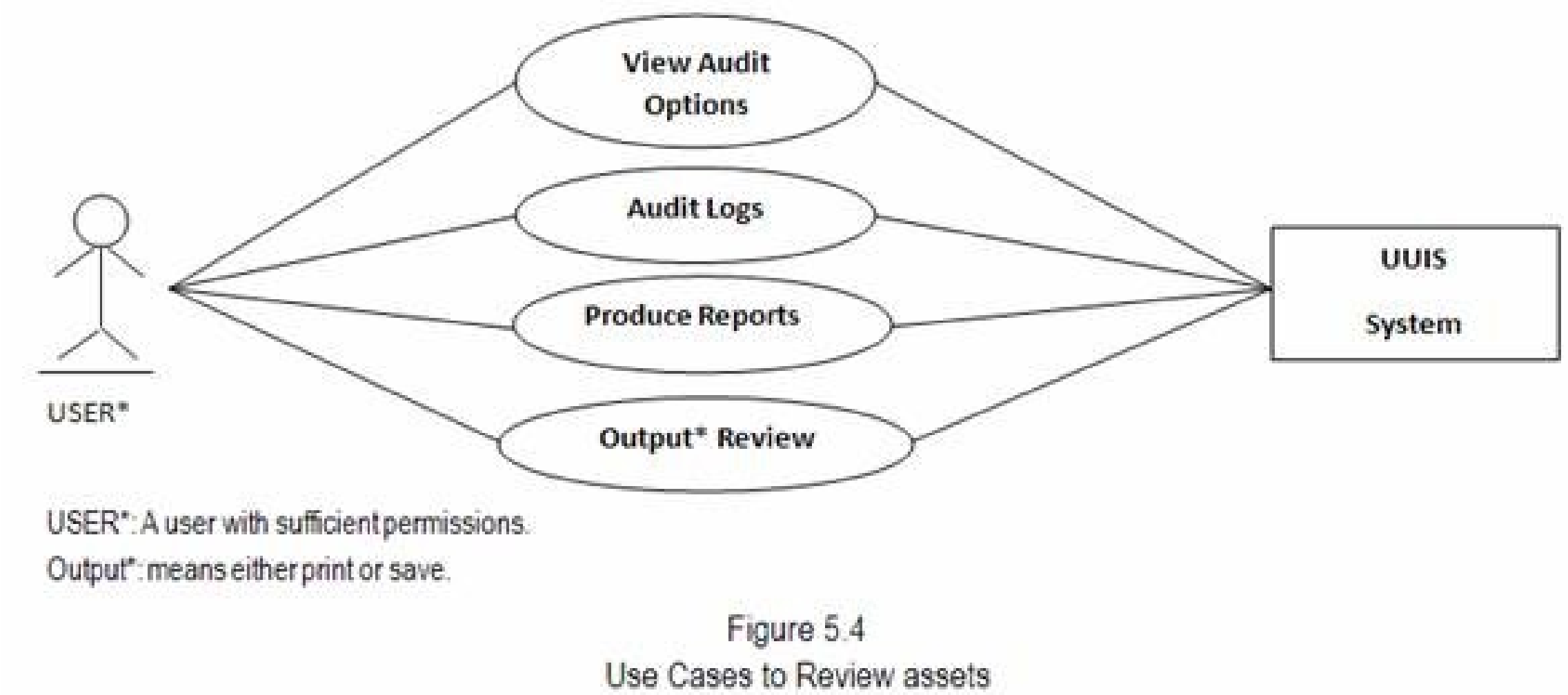

Figure 2.4.0.1 Use Cases to Review. This use-case diagram depicts the different review options.

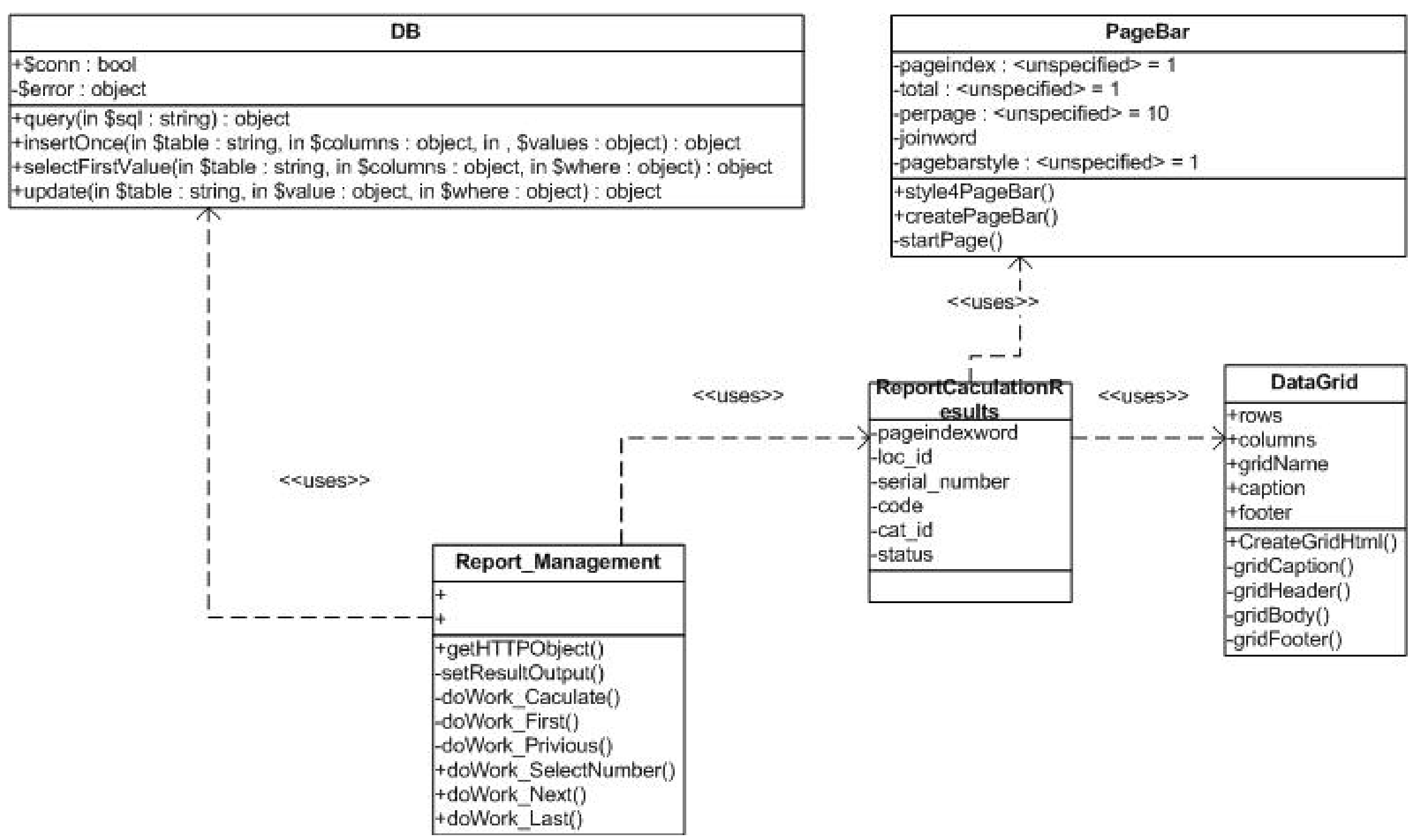

Figure 2.4.0.2 Classes in the Review Module. The above diagram illustrates the classes in the module and their relationships.

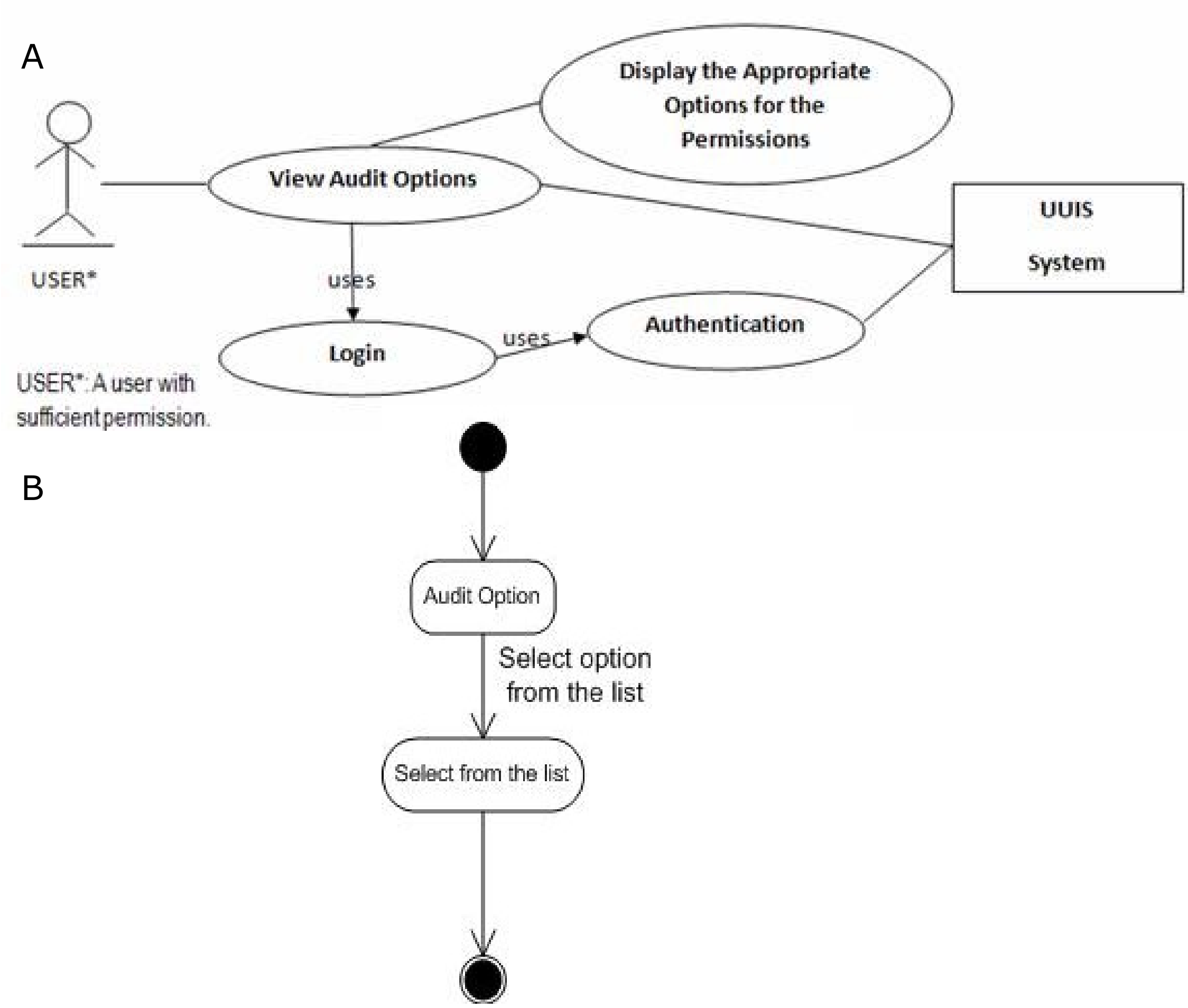

Figure 2.4.1. View Audit Options. A view of the function is shown with a use-case diagram (panel A) and an activity diagram (panel B).

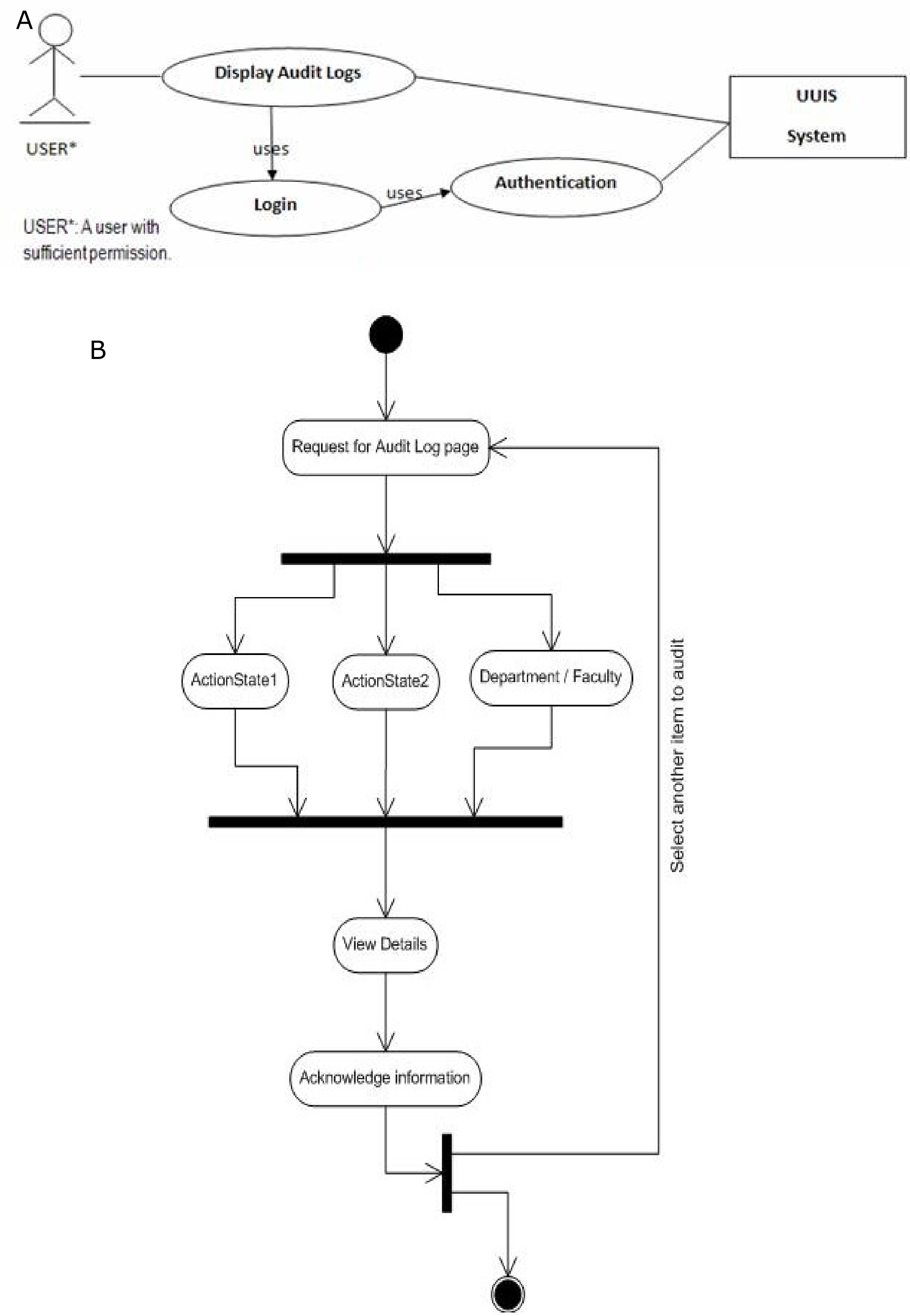

Figure 2.4.2 Audit Logs. A view of the function is shown with a use-case diagram (panel A) and an activity diagram (panel B). The use-case diagram emphasizes the importance of ensuring the user privileges.

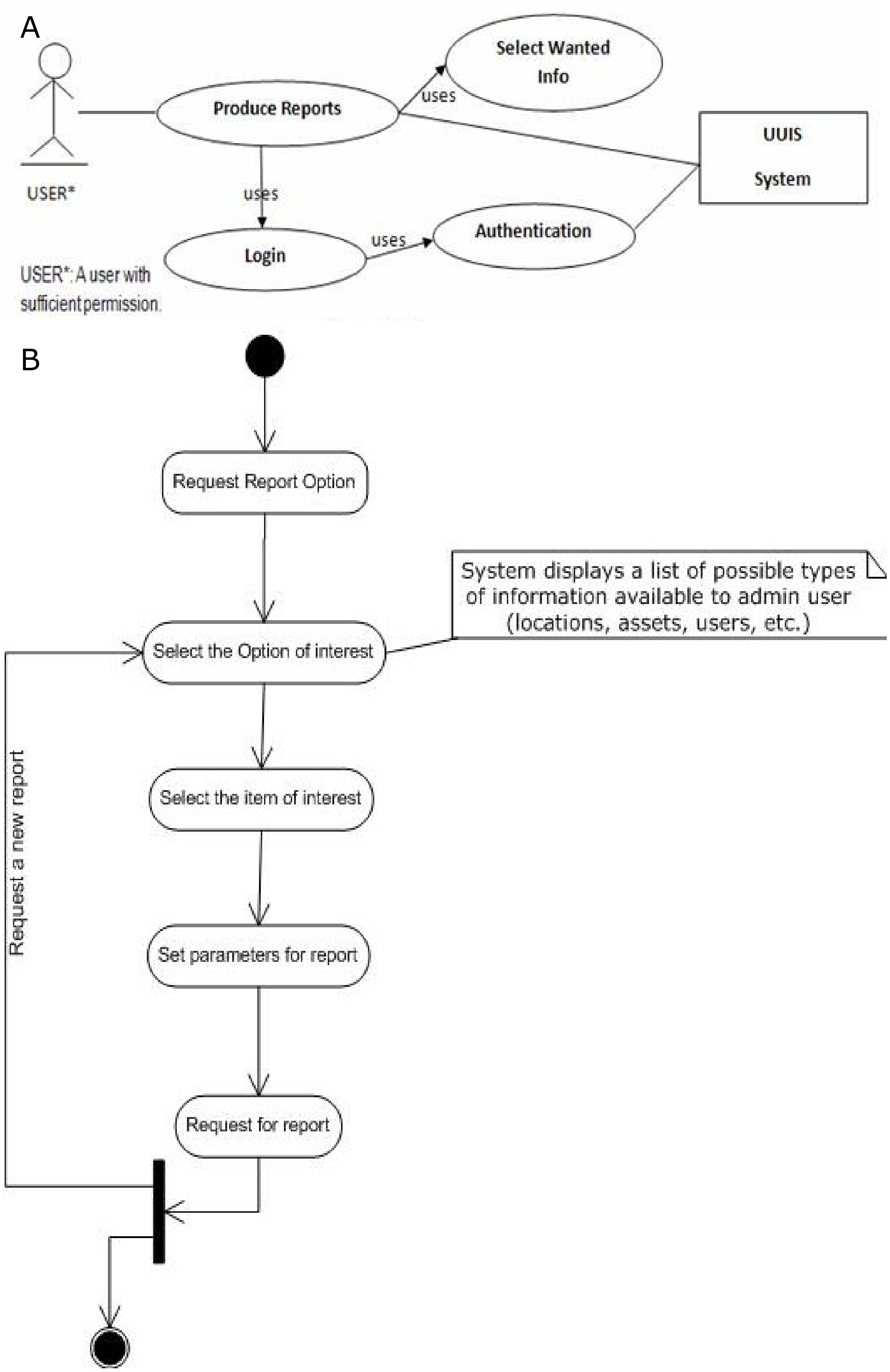

Figure 2.4.3 Produce Reports. A view of the function is shown with a use-case diagram (panel A) and an activity diagram (panel B).

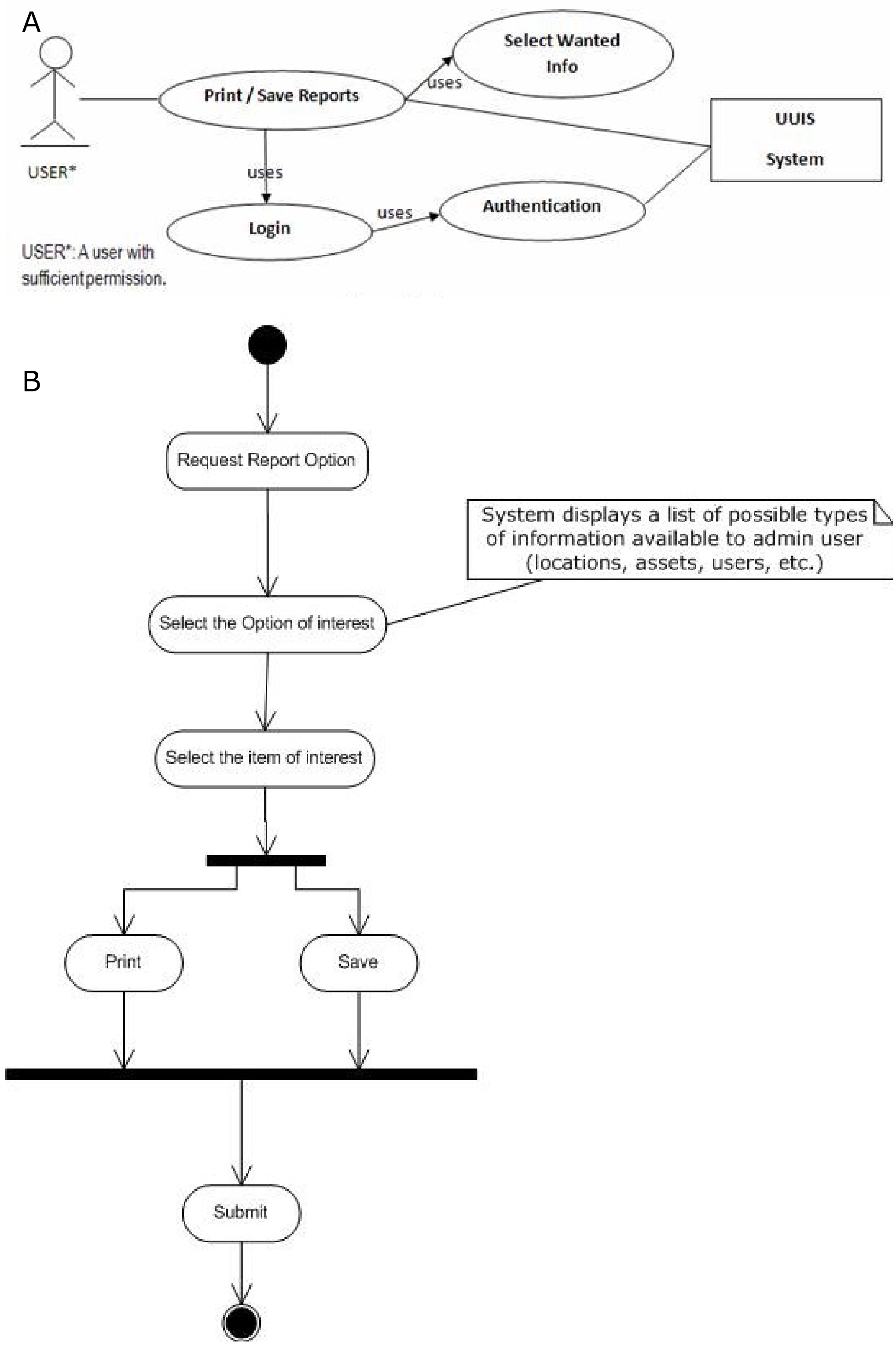

Figure 2.4.4 Output review. A view of the function is shown with a use-case diagram (panel A) and an activity diagram (panel B).

## 2.5. Error Management

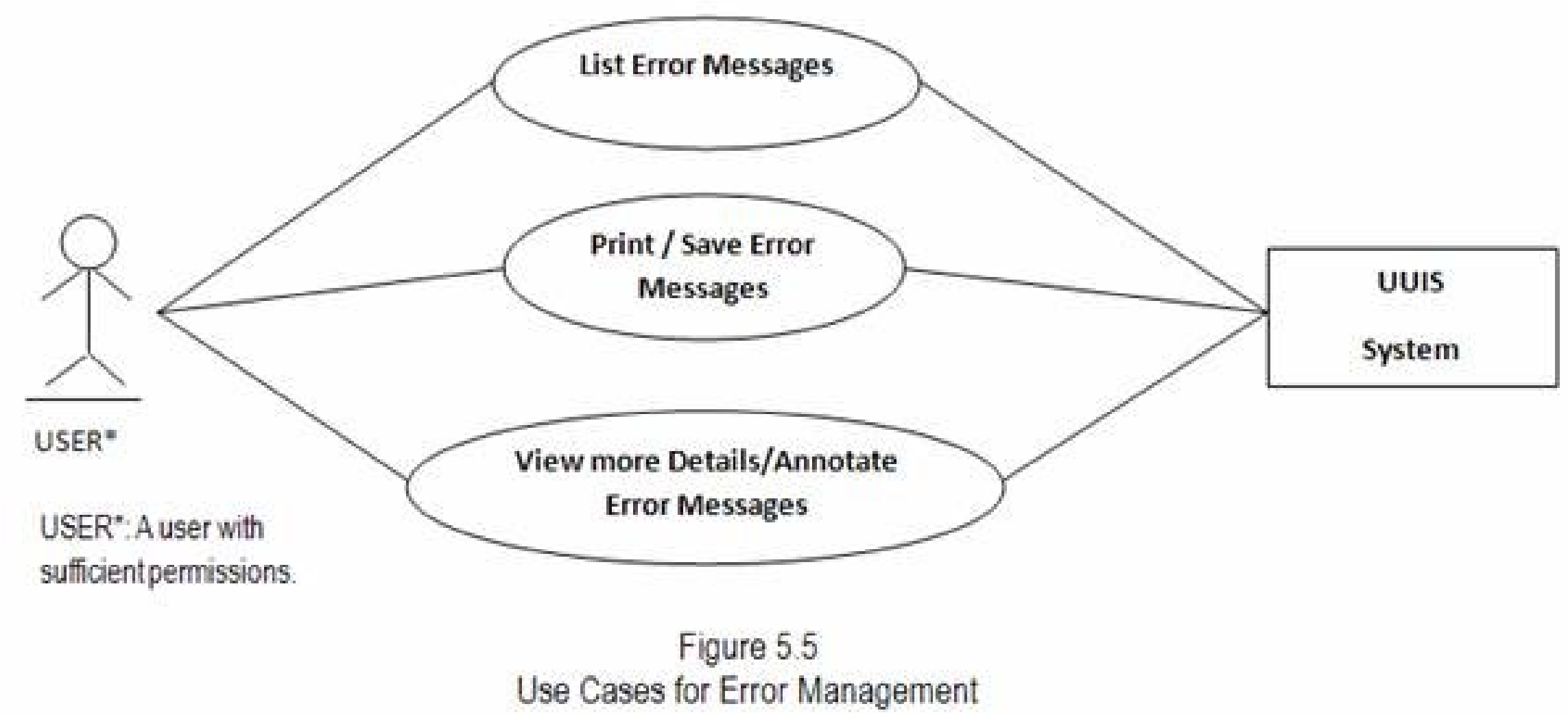

Figure 2.5.0.1 Use Cases for Error Management. The use-case diagram shown above describes the options for error management.

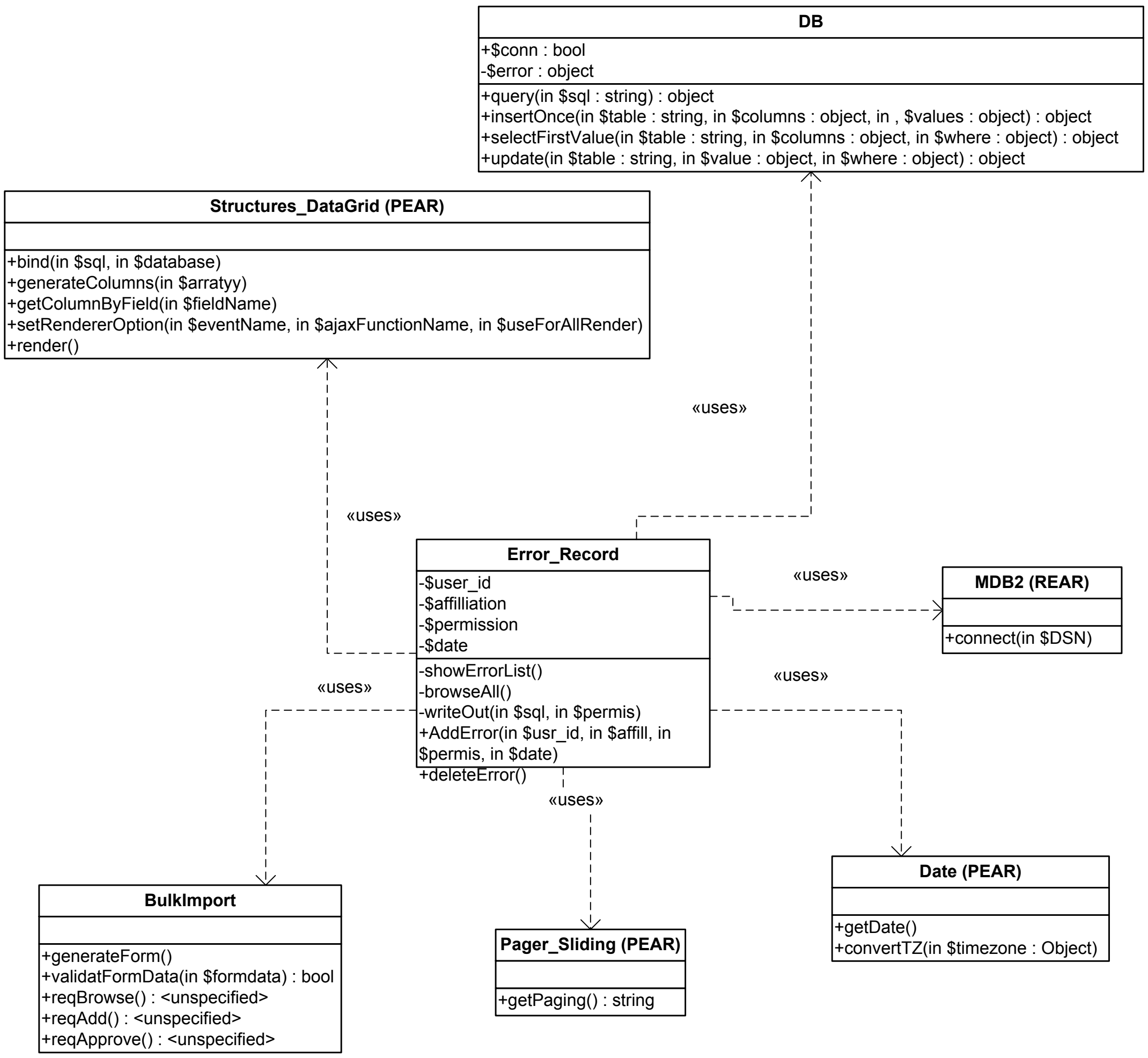

Figure 2.5.0.2 Class Diagram for Error Management Module. The various classes involved in the "Error Management" module are shown above.

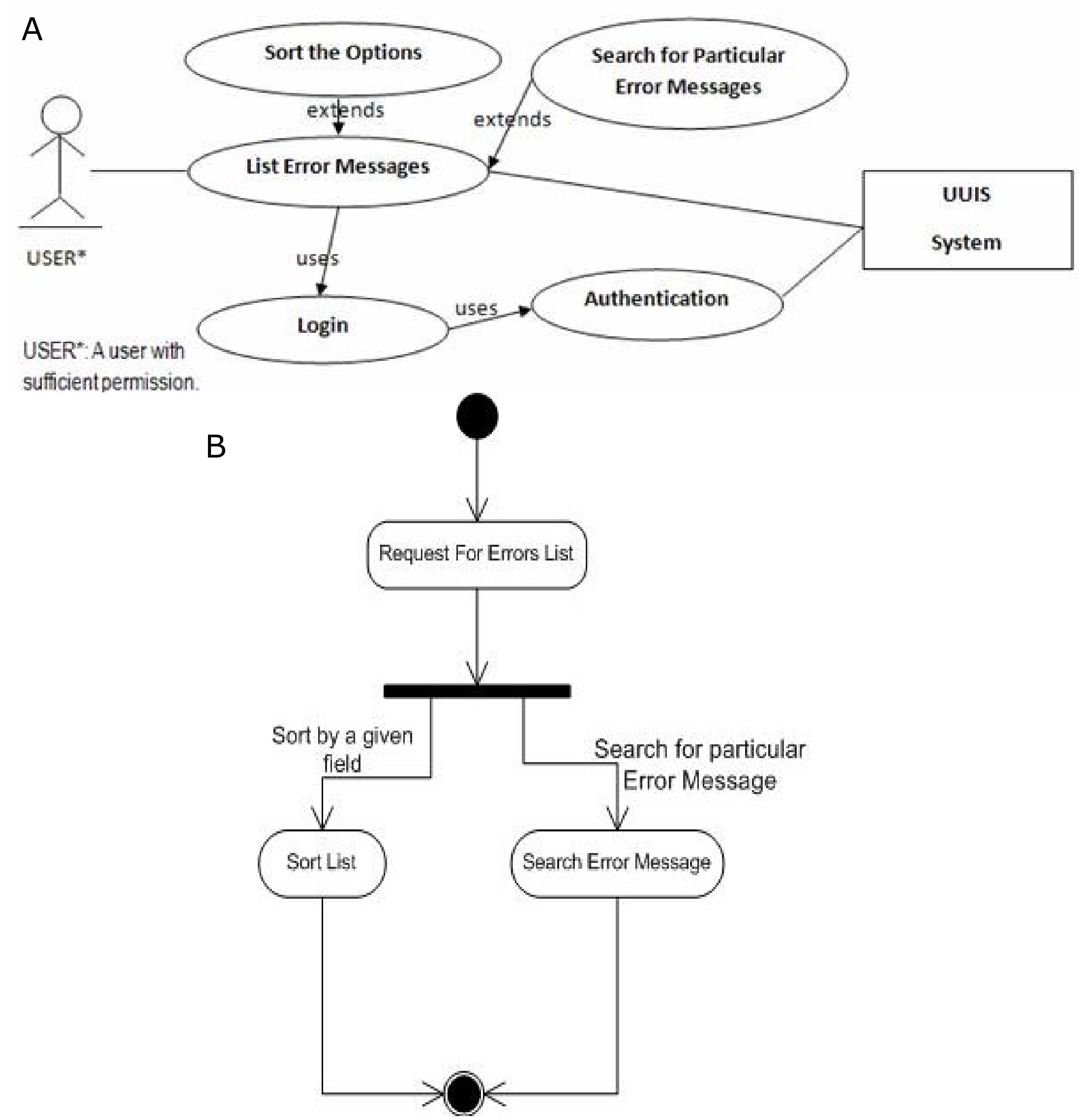

Figure 2.5.1 List Error Messages Based on Searching Conditions. A view of the function is shown with a use-case diagram (panel A) and an activity diagram (panel B).

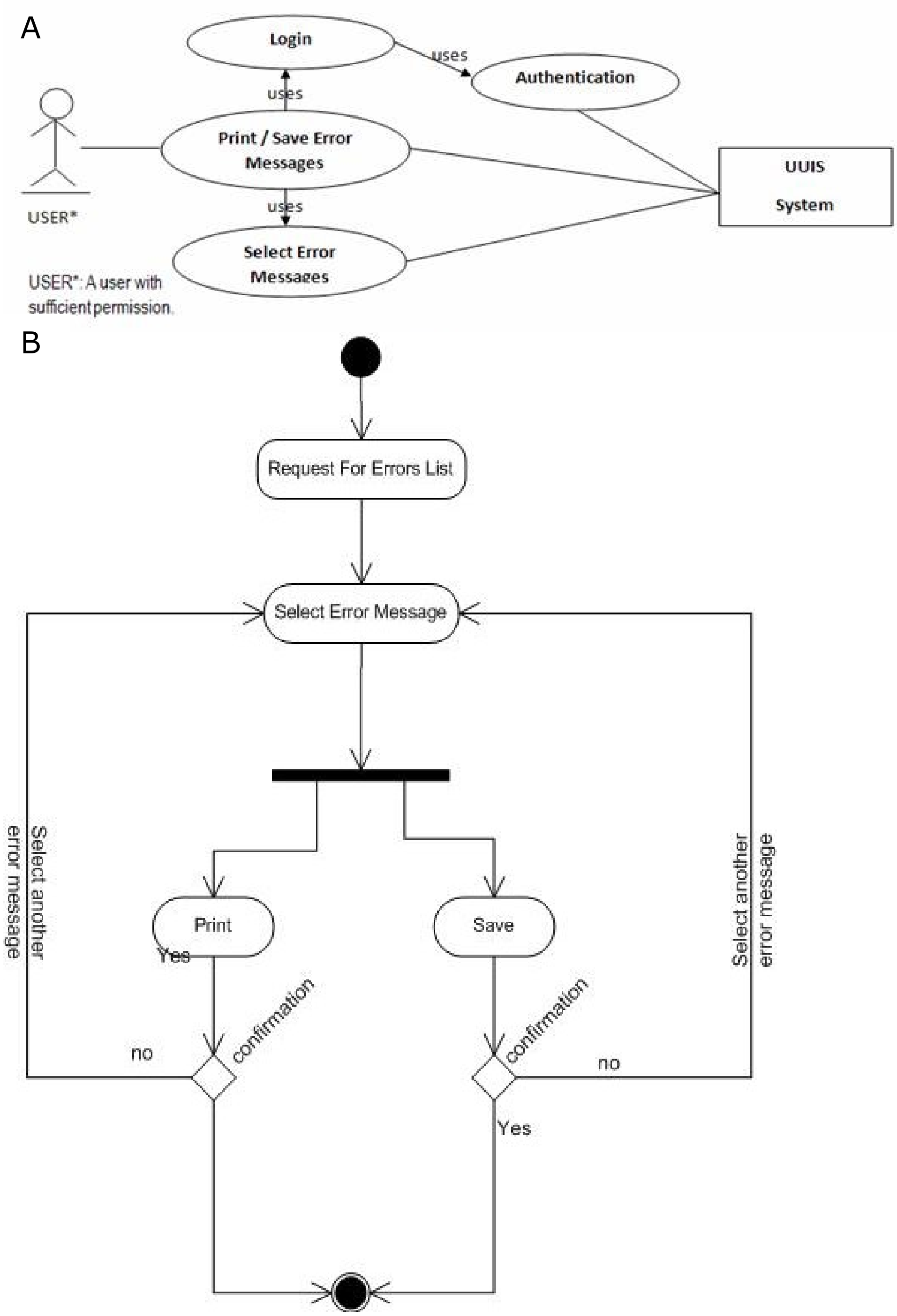

Figure 2.5.2. Print error messages. A view of the function is shown with a use-case diagram (panel A) and an activity diagram (panel B).

A

B

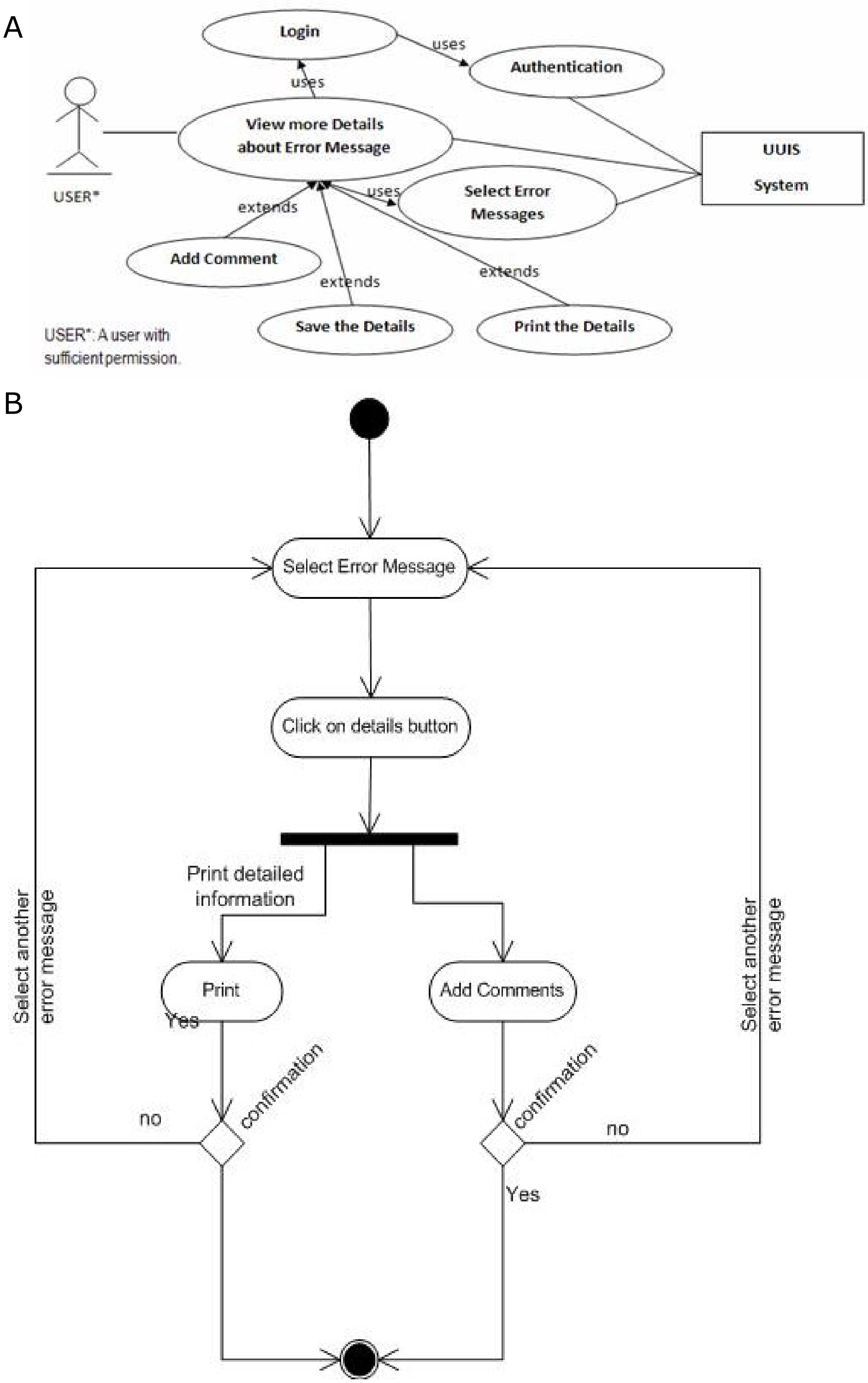

Figure 2.5.3. View More Details/Annotate Error Messages. A view of the function is shown with a use-case diagram (panel A) and an activity diagram (panel B).

### 2.6. Request Management

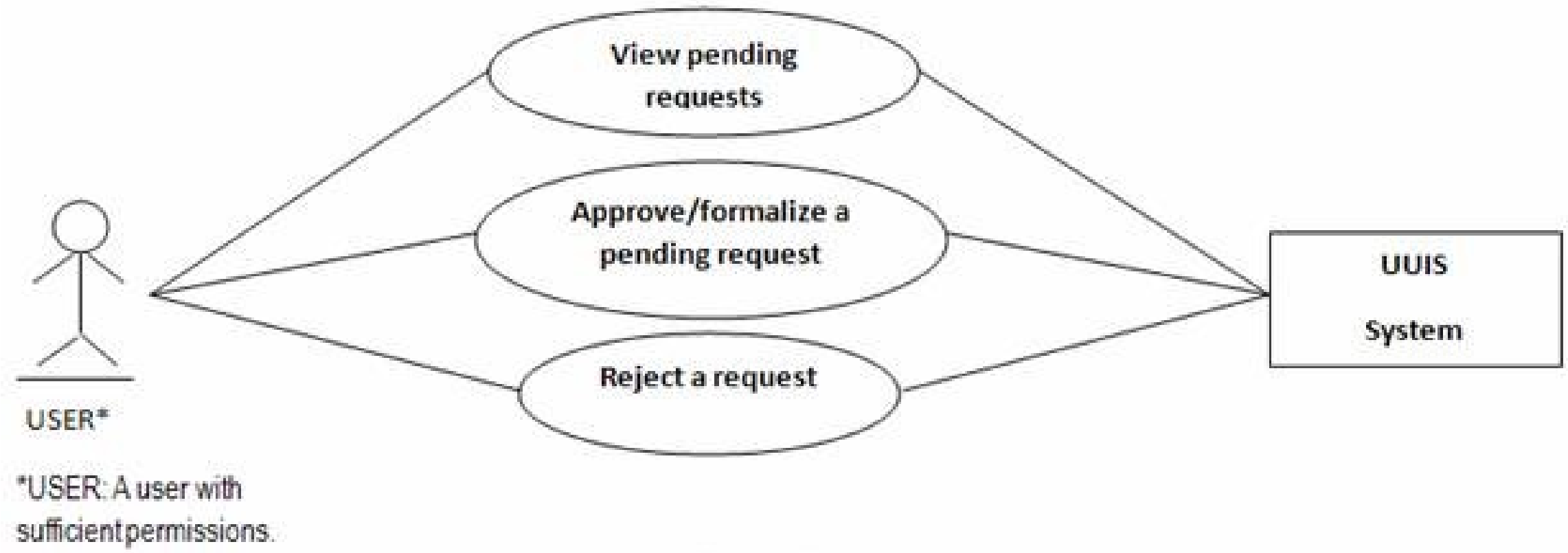

Figure 5.6
Use Cases for Request Management

Figure 2.6.1 Use Cases for Request Management. The use-case diagram above lists the functions provided by the “request management” module.

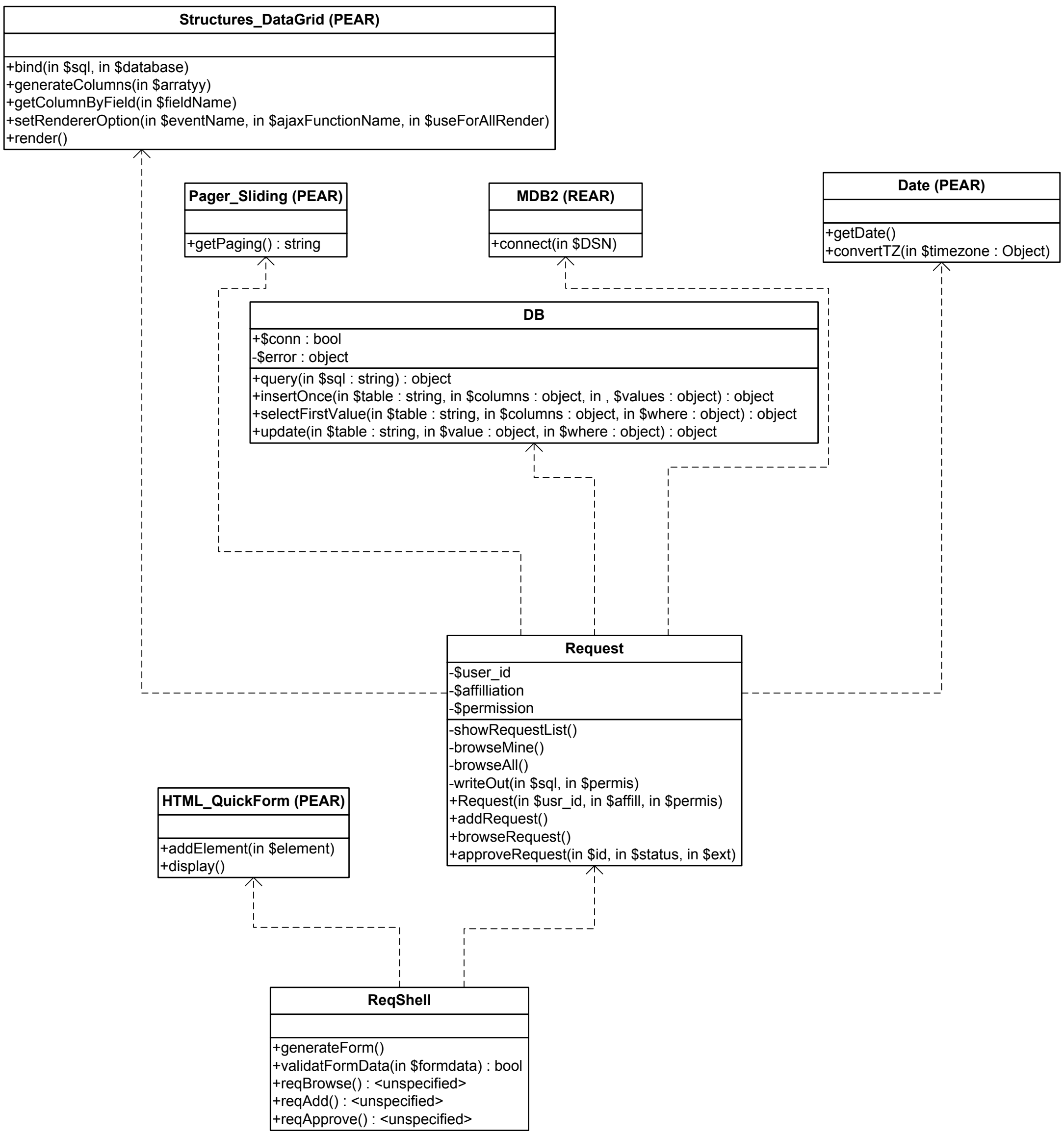

Figure 2.6.2 Request Management Classes. The relationships between the various classes of the “Request Management” module are shown in the above class diagram.

A

B

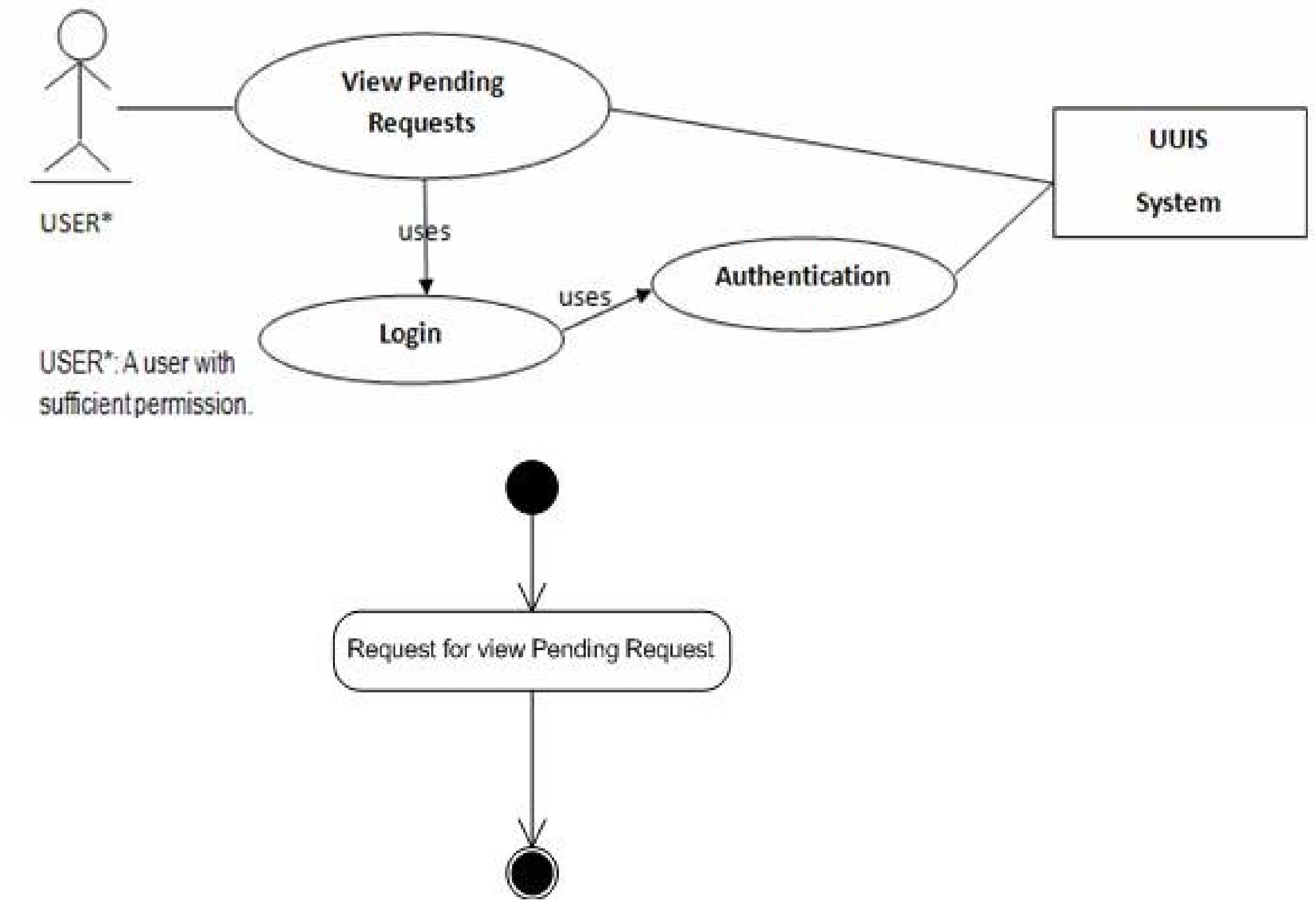

Figure 2.6.1 View Pending Requests. A view of the function is shown with a use-case diagram (panel A) and an activity diagram (panel B).

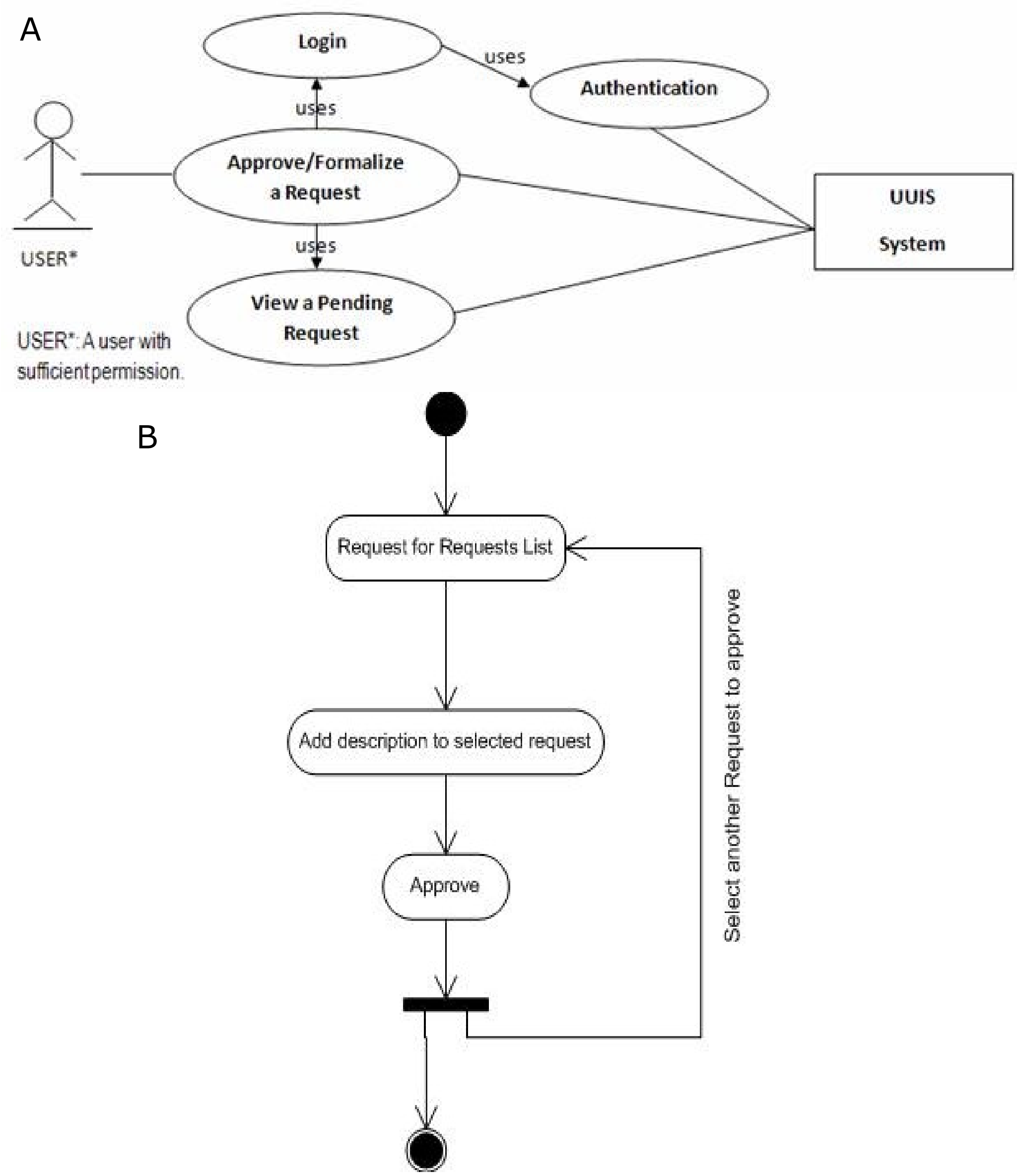

Figure 2.6.2 Approve/Formalize a Pending Request. A view of the function is shown with a use-case diagram (panel A) and an activity diagram (panel B).

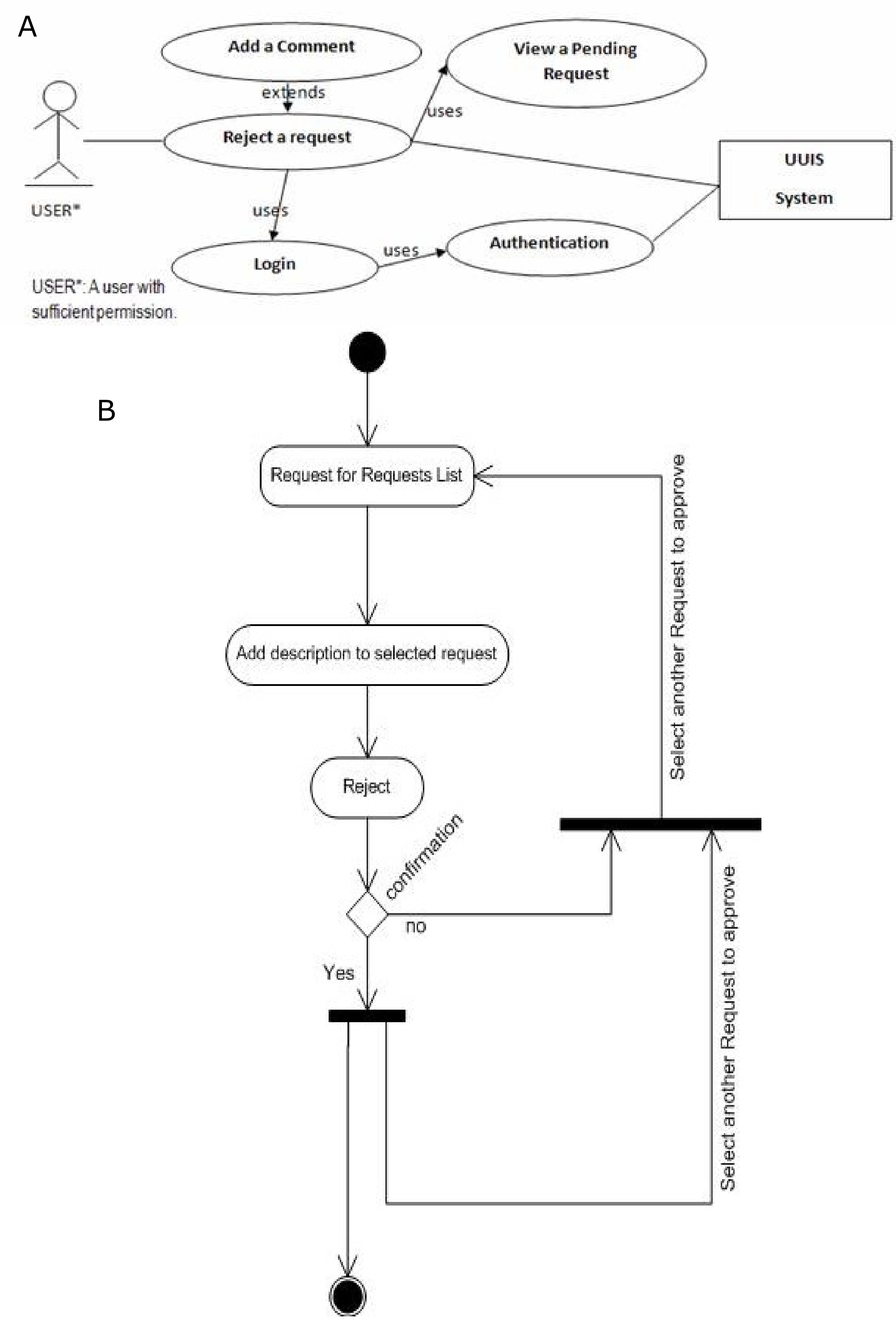

Figure 2.6.3 Reject a Request. A view of the function is shown with a use-case diagram (panel A) and an activity diagram (panel B).

**3. Database Layer**

The database consists of a total of twenty tables. For more details such as a thorough description of each field (data type, unique, etc.) or for a general view of the role each table occupies, please refer to the data dictionary found in the SRS document.

It should be noted that the design for permissions was not captured in the diagrams below. One long integer ($2^{16}$) would store the permission signature wherever applicable. The permission signature is a set of sixteen bits, and a table ("Permissions") would store the relevance of each bit.

In this manner, it is possible to fine-tune the privileges afforded to individuals. For instance, it may be desirable to allow students designated as "archivists" to view assets in a department, but not the requests submitted to the department, while other students may be "secretarial assistants" involved in expediting the request approval process. In this scenario, changing the appropriate bit would be sufficient to ensure that the students receive the appropriate privileges without leaking more information than necessary.

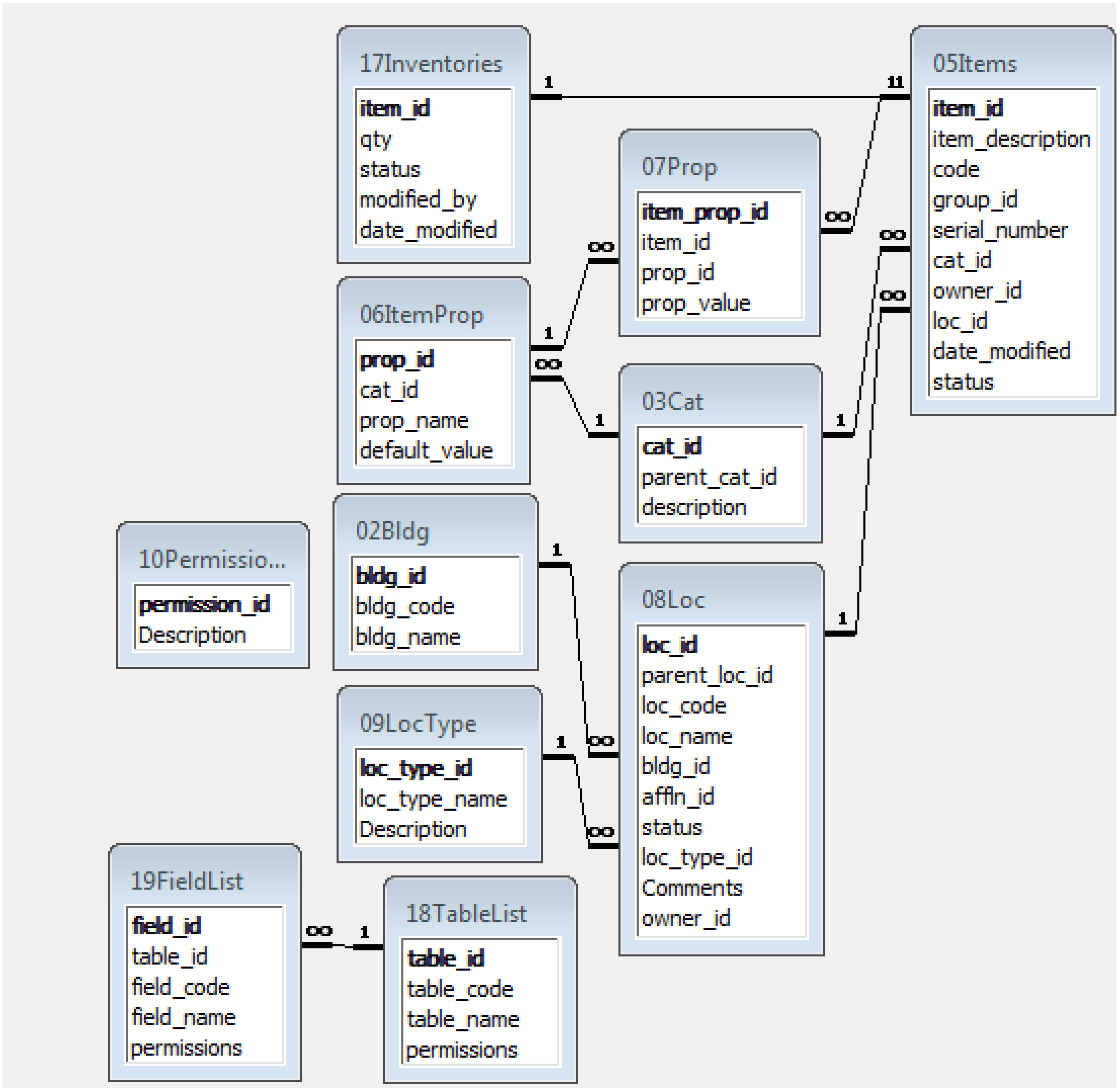

Figure 3.0.1 Items and Locations. The relationships diagram shown above illustrates the relationship between selected tables. Of note are the tables for storing item (asset) properties. Each item belongs in a category (for instance, "desktop"), stored in one table ("cat"). The properties relevant to the category (such as "IP address") would be listed in another table ("item_prop"), and each of the properties would be applied to the item as necessary in another table ("prop").

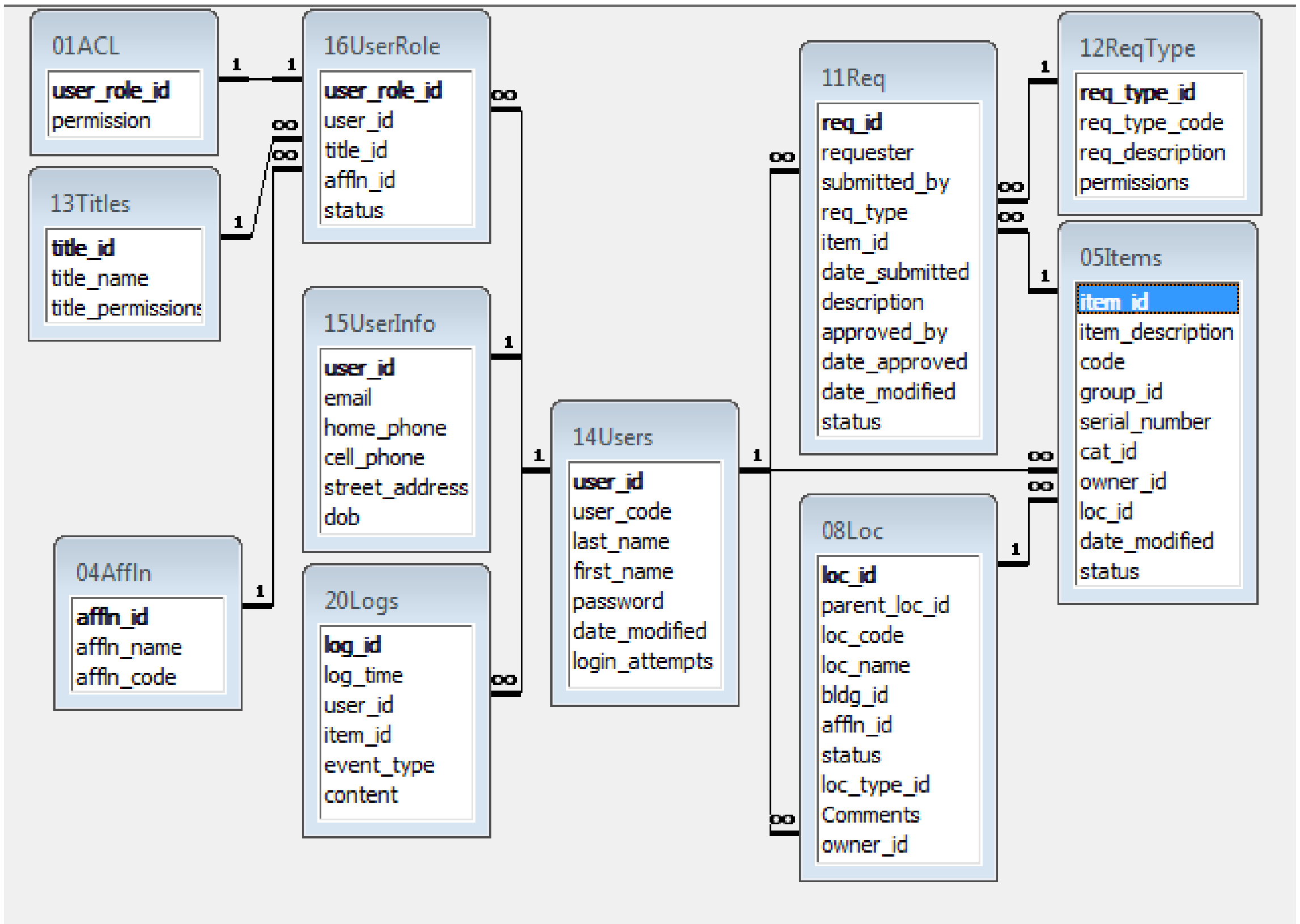

Figure 3.0.2 Users. The above relationships diagram shows the remaining tables in the database, along with two duplicates (locations and items). "Title" refers to the professional title; "affiliation" refers to either Department, Faculty or University, depending on the user. The table "logs" records every transaction the database commits.

## Appendix I: Test Cases [2]

1 **Test cases Common to All UUIS users**
- 1.1 login
- 1.2 logout
- 1.3 change password
- 1.4 view/edit personal information
- 1.5 submit a request
- 1.6 view request status
- 1.7 cancel a request
- 1.8 search
  - 1.8.1 Search for data
  - 1.8.2 Print/save search results

2 **Test cases for University Management**
- 2.1 Create a Department
- 2.2 Create a Faculty
- 2.3 Add a location
- 2.4 Back-up database
- 2.5 Bulk import users from a CSV file
- 2.6 Update user profile

3 **Test cases to Manage Assets**
- 3.1 View assets
- 3.2 Add asset
- 3.3 Update asset(s) information
- 3.4 Bulk add assets
- 3.5 Group assets

4 **Test cases to Review assets**
- 4.1 View audit options
- 4.2 Audit logs
- 4.3 Produce reports
- 4.4 Output review

5 **Test cases for Error Management**
- 5.1 List error messages
- 5.2 Print error messages
- 5.3 View more details/annotate error messages

6 **Test cases for Request Management**
- 6.1 View pending requests
- 6.2 Approve/formalize a pending request
- 6.3 Reject a request

7 **Load Testing**

8 **Browser Consistency Test**

### 1.1 Login

| | |
|---|---|
| **Test Case Number** | 1.1.1 |
| **Test Case Name** | Login (Positive) |
| **Test Case Description** | This test case verifies whether a user can be properly logged in to the System. |
| **Preconditions** | N/A |

| **Process Description** | |
|---|---|
| **Step** | **Action** |
| 1 | User inputs valid userid and password. |
| 2 | System validates userid and password. |
| 3 | System show welcome page corresponding to the user. |
| **Test Case Number** | 1.1.2 |
| **Test Case Name** | Login (Negative) |
| **Test Case Description** | This test case verifies whether unauthorized users are restricted from accessing the system. |
| **Preconditions** | N/A |

| **Process Description** | |
|---|---|
| **Step** | **Action** |
| 1 | User inputs invalid userid and password. |
| 2 | System attempts to validate user. |
| 3 | System redirects user back to login page. |
| 4 | System administrator verifies whether the user is not logged in. |
| **Test Case Number** | 1.1.3 |
| **Test Case Name** | Login (Lock) |
| **Test Case Description** | This test case verifies whether a user cannot attempt more than three consecutive failed attempts at login with a given password. |
| **Preconditions** | N/A |

| **Process Description** | |
|---|---|
| **Step** | **Action** |
| 1 | User inputs a valid userid with an invalid password. |
| 2 | System attempts to validate user. |
| 3 | System redirects user back to login page. |
| 4 | User repeats steps 1-3 (inclusive) two more times. |
| 5 | System locks the userid and redirects user to the appropriate error message. |
| 6 | System administrator verifies whether the information is stored correctly in the database. |

### 1.2 Logout

| | |
|---|---|
| **Test Case Number** | 1.2 |
| **Test Case Name** | Logout |
| **Test Case Description** | This test case verifies whether the current user can log out of the system and verifies whether he/she can no longer access data |
| **Preconditions** | The user is logged in |
| **Process Description** | |
| **Step** | **Action** |
| 1 | User selects option to log out of system. |
| 2 | System updates the system. |
| 3 | System redirects user to login page. |
| 4 | System administrator verifies whether the system has been properly updated. |

### 1.3 Change Password

| | |
|---|---|
| **Test Case Number** | 1.3.1 |
| **Test Case Name** | Change Password (Proper Values) |
| **Test Case Description** | This test case verifies whether the current user can change his/her password |
| **Preconditions** | The user is logged in |
| **Process Description** | |
| **Step** | **Action** |
| 1 | User selects option to change password. |
| 2 | System redirects the current user to the “Change Password” page. |
| 3 | User completes fields with valid input. |
| 4 | System updates the system. |
| 5 | System redirects user to the appropriate “Welcome” page. |
| **Test Case Number** | 1.3.2 |
| **Test Case Name** | Change Password (Incorrect Password) |
| **Test Case Description** | This test case verifies whether the current user may change password with an incorrect confirmation. |
| **Preconditions** | The user is logged in |
| **Process Description** | |

| Step | Action |
|---|---|
| 1 | User selects option to change password. |
| 2 | System redirects the current user to the "Change Password" page. |
| 3 | User completes fields with incorrect password for authorization but valid new passwords. |
| 4 | System fails to update the password. |
| 5 | System notifies user of error and redirects user to the "Change Password" page. |
| 6 | User attempts Test Case 1.3.1 to verify that the new password has not been entered in the system. |

### 1.4 View Personal Information

| | |
|---|---|
| **Test Case Number** | 1.4 |
| **Test Case Name** | View Personal Information |
| **Test Case Description** | This test case verifies whether the current user can view/edit his/her personal information |
| **Preconditions** | The user is logged in |
| **Process Description** | |

| Step | Action |
|---|---|
| 1 | User selects option to view/edit personal information. |
| 2 | System redirects the current user to view/edit personal information page. |
| 3 | User verifies whether the information displayed is correct. |

### 1.5 Submit a Request

| | |
|---|---|
| **Test Case Number** | 1.5 |
| **Test Case Name** | Submit Request |
| **Test Case Description** | This test case verifies whether the user in all levels can submit request |
| **Preconditions** | User is logged in |
| **Process Description** | |

| Step | Action |
|---|---|
| 1 | User fills out the request form correctly. |
| 2 | System saves the request to database. |
| 3 | System returns user to the "Request" page and lists the new request. |
| 4 | System administrator verifies whether the new request has been recorded properly |

### 1.6 View Request Details

| | |
|---|---|
| **Test Case Number** | 1.6 |
| **Test Case Name** | View Request Status (User-Submitted Requests) |
| **Test Case Description** | This test case verifies whether all users can view request status for requests submitted by user. |
| **Preconditions** | User is logged in<br>User has previously submitted requests |
| **Process Description** | |

| Step | Action |
|---|---|
| 1 | User clicks request management button |
| 2 | System shows the "Request" page. |
| 3 | User selects a request of interest. |
| 4 | System displays request details, including status. |
| 5 | User confirms with system administrator whether the information displayed is correct. |

### 1.7 Cancel a Request

| | |
|---|---|
| **Test Case Number** | 1.7 |
| **Test Case Name** | Cancel a Request |
| **Test Case Description** | This test case verifies whether all users may cancel their own requests |
| **Preconditions** | User is logged in |
| **Process Description** | |
| **Step** | **Action** |
| 1 | User views request details as described in test case 1.6, steps 1-4 (inclusive). |
| 2 | User clicks the "Cancel Request" button. |
| 3 | System updates the information in the database. |
| 4 | System redirects user to request list page. |
| 5 | User verifies whether request is cancelled. |
| 6 | System administrator verifies whether request is cancelled in the database. |

## 1.8 Search for Data

### 1.8.1. Basic Search

| | |
|---|---|
| **Test Case Number** | 1.8.1.1 |
| **Test Case Name** | Capture of Input String (Basic Search) |
| **Test Case Description** | This test case verifies whether the System is able to record the input string for subsequent uses (such as translating the input string into an SQL query). |
| **Preconditions** | N/A |
| **Process Description** | |
| **Step** | **Action** |
| 1. | User inputs valid String for Search. |
| 2. | System accepts the string along with appropriate information (such as user permission level and affiliation). |
| 3. | System locally displays the input string from the search box and all other pieces of data collected. |
| 4. | User verifies whether all data collected are correct. |
| **Test Case Number** | 1.8.1.2 |
| **Test Case Name** | SQL Query for Basic Search |
| **Test Case Description** | This test case verifies the generation of an SQL query for a Basic Search. |
| **Preconditions** | N/A |
| **Process Description** | |

| Step | Action |
| --- | --- |
| 1. | User enters query as in Test Case 1.8.1.1. |
| 2. | System accepts the string along with all relevant information (c.f. 1.8a, Step 2) and translates it into an SQL query. |
| 3. | System displays the SQL query. |
| 4. | User verifies whether the SQL query is valid. |
| **Test Case Number** | 1.8.1.3 |
| **Test Case Name** | Data Retrieval for Basic Search |
| **Test Case Description** | This test case verifies the validity of the Basic Search. |
| **Preconditions** | Users with permission level 1, 2 and 3 are required, preferably with users from different affiliations. |
| **Process Description** | |
| **Step** | **Action** |
| 1. | User enters query as in Test Case 1.8.1.1. The input string will contain a string corresponding to at least one record. |
| 2. | System accepts the string along with all relevant information (c.f. 1.8a, Step 2) and translates it into an SQL query. |
| 3. | System performs the SQL search. |
| 4. | System locally displays the record(s) retrieved. |
| 5. | User verifies whether the record(s) retrieved are correct. |
| 6. | User repeats steps 1 through 4 (inclusive), but using a nonsense string (one for which it is known in advance to correspond to no record). |
| 7. | System retrieves no data and displays results locally. |
| 8. | User confirms that no data was retrieved. |
| **Test Case Number** | 1.8.1.4 |
| **Test Case Name** | Basic Search |
| **Test Case Description** | This test case verifies the validity of the Basic Search. |
| **Preconditions** | Users with permission level 1, 2 and 3 are required, preferably with users from different affiliations. |
| **Process Description** | |

| Step | Action |
|---|---|
| 1. | User enters query as in Test Case 1.8.1.3, steps 1 through 4 (inclusive). |
| 2. | System performs search and outputs results. |
| 3. | User verifies results and performs a query as in Test Case 1.8.1.3, step 6. |
| 4. | System performs search and outputs results to the user's machine. |
| 5. | User verifies whether the output format is correct, and also verifies whether search results are correct. |

### 1.8.2 Advanced Search

| | |
|---|---|
| **Test Case Number** | 1.8.2.1 |
| **Test Case Name** | Capture of a Search Parameter (Advanced Search) |
| **Test Case Description** | This test case verifies whether the System is able to capture a single search parameter. |
| **Preconditions** | 1. User clicks the "Advanced Search" tab in the Search Page.<br>2. System shows the "Advanced Search" page. |
| **Process Description** | |

| Step | Action |
|---|---|
| 1. | User inputs a valid parameter for Advanced Search. |
| 2. | System accepts the parameter along with appropriate information (c.f. 1.8.1.1, Step 2). |
| 3. | System locally displays the parameter from the search boxes for Advanced Search, along with the extraneous information collected. |
| 4. | User verifies all data displayed. |
| **Test Case Number** | 1.8.2.2 |
| **Test Case Name** | Capture of a Search Parameter (Advanced Search) |
| **Test Case Description** | This test case verifies whether the System is able to concatenate search parameters into a single query. |
| **Preconditions** | 1. User clicks the "Advanced Search" tab in the Search Page.<br>2. System shows the "Advanced Search" Page. |

| Process Description | |
|---|---|
| **Step** | **Action** |
| 1 | User inputs a valid parameter for Advanced Search. |
| 2 | System accepts the parameter. |
| 3 | User inputs the next (valid) parameter for Advanced Search, along with the appropriate logical expression. |
| 4. | System accepts the parameter and concatenates the expression. |
| 5. | Steps 3 and 4 are repeated as often as the user wishes. If the logical expression requires further refinement (such as adding parentheses to remove ambiguity), the System will request such refinements from the user. |
| 6. | User completes the entry of search parameters for Advanced Search. |
| 7. | System collects all relevant data extraneous to the search parameters (c.f. 1.8.1.1, Step 2) and locally displays the query of concatenated search parameters for Advanced Search along with the other pieces of data collected. |
| 8. | User verifies all information collected is correct, and that concatenation is also correct. |

| | | |
|---|---|---|
| **Test Case Number** | 1.8.2.3 | |
| **Test Case Name** | SQL Query for Advanced Search | |
| **Test Case Description** | This test case verifies the generation of an SQL query for an Advanced Search. | |
| **Preconditions** | N/A | |
| **Process Description** | | |
| **Step** | **Action** | |
| 1. | User enters query as in Test Case 1.8.2.2. | |
| 2. | System accepts the string along with all relevant information (c.f. 1.8.1.1, Step 2) and translates it into an SQL query. | |
| 3. | System displays the SQL query locally. | |
| 4. | User verifies whether the SQL query is correct. | |

| | |
|---|---|
| **Test Case Number** | 1.8.2.4 |
| **Test Case Name** | Data Retrieval for Advanced Search |

| | | |
|---|---|---|
| **Test Case Description** | This test case verifies proper data retrieval during an Advanced Search. | |
| **Preconditions** | N/A | |
| **Process Description** | | |
| **Step** | **Action** | |
| 1. | User enters query as in Test Case 1.8.2.2, making sure that there exists at least one record corresponding to the search parameter(s). | |
| 2. | System accepts the search parameter(s) along with all relevant information (c.f. 1.8a, Step 2) and performs the search. | |
| 3. | System locally displays the search results. | |
| 4. | User verifies whether the record(s) retrieved is (are) correct. | |
| 5. | User enters query as in Test Case 2aii, making sure that no records correspond to the search parameter(s) | |
| 6. | System repeats steps 2 and 3. | |
| 7. | User verifies whether no record was retrieved. | |

| | |
|---|---|
| **Test Case Number** | 1.8.2.5 |
| **Test Case Name** | Advanced Search |
| **Test Case Description** | This test case verifies the validity of the Advanced Search. |
| **Preconditions** | N/A |
| **Process Description** | |
| **Step** | **Action** |
| 1. | User enters query as in Test Case 1.8.2.4 |
| 2. | System accepts the search parameter(s) along with all relevant information (c.f. 1.8.1.1, Step 2) and performs the search. |
| 3. | System outputs the search results to the user's machine. |
| 4. | User verifies whether the search results are correct and that output is also correct. |

### 1.8.3 Boundary Function Testing

The following tests are designed to ensure proper system behaviour when the search input is at or exceeds the conditions accepted by the system.

| | | |
|---|---|---|
| **Test Case Number** | 1.8.3.1 | |
| **Test Case Name** | Empty String Search –Basic Search | |
| **Test Case Description** | This test case verifies the behaviour Basic Search Function when no string is inputted. | |
| **Preconditions** | N/A | |
| **Process Description** | | |
| **Step** | **Action** | |
| 1. | User does not input a string for Search. | |
| 2. | System detects an empty string and displays error message locally. | |
| 3. | User verifies error message. | |
| **Test Case Number** | 1.8.3.2 | |
| **Test Case Name** | Long String Search –Basic Search | |
| **Test Case Description** | This test case verifies the Basic Search Function | |
| **Preconditions** | N/A | |
| **Process Description** | | |
| **Step** | **Action** | |
| 1. | User inputs a long string (>30 characters) for Search. Textbox ceases to accept characters after 30 have been entered | |
| 2. | System detects the long string and truncates the string if necessary. System performs the search on the truncated string. | |
| 3. | User verifies whether the string has been properly truncated. | |

| | |
|---|---|
| **Test Case Number** | 1.8.3.3 |
| **Test Case Name** | Long String Search –Advanced Search |
| **Test Case Description** | This test case verifies the Advanced Search Function |
| **Preconditions** | 1. User clicks the “Advanced Search” tab in the Search |
| | Page<br>2. System shows the "Advanced Search" Page |
| **Process Description** | |
| **Step** | **Action** |
| 1. | User inputs a long string (>30 characters) in the "Value" field for Advanced Search as in 1.8.3b, step 1, and enters the search parameter |
| 2. | System accepts the long string as in 1.8.3.2, step 2, and initializes the aggregation of search parameters. |
| 3. | User inputs a large number (>20) of search parameters (as in 1.8.2bii), each with a long string. |
| 4. | System concatenates the search parameters until the $20^{th}$ parameter, but stops adding parameters afterwards. |
| 5. | User completes the Advanced Search. |
| 6. | System collects relevant information (c.f. 1.8.1.1, step 2) and locally displays both all data collected. |
| 7. | User verifies the data displayed. |
| **Test Case Number** | 1.8.3.4 |
| **Test Case Name** | Empty String Advanced Search |
| **Test Case Description** | This test case verifies the Advanced Search Function |
| **Preconditions** | 1. User clicks the Advanced Search Tab in the Search Page<br>2. System shows the Advanced Search Page |
| **Process Description** | |
| **Step** | **Action** |
| 1 | User inputs empty string for the "Value" field in the Advanced Search. |
| 2 | System detects the empty string and locally displays an error message. |
| 3 | User verifies the error message. |

### 1.1.8.4 Print/Save Search Results

| | |
|---|---|
| **Test Case Number** | 1.8.4 |
| **Test Case Name** | Print/Save Search Results |
| **Test Case Description** | This test case verifies whether all users can Print/save search results |
| **Preconditions** | 1. User is logged in<br>2. User has done a Search job. |

| **Process Description** | |
|---|---|
| **Step** | **Action** |
| 1 | User clicks Search or Advance Search button |
| 2 | System shows the Search result page. |
| 3 | User clicks Print/Save from the browser. |

# 2 University Management

## 2.1 Create a Department

| | |
|---|---|
| **Test Case Number** | 2.1 |
| **Test Case Name** | Create Department |
| **Test Case Description** | This test case verifies whether users Level-2 permission can create a new Department |
| **Preconditions** | 1. User is logged in<br>2. User has sufficient privileges. |

| **Process Description** | |
|---|---|
| **Step** | **Action** |
| 1 | User clicks the "Create a Department" option. |
| 2 | System shows the "Create Department" page. |
| 3 | User enters required fields and submits the information to the system. |
| 4 | System validates the input and updates the database. |
| 5 | System notifies user of successful update and redirects user to "University Management" page. |
| 6 | System administrator verifies whether the database was properly updated. |

### 2.2 Create a Faculty

| | |
|---|---|
| **Test Case Number** | 2.2 |
| **Test Case Name** | Create Faculty |
| **Test Case Description** | This test case verifies whether users with Level-3 permission can create a Faculty |
| **Preconditions** | 1. User is logged in.<br>2. User has sufficient privileges. |
| **Process Description** | |

| Step | Action |
|---|---|
| 1 | User clicks the "Create a Faculty" option |
| 2 | System shows the "Create Faculty" page |
| 3 | User enters required fields and submits the information to the system. |
| 4 | System validates the input and updates the database. |
| 5 | System notifies user of successful update and redirects user to "University Management" page. |
| 6 | System administrator verifies whether the database was properly updated. |

### 2.3 Add a Location

| | |
|---|---|
| **Test Case Number** | 2.3.1 |
| **Test Case Name** | Add a Location (Positive) |
| **Test Case Description** | This test case verifies whether user with Level-2 permission (or higher) can add a location |
| **Preconditions** | 1. User is logged in.<br>2. User has sufficient privileges. |
| **Process Description** | |

| Step | Action |
|---|---|
| 1 | User clicks "Add a location" option |
| 2 | System shows the "Add Location" page. |
| 3 | User enters valid information into corresponding fields and submits the information to the system. |
| 4 | System validates all data submitted. |
| 5 | System updates the database. |
| 6 | System notifies user of successful update and returns user to "University Management" page. |
| 7 | System administrator verifies whether the location was successfully added, and whether all properties are correct. |

### 2.4 Back-up Database

| | | |
|---|---|---|
| **Test Case Number** | | 2.4.1 |
| **Test Case Name** | | Manual Database Back-Up |
| **Test Case Description** | | This test case verifies whether user with sufficient privileges can back up database |
| **Preconditions** | | 1. User is logged in.<br>2. User has sufficient privileges. |
| **Process Description** | | |
| **Step** | **Action** | |
| 1 | User clicks "Backup database" button | |
| 2 | System shows the "Confirmation backup" window. | |
| 3 | User clicks "Confirm " button | |
| 4 | System administrator restores back-up data to the secondary back-up database and verifies whether main database and secondary database are identical. | |
| **Test Case Number** | | 2.4.2 |
| **Test Case Name** | | Automatic Database Back-Up |
| **Test Case Description** | | This test case verifies whether the database automatically backs up the data correctly. |
| **Preconditions** | | N/A |
| **Process Description** | | |
| **Step** | **Action** | |
| 1 | System administrator sets time to one second prior to normal back-up date and time. | |
| 2 | System automatically initiates back-up. | |
| 3 | System administrator verifies correctness of back-up data (c.f. 2.4.1, step 4) | |

### 2.5 Bulk import users from a CSV file

| | |
|---|---|
| **Test Case Number** | 2.5.1 |
| **Test Case Name** | Bulk Import Users (Positive) |
| **Test Case Description** | This test case verifies whether users with sufficient privileges can import users from a |
| | CSV file |
| **Preconditions** | 1. User is logged in.<br>2. User has sufficient privileges. |

| **Process Description** | |
|---|---|
| **Step** | **Action** |
| 1 | User initiates Bulk Import (please refer to Use Case 2.5 for details). |
| 2 | System parses data for valid input. |
| 3 | System verifies whether input conflicts with database entries. |
| 4 | System notifies user of successful update. |
| 5 | System administrator verifies whether records were correctly added. |

| **Test Case Number** | 2.5.2 |
|---|---|
| **Test Case Name** | Bulk Import Users (Conflicting Entries) |
| **Test Case Description** | This test case verifies whether user with sufficient privileges can import users from a CSV file |
| **Preconditions** | 1. User is logged in.<br>2. User has sufficient privileges. |

| **Process Description** | |
|---|---|
| **Step** | **Action** |
| 1 | User initiates Bulk Import (please refer to Use Case 2.5 for details). |
| 2 | System parses data for valid input. |
| 3 | System verifies whether input conflicts with database entries. |
| 4 | System does not update database. System notifies user of conflict and displays conflicting data. |
| 5 | System administrator verifies whether no data was inserted into database. |

| **Test Case Number** | 2.5.3 |
|---|---|
| **Test Case Name** | Bulk Import Users (Invalid Input) |
| **Test Case Description** | This test case verifies whether system administrators (permission level 3) can import users from a CSV file |
| **Preconditions** | 1. User is logged in. |
| | 2. User has sufficient privileges. |
| **Process Description** | |
| **Step** | **Action** |
| 1 | User initiates Bulk Import (please refer to Use Case 2.5 for details). |
| 2 | System parses data for valid input. |
| 3 | System notifies user of invalid data and returns user to the previous page. |
| 4 | System administrator verifies whether no records were added. |

### 2.6 Update User Profile

| | |
|---|---|
| **Test Case Number** | 2.6.1 |
| **Test Case Name** | Update User Role Profile (Positive) |
| **Test Case Description** | This test case verifies whether users with sufficient privileges can update user profiles for users with lower permission levels. |
| **Preconditions** | 1. User is logged in.<br>2. User has sufficient privileges.<br>3. There exist users with lower privileges in the database. |
| **Process Description** | |
| **Step** | **Action** |
| 1 | User updates user role profile with valid input (c.f. use case 2.6 for details). |
| 2 | System validates input. |
| 3 | System updates the database as requested. |
| 4 | System notifies user of successful update and returns user to previous page. |
| 5 | System administrator verifies whether database was properly updated. |
| 6 | Target user logs in to verify whether the update was successful. |
| **Test Case Number** | 2.6.2 |
| **Test Case Name** | Update User Role Profile (Negative) |
| **Test Case Description** | This test case verifies whether users with insufficient privileges can update user profiles for users with lower permission |
| | levels. |
| **Preconditions** | 1. User is logged in.<br>2. User has sufficient privileges.<br>3. There exist users with lower privileges in the database. |

| **Process Description** | |
|---|---|
| **Step** | **Action** |
| 1 | Admin user updates user role profile (c.f. use case 2.6 for details), but requests a privilege level for target user above that which admin user may grant. |
| 2 | System validates input. |
| 3 | System does not update database. System notifies user of error and returns user to previous page. |
| 4 | System administrator verifies whether database was properly updated. |
| 5 | Target user logs in to verify whether the update was successful. |

# 3 Test Cases to Manage Assets

## 3.1 View Assets

| | |
|---|---|
| **Test Case Number** | 3.1 |
| **Test Case Name** | View Assets |
| **Test Case Description** | This test case verifies whether users at level 1 & above may view assets |
| **Preconditions** | 1. User is logged in.<br>2. User has sufficient privileges (preferably, several users with different affiliations, and with different permission levels).<br>3. There exist assets viewable by the user in the database. |

| **Process Description** | |
|---|---|
| **Step** | **Action** |
| 1 | User views a list of assets (c.f. use case 3.1 for details). |
| 2 | User verifies whether all assets should be viewable with his/her permission level and affiliation. |
| 3 | System administrator verifies whether no data that should be viewable is filtered out. |

### 3.2 Add Asset

| | |
|---|---|
| **Test Case Number** | 3.2 |
| **Test Case Name** | Add Assets (Positive) |
| **Test Case Description** | This test case verifies whether user with permission level 1 & above may add assets |
| **Preconditions** | 1. User is logged in.<br>2. User has sufficient privileges. |

| **Process Description** | |
|---|---|
| **Step** | **Action** |
| 1 | User adds assets with valid input (c.f. use case 3.2 for details). |
| 2 | System validates input. |
| 3 | System updates database. |
| 4 | System notifies user of successful update and returns user to previous page. |
| 5 | System administrator verifies whether assets are correctly inserted into the database. |
| **Test Case Number** | 3.2 |
| **Test Case Name** | Add Assets (Negative) |
| **Test Case Description** | This test case verifies whether user with permission level 1 & above may add assets |
| **Preconditions** | 1. User is logged in.<br>2. User has sufficient privileges to add assets, but insufficient privileges for the data specified. |

| **Process Description** | |
|---|---|
| **Step** | **Action** |
| 1 | User adds assets (c.f. use case 3.2 for details), but inputs data above his permission level. |
| 2 | System validates input. |
| 3 | System does not update database. |
| 4 | System notifies user of error and returns user to previous page. |
| 5 | System administrator verifies whether assets are not inserted into the database. |

### 3.3 Update Asset(s) Information

| | | |
|---|---|---|
| **Test Case Number** | 3.3.1 | |
| **Test Case Name** | Update asset(s) (Positive) | |
| **Test Case Description** | This test case verifies whether user with permission level 1 & above may update asset information | |
| **Preconditions** | 1. User is logged in<br>2. User has sufficient privileges.<br>3. There exist assets to update. | |
| **Process Description** | | |
| **Step** | **Action** | |
| 1 | User updates data with valid data (c.f. use case 3.3 for details). | |
| 2 | System validates input. | |
| 3 | System updates database. | |
| 4 | System notifies user of successful update and returns user to the "View Assets" page. | |
| **Test Case Number** | 3.3.1 | |
| **Test Case Name** | Update asset(s) (Negative) | |
| **Test Case Description** | This test case verifies whether user with permission level 1 & above may update asset information | |
| **Preconditions** | 1. User is logged in<br>2. User has sufficient privileges.<br>3. There exist assets to update. | |
| **Process Description** | | |
| **Step** | **Action** | |
| 1 | User updates data (c.f. use case 3.3 for details) with data above user's permission level. | |
| 2 | System validates input. | |
| 3 | System does not update database. | |
| 4 | System notifies user of error and returns user to the previous page. | |

### 3.4 Bulk Add Assets

| | |
|---|---|
| **Test Case Number** | 3.4.1 |
| **Test Case Name** | Bulk Add Assets (Positive) |
| **Test Case Description** | This test case verifies whether user with permission level 1 & above may bulk add assets |
| **Preconditions** | 1. User is logged in<br>2. User has sufficient privileges. |
| **Process Description** | |
| **Step** | **Action** |
| 1 | User adds data from CSV file(s) (c.f. use case 3.4 for details). |
| 2 | System parses data for valid input. |
| 3 | System updates database. |
| 4 | System notifies user of successful operation and returns user to "View Assets" page. |
| 5 | System administrator verifies whether data was correctly entered. |
| **Test Case Number** | 3.4.2 |
| **Test Case Name** | Bulk Add Assets (Negative) |
| **Test Case Description** | This test case verifies whether user with permission level 1 & above may bulk add assets |
| **Preconditions** | 1. User is logged in<br>2. User has sufficient privileges. |
| **Process Description** | |
| **Step** | **Action** |
| 1 | User adds data from CSV file(s) (c.f. use case 3.4 for details), but CSV files contain invalid data. |
| 2 | System parses data for valid input. |
| 3 | System does not update database. |
| 4 | System notifies user of error and returns user to "View Assets" page. |
| 5 | System administrator verifies no data was entered. |

### 3.5 Group Assets

| | |
|---|---|
| **Test Case Number** | 3.5.1 |
| **Test Case Name** | Group Assets (Positive) |
| **Test Case Description** | This test case verifies whether user with permission level 1 & above may group assets |
| **Preconditions** | 1. User is logged in.<br>2. User has sufficient privileges.<br>3. There exist assets in the database to group. |

**Process Description**

| Step | Action |
|---|---|
| 1 | User groups assets (c.f. use case 3.5 for details). |
| 2 | System verifies whether operation is within user's privileges. |
| 3 | System updates database. |
| 4 | System notifies user of successful update and returns user to the "View Assets" page. |
| 5 | System administrator verifies whether all relevant tables were updated. |
| **Test Case Number** | 3.5.2 |
| **Test Case Name** | Group Assets (Negative) |
| **Test Case Description** | This test case verifies whether user with permission level 1 & above may group assets |
| **Preconditions** | 1. User is logged in.<br>2. User has sufficient privileges. |

**Process Description**

| Step | Action |
|---|---|
| 1 | User groups assets (c.f. use case 3.5 for details), but selects assets which are beyond user's scope. |
| 2 | System verifies whether operation is within user's privileges. |
| 3 | System does not update database. |
| 4 | System notifies user of error and returns user to the previous page. |
| 5 | System administrator confirms that data was added. |

# 4. Test Cases to Review assets

## 4.1 View Audit Options

| | | |
|---|---|---|
| **Test Case Number** | 4.1.1 | |
| **Test Case Name** | View Audit Options (Positive) | |
| **Test Case Description** | This test case verifies whether user with permission level 1 & above may audit | |
| **Preconditions** | 1. User is logged in.<br>2. User has sufficient privileges. | |
| **Process Description** | | |
| **Step** | **Action** | |
| 1 | User views audit options (c.f. use case 4.1 for details). | |
| 2 | System verifies whether user has sufficient privileges. | |
| 3 | System displays the audit options. | |
| **Test Case Number** | 4.1.2 | |
| **Test Case Name** | View Audit Options (Negative) | |
| **Test Case Description** | This test case verifies whether user with permission level 1 & above may audit | |
| **Preconditions** | 1. User is logged in.<br>2. User has sufficient privileges. | |
| **Process Description** | | |
| **Step** | **Action** | |
| 1 | User requests to view audit options from without the system. | |
| 2 | System verifies whether user has sufficient privileges. | |
| 3 | System informs user of error and redirects user to welcome page. | |

## 4.2 Audit Logs

| | |
|---|---|
| **Test Case Number** | 4.2 |
| **Test Case Name** | Audit Logs |
| **Test Case Description** | This test case verifies whether user with permission level 1 & above may audit logs |
| **Preconditions** | 1. User is logged in.<br>2. User has sufficient privileges. |
| **Process Description** | |

| Step | Action |
|---|---|
| 1 | User clicks "Audit logs" button |
| 2 | System filters data according to user's privileges. |
| 3 | System displays filtered data. |
| 4 | User verifies that no data displayed is beyond those corresponding to his/her privileges and affiliation. |
| 5 | System administrator verifies whether data displayed is complete. |

### 4.3 Produce Reports

| | |
|---|---|
| **Test Case Number** | 4.3 |
| **Test Case Name** | Produce Reports |
| **Test Case Description** | This test case verifies whether user with permission level 1 & above may produce reports |
| **Preconditions** | 1. User is logged in.<br>2. User has sufficient privileges. |
| **Process Description** | |
| **Step** | **Action** |
| 1 | User produces report (c.f. use case 4.3 for details). |
| 2 | System filters data and produces requested report. |
| 3 | Data is verified for correctness (c.f. test case 4.2, steps 4 and 5) |

### 4.4 Print/Save Audit Logs

| | |
|---|---|
| **Test Case Number** | 4.4.1 |
| **Test Case Name** | Print Audit Logs/Reports |
| **Test Case Description** | This test case verifies whether user with permission level 1 & above may print relevant data. |
| **Preconditions** | User has relevant data displayed on the screen. |
| **Process Description** | |
| **Step** | **Action** |
| 1 | User clicks the "Print" button. |
| 2 | System produces a printable version of the page and sends request to client's browser. |
| 3 | User completes the "Print" request. |
| 4 | User verifies whether the printed data is identical to that displayed. |

| Test Case Number | 4.4.2 |
|---|---|
| Test Case Name | Save Audit Logs/Reports |
| Test Case Description | This test case verifies whether user with permission level 1 & above may save relevant data. |
| Preconditions | User has relevant data displayed on the screen. |
| **Process Description** | |
| **Step** | **Action** |
| 1 | User clicks "Save" button. |
| 2 | System outputs the data into proper format and sends request to client's browser. |
| 3 | Client completes save request. |
| 4 | Client verifies whether saved file corresponds to the data requested. |

## 5. Test cases for Error Management

### 5.1 List Error Messages

| Test Case Number | 5.1 |
|---|---|
| Test Case Name | List Error Messages |
| Test Case Description | This test case verifies whether system administrators (permission level 3) may view a list of error messages |
| Preconditions | 1. User is logged in<br>2. User has sufficient privileges. |
| **Process Description** | |
| **Step** | **Action** |
| 1 | User views a list of error messages, filtered according to the user's permissions and affiliation. |
| 2 | System shows the filtered list of error messages. |
| 3 | User confirms that all entries displayed fall within his/her privileges. |
| 4 | System administrator verifies whether any error message viewable with user's privileges is not visible. |

### 5.2 Print Error Messages

| | |
|---|---|
| **Test Case Number** | 5.2 |
| **Test Case Name** | Print Error Messages |
| **Test Case Description** | This test case verifies whether system administrators (permission level 3) may print error messages |
| **Preconditions** | User has selected the error message(s) of interest |

**Process Description**

| Step | Action |
|---|---|
| 1 | User selects error message(s) and chooses to print them (c.f. use case 5.2 for details). |
| 2 | System compiles the error message(s) and transmits the request to the client's browser. |
| 3 | User completes the request and verifies that all details are correct. |

### 5.3 View More Details/Annotate Error Messages

| | |
|---|---|
| **Test Case Number** | 5.3 |
| **Test Case Name** | View More Details/Annotate Error Messages |
| **Test Case Description** | This test case verifies whether system administrators (permission level 3) can view more details/annotate error messages |
| **Preconditions** | Administrator is viewing error messages. |

**Process Description**

| Step | Action |
|---|---|
| 1 | User clicks "More details/Edit" button |
| 2 | System shows required information, and an additional field for comments is shown. |
| 3 | User verifies that the details correspond to those stored in the database. |
| 4 | User enters comments and submits comments. |
| 5 | System accepts comments and returns user to previous page. |
| 6 | Administrator verifies whether comment was successfully stored in the system. |

## 6. Test cases for Request Management

### 6.1 View Pending Requests

| | |
|---|---|
| **Test Case Number** | 6.1 |
| **Test Case Name** | View Pending Requests |
| **Test Case Description** | This test case verifies whether the system is able to filter pending requests according to the user. |
| **Preconditions** | 1. User is logged in.<br>2. User has sufficient permission.<br>3. There exist requests in the database corresponding to the user's privileges. |

**Process Description**

| Step | Action |
|---|---|
| 1 | User views requests (c.f. use case 6.1 for details). |
| 2 | System filters requests according to user's permission level and affiliation. |
| 3 | System displays the filtered data. |
| 4 | User verifies that all data shown is within the scope allowed by user's privileges. |
| 5 | System administrator verifies that no request is withheld that should be displayed. |

### 6.2 Approve/Formalize a Pending Request

| | |
|---|---|
| **Test Case Number** | 6.2.1 |
| **Test Case Name** | Approve/Formalize a Request (Positive) |
| **Test Case Description** | This test case verifies whether user with permission level 1 & above may approve/formalize pending requests |
| **Preconditions** | User has selected a request to approve/formalize |

**Process Description**

| Step | Action |
|---|---|
| 1 | User completes fields as necessary (c.f. use case 6.2 for details) and submits the information. |
| 2 | System validates the information. |
| 3 | System updates the information. |
| 4 | System notifies user of successful update and returns |
| | user to viewing requests. |
| 5 | System administrator verifies whether the request has been successfully approved/formalized. |

| **Test Case Number** | 6.2.2 |
|---|---|
| **Test Case Name** | Approve/Formalize a Request (Negative) |
| **Test Case Description** | This test case verifies whether the system rejects attempts at approving/formalizing requests when errors are detected. |
| **Preconditions** | User has selected a request to approve/formalize |
| **Process Description** | |
| **Step** | **Action** |
| 1 | User completes fields as necessary (c.f. use case 6.2 for details) but with some errors (wrong location ID, etc.) and submits the information. |
| 2 | Validation is not successful. |
| 3 | System does not update the information. |
| 4 | System notifies user of error and returns user to previous page. |
| 5 | System administrator verifies whether the request has been successfully modified. |

### 6.3 Reject a Request

| **Test Case Number** | 6.3.1 |
|---|---|
| **Test Case Name** | Reject a Request (Positive) |
| **Test Case Description** | This test case verifies whether user with permission level 1 & above may reject a pending request |
| **Preconditions** | User selects a pending request. |
| **Process Description** | |
| **Step** | **Action** |
| 1 | User rejects a request (c.f. use case 6.3 for details). |
| 2 | System verifies whether user has the authority to reject the request. |
| 3 | System updates the request status. |
| 4 | System informs user of successful update and returns user to viewing requests. |
| 5 | System administrator verifies whether rejection was |
| | successful. |

| Test Case Number | 6.3.1 |
|---|---|
| **Test Case Name** | Reject a Request (Negative) |
| **Test Case Description** | This test case verifies whether user with permission level 1 & above may reject a pending request |
| **Preconditions** | User selects a pending request. |
| **Process Description** | |
| **Step** | **Action** |
| 1 | User without sufficient authority attempts to reject a request (c.f. use case 6.3 for details). |
| 2 | System verifies whether user has the authority to reject the request. |
| 3 | System updates the request status, marking the rejection as a comment. |
| 4 | System informs user of both error and action taken, then returns user to viewing requests. |
| 5 | System administrator verifies whether comment was added, and whether rejection was successful. |

## 7. Load Testing

The test outlined below verifies whether the performance of the test is at an acceptable level under increased server loads.

| Test Case Number | 7.1 |
|---|---|
| **Test Case Name** | Response Time During Stress |
| **Test Case Description** | This test case verifies whether the UUIS provides the required response time of less than five seconds for a transaction even when there is an increased load on the server. |
| **Preconditions** | 50 users must be logged in. |
| **Process Description** | |
| **Step** | **Action** |
| 1. | Fifty users enter their requests (excluding back-up requests) simultaneously and initiate the transactions. |
| 2. | System performs all transactions (except for back-up) and outputs the results within five seconds. |

## 8. Browser Consistency Test

| | |
|---|---|
| **Test Case Number** | 8.1 |
| **Test Case Name** | Support for IE 8, Firefox 3.6.3, Chrome 4.1, Safari 4, and Opera10.51 |
| **Test Case Description** | This test case verifies whether the UUIS provides the same or an equivalent interface in when rendered by the following five browsers: IE, FIREFOX,CHROME,SAFARI AND OPERA. |
| **Preconditions** | An Internet Explorer, a Firefox, a Chrome, a Safari and Opera browser open to the appropriate page. |

| **Process Description** | |
|---|---|
| **Step** | **Action** |
| 1 | Title, button, tables, data and description are verified for the adherence to standards. |
| 2 | The tab key is used to toggle between widgets. |
| 3 | The user's expectations and actual behaviour are noted down for each of the browsers. |

## Appendix I: Deployment and Configuration

The deployment (Figure AI.0.1) illustrates the basic system used during development. Scientific Linux 5.4 was installed on the server spec106.encs.concordia.ca located in Room H-833 at Concordia University, Montreal, and updated the system in February 2010 by running "yum" from the command prompt. A developer group ("team2") was created and an account opened for each member of Team 2.

MySQL 6.0 was downloaded from http://dev.mysql.com/donloads (file: MySQL-server-community-6.0.11-0.rhel5.i386.rpm) and installed (command prompt: "rpm -ivh  MySQL-server-community-6.0.11-0.rhel5.i386.rpm"). We selected the option to automatically start running MySQL upon system reboot (command prompt: "chkconfig mysqld on"). Afterwards, we set the root password (omitted for security reasons) for the database server and created the "uuisdb" database (section Appendix III.1, "uuisdb.sql"). We created a user account ("dbuser") and granted the permissions to access "uuisdb".

Apache 2.2.3 was pre-installed on Scientific Linux. We modified httpd.conf (section Appendix III.2, "httpd.conf") to set document root at "/var/www/html" and ssl.conf to ensure "/etc/pki/tls/certs/localhost.crt" and "/etc/pki/tls/localhost.key" exist. We tested the settings by visiting "https://spec106.encs.concordia.ca". Similarly to MySQL configuration, we elected to have Apache run automatically upon system reboot (command prompt: "chkconfig httpd on").

We installed the necessary components for PHP (php-mysql, php-gd, php-mbstring, php-mcrypt) and edited the php.ini file ("/etc/php.ini", provided in Appendix III.3).

We then downloaded phpMySQLAdmin (http://www.phpmyadmin.net/home_page/downloads.php) and extracted it to "/var/www/phpMyAdmin". We configured "config.inc.php" and "config.php" ("/var/www/html/includes/") to set the hostname, database, users and other details (provided in Appendix III.4).

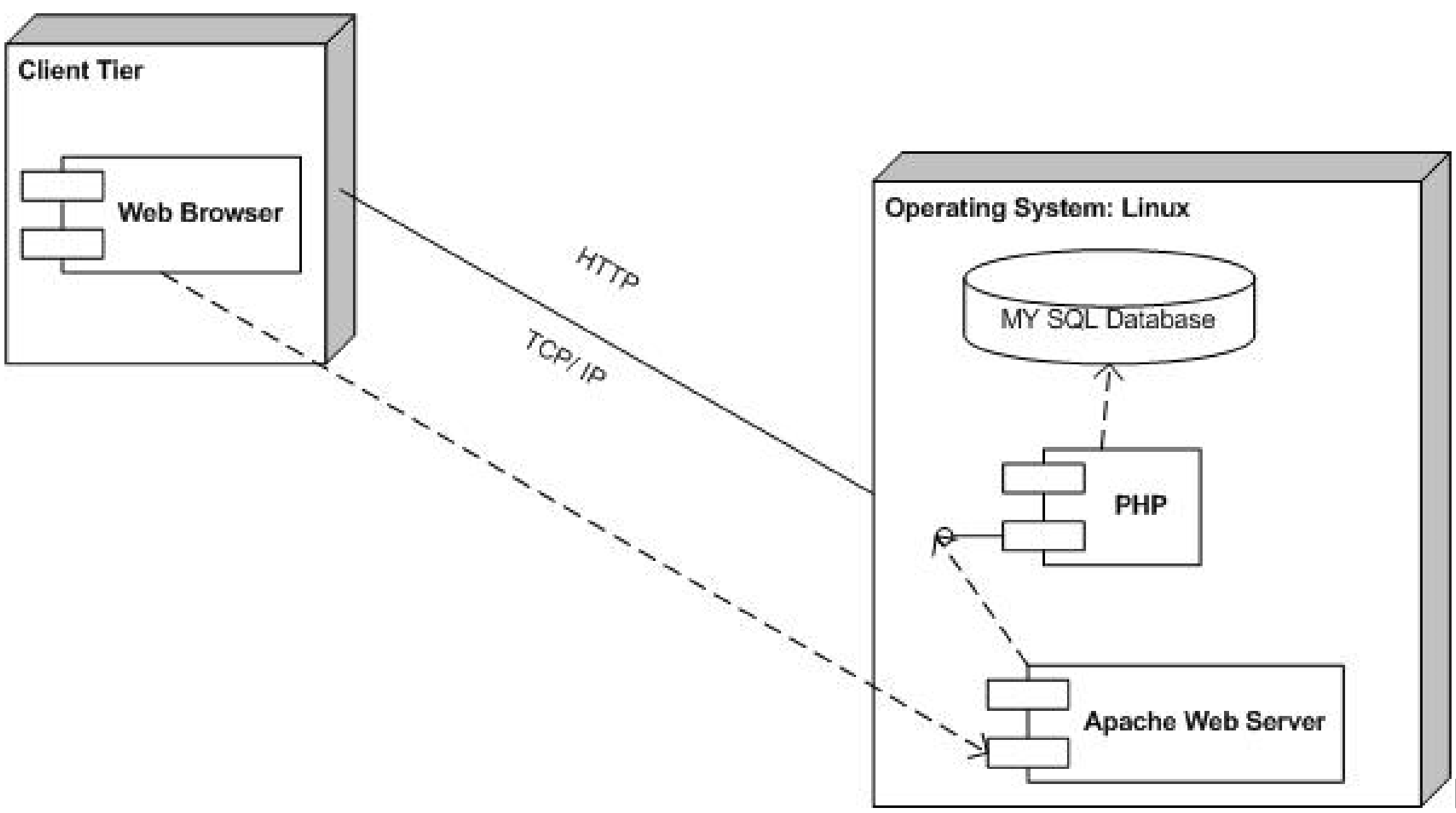

Figure AI.0.1 Deployment.

**Appendix III.1. uuisdb.sql**

```
-- phpMyAdmin SQL Dump
-- version 2.11.10
-- http://www.phpmyadmin.net
--
-- Host: spec106.encs.concordia.ca
-- Generation Time: May 02, 2010 at 08:41 PM
-- Server version: 6.0.11
-- PHP Version: 5.1.6

SET SQL_MODE="NO_AUTO_VALUE_ON_ZERO";


/*!40101 SET
@OLD_CHARACTER_SET_CLIENT=@@CHARACTER_SET_CLIENT */;
/*!40101 SET
@OLD_CHARACTER_SET_RESULTS=@@CHARACTER_SET_RESULTS */;
/*!40101 SET
@OLD_COLLATION_CONNECTION=@@COLLATION_CONNECTION */;
/*!40101 SET NAMES utf8 */;

--
-- Database: `uuisdb`
--

-- --------------------------------------------------------

--
-- Table structure for table `acls`
--

CREATE TABLE IF NOT EXISTS `acls` (
  `user_role_id` int(11) NOT NULL COMMENT 'Role ID',
  `permission` bigint(20) DEFAULT NULL COMMENT 'Permission',
  PRIMARY KEY (`user_role_id`)
) ENGINE=InnoDB DEFAULT CHARSET=latin1;

--
-- Dumping data for table `acls`
--

INSERT INTO `acls` (`user_role_id`, `permission`) VALUES
(1, 2048);
-- --------------------------------------------------------
```

```
--
-- Table structure for table `affiliations`
--

CREATE TABLE IF NOT EXISTS `affiliations` (
  `affln_id` int(11) NOT NULL COMMENT 'Affiliation ID',
  `affln_name` varchar(255) DEFAULT NULL COMMENT 'Affiliation Name',
  `alffln_code` varchar(100) DEFAULT NULL COMMENT 'Affiliation Code',
  PRIMARY KEY (`affln_id`)
) ENGINE=InnoDB DEFAULT CHARSET=latin1;

--
-- Dumping data for table `affiliations`
--

INSERT INTO `affiliations` (`affln_id`, `affln_name`, `alffln_code`)
VALUES
(0, 'UUIS', 'UUIS'),
(1, 'Arts and Science', 'ASF'),
(2, 'Computer Science', 'CSF'),
(3, 'Engineering', 'EN'),
(10, 'History', 'HIS'),
(11, 'Religion', 'REL'),
(12, 'Visual Arts', 'VA'),
(13, 'Math', 'MA'),
(20, 'SOEN', 'SOEN'),
(21, 'CS', 'CS'),
(30, 'ECE', 'ECE'),
(31, 'MIE', 'MIE');

-- --------------------------------------------------------

--
-- Table structure for table `building`
--

CREATE TABLE IF NOT EXISTS `building` (
  `bldg_id` int(11) NOT NULL AUTO_INCREMENT COMMENT 'Building ID',
  `bldg_code` varchar(100) DEFAULT NULL COMMENT 'Building Code',
  `bldg_name` varchar(500) DEFAULT NULL COMMENT 'Building Name',
  PRIMARY KEY (`bldg_id`)
) ENGINE=InnoDB  DEFAULT CHARSET=latin1 AUTO_INCREMENT=6 ;
```

```
--
-- Dumping data for table `building`
--

INSERT INTO `building` (`bldg_id`, `bldg_code`, `bldg_name`) VALUES
(0, 'N/A', 'N/A'),
(1, 'Hall', 'Hall Building'),
(2, 'EV', NULL),
(3, 'FB', 'Fabien'),
(4, 'MB', NULL);

-- --------------------------------------------------------

--
-- Table structure for table `categories`
--

CREATE TABLE IF NOT EXISTS `categories` (
  `cat_id` int(11) NOT NULL AUTO_INCREMENT COMMENT 'Category ID',
  `parent_cat_id` varchar(500) DEFAULT NULL COMMENT 'Parent Category',
  `description` varchar(100) DEFAULT NULL COMMENT 'Category
Description',
  PRIMARY KEY (`cat_id`)
) ENGINE=InnoDB  DEFAULT CHARSET=latin1 AUTO_INCREMENT=5 ;

--
-- Dumping data for table `categories`
--

INSERT INTO `categories` (`cat_id`, `parent_cat_id`, `description`)
VALUES
(0, 'N/A', 'N/A'),
(1, '1', 'Computer'),
(2, '1', 'Mouse'),
(3, NULL, 'Printer');

-- --------------------------------------------------------

--
-- Table structure for table `fieldlist`
--
```

```
CREATE TABLE IF NOT EXISTS `fieldlist` (
  `field_id` int(11) NOT NULL AUTO_INCREMENT COMMENT 'Field ID',
  `table_id` int(11) NOT NULL DEFAULT '0' COMMENT 'Table ID',
  `field_code` varchar(100) DEFAULT NULL COMMENT 'Field Code',
  `field_name` varchar(500) DEFAULT NULL COMMENT 'Field Name',
  `permissions` int(11) DEFAULT NULL COMMENT 'Permission',
  PRIMARY KEY (`field_id`,`table_id`)
) ENGINE=InnoDB DEFAULT CHARSET=latin1 AUTO_INCREMENT=1 ;

--
-- Dumping data for table `fieldlist`
--


-- --------------------------------------------------------

--
-- Table structure for table `inventories`
--

CREATE TABLE IF NOT EXISTS `inventories` (
  `item_id` int(11) NOT NULL COMMENT 'Inventory ID',
  `qty` int(11) DEFAULT NULL COMMENT 'Quantity',
  `status` varchar(10) DEFAULT NULL COMMENT 'Inventory Status',
  `modified_by` int(11) DEFAULT NULL COMMENT 'Modifiedby',
  `date_modified` datetime DEFAULT NULL COMMENT 'Modified Date',
  PRIMARY KEY (`item_id`)
) ENGINE=InnoDB DEFAULT CHARSET=latin1;

--
-- Dumping data for table `inventories`
--

INSERT INTO `inventories` (`item_id`, `qty`, `status`, `modified_by`,
`date_modified`) VALUES
(20, 2, NULL, NULL, NULL),
(21, 3, NULL, NULL, NULL),
(22, 4, NULL, NULL, NULL),
(23, 5, NULL, NULL, NULL),
(329, 1, NULL, NULL, NULL);

-- --------------------------------------------------------
```

```
--
-- Table structure for table `itemproperties`
--

CREATE TABLE IF NOT EXISTS `itemproperties` (
  `item_prop_id` int(11) NOT NULL AUTO_INCREMENT COMMENT 'Item
Property ID',
  `item_id` int(11) DEFAULT NULL COMMENT 'Item ID',
  `prop_id` int(11) DEFAULT NULL COMMENT 'Property ID',
  `prop_value` varchar(200) DEFAULT NULL COMMENT 'Property Value',
  PRIMARY KEY (`item_prop_id`)
) ENGINE=InnoDB  DEFAULT CHARSET=latin1 AUTO_INCREMENT=6 ;

--
-- Dumping data for table `itemproperties`
--

INSERT INTO `itemproperties` (`item_prop_id`, `item_id`, `prop_id`,
`prop_value`) VALUES
(1, 20, 1, 'Dell9000'),
(2, 21, 1, 'Dell9000'),
(3, 22, 1, 'Dell9000'),
(4, 23, 2, 'HP 4200'),
(5, 329, 1, 'Dell9000');

-- --------------------------------------------------------

--
-- Table structure for table `itempropertylist`
--

CREATE TABLE IF NOT EXISTS `itempropertylist` (
  `prop_id` int(11) NOT NULL AUTO_INCREMENT COMMENT 'Property ID',
  `cat_id` int(11) DEFAULT NULL COMMENT 'Category ID',
  `prop_name` varchar(200) DEFAULT NULL COMMENT 'Property Name',
  `default_value` varchar(200) DEFAULT NULL COMMENT 'Default Value',
  PRIMARY KEY (`prop_id`)
) ENGINE=InnoDB  DEFAULT CHARSET=latin1 AUTO_INCREMENT=3 ;

--
-- Dumping data for table `itempropertylist`
--
```

```
INSERT INTO `itempropertylist` (`prop_id`, `cat_id`, `prop_name`,
`default_value`) VALUES
(1, 1, 'Desktop', NULL),
(2, 3, 'Desktop Laser', NULL);

-- --------------------------------------------------------

--
-- Table structure for table `items`
--

CREATE TABLE IF NOT EXISTS `items` (
  `item_id` int(11) NOT NULL AUTO_INCREMENT COMMENT 'Item ID',
  `item_description` varchar(500) DEFAULT NULL COMMENT 'Item
Description',
  `code` varchar(200) DEFAULT NULL COMMENT 'Item Code',
  `group_id` int(11) DEFAULT NULL COMMENT 'Item Group ID',
  `serial_number` varchar(200) DEFAULT NULL COMMENT 'Item Serial
Number',
  `cat_id` int(11) DEFAULT NULL COMMENT 'Item Category ID',
  `owner_id` int(11) DEFAULT NULL COMMENT 'Item Own ID',
  `loc_id` int(11) DEFAULT NULL COMMENT 'Item Location ID',
  `date_modified` datetime DEFAULT NULL COMMENT 'Item Modified Date',
  `status` varchar(10) DEFAULT NULL COMMENT 'Item Status',
  PRIMARY KEY (`item_id`)
) ENGINE=InnoDB  DEFAULT CHARSET=latin1 AUTO_INCREMENT=37 ;

--
-- Dumping data for table `items`
--

INSERT INTO `items` (`item_id`, `item_description`, `code`, `group_id`,
`serial_number`, `cat_id`, `owner_id`, `loc_id`, `date_modified`,
`status`) VALUES
(0, 'N/A', 'N/A', 576, 'N/A', 0, 0, 3, '2010-05-02 19:50:24', 'stolen'),
(3, 'desktop', 'UUIS000001', 333, 'abcdefg', 1, 1, 1, '0000-00-00 00:00:00',
'inactive'),
(4, 'Dell00001', 'UUIS000002', 5555555, 'abcdefg', 2, 1, 2, '0000-00-00
00:00:00', 'inactive'),
(5, 'aa', 'aa', 576, 'aa', 1, 0, 3, '2010-05-02 19:50:24', 'stolen'),
(6, 'desktop', 'UUIS000001', 5555555, 'abcdefg', 1, 1, 1, '0000-00-00
00:00:00', 'inactive'),
```

```
(7, 'demo1', 'DEMO0001', 5555555, '1000demo', 2, 0, 1, '0000-00-00
00:00:00', 'inactive'),
(8, 'demo2', 'DEMO0001', 576, '2000demo', 2, 0, 3, '2010-05-02 19:50:24',
'stolen'),
(9, 'demo3', 'DEMO0003', 0, '3000demo', 2, 0, 1, '0000-00-00 00:00:00',
'inactive'),
(10, 'demo4', 'DEMO0004', 0, '4000demo', 2, 0, 3, '0000-00-00 00:00:00',
'inactive'),
(11, 'demo5', 'DEMO0005', 2147483647, '5000demo', 2, 0, 3, '2010-05-02
20:14:13', 'active'),
(12, 'demo6', 'DEMO0006', 2147483647, '6000demo', 2, 0, 1, '0000-00-00
00:00:00', 'inactive'),
(13, 'demo7', 'DEMO0007', 0, '7000demo', 2, 0, 2, '0000-00-00 00:00:00',
'stolen'),
(14, 'demo8', 'DEMO0008', 0, '8000demo', 2, 0, 2, '0000-00-00 00:00:00',
'stolen'),
(15, 'demo10', 'DEMO00010', 0, '10000demo', 2, 0, 2, '0000-00-00
00:00:00', 'stolen'),
(16, 'demo3', 'DEMO0003', 0, '3000demo', 2, 0, 2, '0000-00-00 00:00:00',
'stolen'),
(17, '3333333333', '333333333', 2147483647, '33333333', 1, 333333, 2,
'0000-00-00 00:00:00', 'stolen'),
(18, 'qqqqq', 'qqqqq', 0, 'qqqq', 1, 0, 1, '0000-00-00 00:00:00', 'active'),
(19, '', '', 0, '          ', 0, 0, 0, '0000-00-00 00:00:00', ''),
(20, 'Dell tower 1', 'UUIS000002', NULL, 'a0002', 1, 3, 3, '2010-05-02
13:37:34', NULL),
(21, 'Dell tower 2', 'UUIS000003', NULL, 'a0003', 1, 21, 4, '2010-05-02
13:37:34', NULL),
(22, 'Dell tower 3', 'UUIS000004', NULL, 'a0004', 1, 20, 5, '2010-05-02
13:37:34', NULL),
(23, 'Marker', 'UUIS000005', NULL, 'a0005', 3, 21, 6, '2010-05-02
13:37:34', NULL),
(24, '66666', '66666', 66666, '666666    ', 2, 6666, 4, '0000-00-00
00:00:00', 'active'),
(25, 'uuuu', 'uuuu', 0, 'uuuu          ', 1, 0, 1, '0000-00-00 00:00:00',
'active'),
(27, '999', '999', 999, '999', 1, 999, 2, '0000-00-00 00:00:00', 'lent'),
(28, '22222', '22222', 22222, '22  ', 1, 222222, 2, '0000-00-00 00:00:00',
'inactive'),
(29, '', '', 0, '          ', 0, 0, 2, '0000-00-00 00:00:00', ''),
(30, '', '', 0, '2222DEMO', 0, 0, 0, '0000-00-00 00:00:00', ''),
(31, 'Table1', '', 0, '11111DEMO', 1, 0, 2, '0000-00-00 00:00:00', 'active'),
```

```
(32, 'speaker', 'DEMO0002022', 3, '222222DEMO', 1, 222, 0, '2010-05-02
00:08:26', ''),
(33, 'mobile', 'DEMO33333', 5, '222222DEMO', 1, 433, 2, '2010-05-02
00:17:03', 'active'),
(34, 'desktop Dell 1', 'UUIS000001', NULL, 'a0001', NULL, NULL, NULL,
NULL, NULL),
(35, 'ppppp', 'DEMO234445', 5, '6778899DEMO', 1, 555, 2, '2010-05-02
18:37:30', 'active'),
(36, 'wwww', 'DEMOwwwww', 3, 'wwwwwDEMO', 1, 2222, 2, '2010-05-02
18:36:46', 'inactive');

-- --------------------------------------------------------

--
-- Table structure for table `locations`
--

CREATE TABLE IF NOT EXISTS `locations` (
  `loc_id` int(11) NOT NULL AUTO_INCREMENT COMMENT 'Location ID',
  `parent_loc_id` int(11) DEFAULT NULL COMMENT 'Parent Location',
  `loc_code` varchar(100) DEFAULT NULL COMMENT 'Location Code',
  `loc_name` varchar(500) DEFAULT NULL COMMENT 'Location Name',
  `bldg_id` int(11) DEFAULT NULL COMMENT 'Build ID',
  `affln_id` int(11) DEFAULT NULL COMMENT 'Affiliation ID',
  `Status` varchar(10) DEFAULT NULL COMMENT 'Location Status',
  `loc_type_id` int(11) DEFAULT NULL COMMENT 'Location Type ID',
  `Comment` varchar(500) DEFAULT NULL COMMENT 'Location Comment',
  PRIMARY KEY (`loc_id`)
) ENGINE=InnoDB  DEFAULT CHARSET=latin1 AUTO_INCREMENT=8 ;

--
-- Dumping data for table `locations`
--

INSERT INTO `locations` (`loc_id`, `parent_loc_id`, `loc_code`,
`loc_name`, `bldg_id`, `affln_id`, `Status`, `loc_type_id`, `Comment`)
VALUES
(1, 1, 'H-613', 'H-613 Classroom', 1, 1, 'available', 1, NULL),
(2, 2, 'H-627', 'H-627 Classroom', 1, 1, 'available', 1, NULL),
(3, NULL, 'H-011', 'H-011 Classromm', NULL, NULL, NULL, NULL, NULL),
(4, NULL, 'EV-011', 'EV-011 Classromm', NULL, NULL, NULL, NULL, NULL),
(5, NULL, 'H-833', 'On hand Lab', NULL, NULL, NULL, NULL, NULL),
(6, NULL, 'H-866', 'Printer Room', NULL, NULL, NULL, NULL, NULL),
```

```
(7, NULL, 'MB-011', 'MB.011 Classromm', NULL, NULL, NULL, NULL, NULL);

-- --------------------------------------------------------

--
-- Table structure for table `locationtypes`
--

CREATE TABLE IF NOT EXISTS `locationtypes` (
  `loc_type_id` int(11) NOT NULL AUTO_INCREMENT COMMENT 'Location
Type ID',
  `loc_type_name` varchar(200) DEFAULT NULL COMMENT 'Location Type
Name',
  `Description` varchar(500) DEFAULT NULL COMMENT 'Location
Description',
  PRIMARY KEY (`loc_type_id`)
) ENGINE=InnoDB DEFAULT CHARSET=latin1 AUTO_INCREMENT=1 ;

--
-- Dumping data for table `locationtypes`
--


-- --------------------------------------------------------

--
-- Table structure for table `logs`
--

CREATE TABLE IF NOT EXISTS `logs` (
  `log_id` int(11) NOT NULL AUTO_INCREMENT COMMENT 'Log ID',
  `log_time` timestamp NULL DEFAULT CURRENT_TIMESTAMP COMMENT
'Log Time',
  `user_id` int(11) DEFAULT NULL COMMENT 'User ID',
  `item_id` int(11) DEFAULT NULL COMMENT 'Item ID',
  `event_type` varchar(100) DEFAULT NULL COMMENT 'Event Type',
  `content` varchar(2000) DEFAULT NULL COMMENT 'Content',
  PRIMARY KEY (`log_id`)
) ENGINE=InnoDB  DEFAULT CHARSET=latin1 AUTO_INCREMENT=1 ;

-- --------------------------------------------------------

--
```

```
-- Table structure for table `permissions`
--

CREATE TABLE IF NOT EXISTS `permissions` (
  `permission_id` int(11) NOT NULL COMMENT 'Permission ID',
  `Description` varchar(500) DEFAULT NULL COMMENT 'Permission
Description',
  PRIMARY KEY (`permission_id`)
) ENGINE=InnoDB DEFAULT CHARSET=latin1;

--
-- Dumping data for table `permissions`
--

INSERT INTO `permissions` (`permission_id`, `Description`) VALUES
(1, 'reserved for level 0'),
(2, 'reserved for level 0'),
(4, 'reserved for level 0'),
(8, 'reserved for level 1'),
(16, 'reserved for level 1'),
(32, 'reserved for level 1'),
(64, 'reserved for level 2'),
(128, 'reserved for level 2'),
(256, 'reserved for level 2'),
(512, 'reserved for level 3'),
(1024, 'reserved for level 3'),
(2048, 'reserved for level 3');

-- --------------------------------------------------------

--
-- Table structure for table `professionaltitles`
--

CREATE TABLE IF NOT EXISTS `professionaltitles` (
  `title_id` int(11) NOT NULL COMMENT 'Title ID',
  `title_name` varchar(500) DEFAULT NULL COMMENT 'Title Name',
  `permission` int(11) DEFAULT NULL COMMENT 'Title Permission',
  PRIMARY KEY (`title_id`)
) ENGINE=InnoDB DEFAULT CHARSET=latin1;

--
-- Dumping data for table `professionaltitles`
```

```
--

INSERT INTO `professionaltitles` (`title_id`, `title_name`, `permission`)
VALUES
(1, 'Inventory staff - Common / administrative', 512),
(2, 'Inventory staff - Per department', 8),
(3, 'Inventory staff - Per faculty', 64),
(4, 'Full-time Faculty', 1),
(5, 'Part-time Faculty', 1),
(6, 'University Administration', 1024),
(7, 'IT Group', 2048),
(8, 'Research assistants', 1),
(9, 'Research associates', 1),
(10, 'Students - diploma', 1),
(11, 'Students - master''s thesis option', 1),
(12, 'Students - master''s course option', 1),
(13, 'Students - PhD', 1),
(14, 'Security', 1);

-- --------------------------------------------------------

--
-- Table structure for table `requests`
--

CREATE TABLE IF NOT EXISTS `requests` (
  `req_id` int(11) NOT NULL AUTO_INCREMENT COMMENT 'Request ID',
  `requester` int(11) DEFAULT NULL COMMENT 'Requester ',
  `request_type` int(11) DEFAULT NULL COMMENT 'Request Type',
  `submitted_by` int(11) DEFAULT NULL COMMENT 'Submittedby',
  `item_id` int(11) DEFAULT NULL COMMENT 'Item ID',
  `description` varchar(500) DEFAULT NULL COMMENT 'Request
Description',
  `date_submitted` datetime DEFAULT NULL COMMENT 'Submitted Date',
  `approved_by` int(11) DEFAULT NULL COMMENT 'Approvedby',
  `date_approved` datetime DEFAULT NULL COMMENT 'Approved Date',
  `Status` varchar(10) DEFAULT NULL COMMENT 'Request Status',
  `date_modified` datetime DEFAULT NULL COMMENT 'Modified Date',
  PRIMARY KEY (`req_id`)
) ENGINE=InnoDB  DEFAULT CHARSET=latin1 AUTO_INCREMENT=1 ;


-- --------------------------------------------------------
```

```
--
-- Table structure for table `requesttypes`
--

CREATE TABLE IF NOT EXISTS `requesttypes` (
  `req_type_id` int(11) NOT NULL COMMENT 'Request Type ID',
  `req_type_code` varchar(100) DEFAULT NULL COMMENT 'Request Type
Code',
  `description` varchar(500) DEFAULT NULL COMMENT 'Request
Description',
  `permission` int(11) DEFAULT NULL COMMENT 'Request permission',
  PRIMARY KEY (`req_type_id`)
) ENGINE=InnoDB DEFAULT CHARSET=latin1;

--
-- Dumping data for table `requesttypes`
--

INSERT INTO `requesttypes` (`req_type_id`, `req_type_code`,
`description`, `permission`) VALUES
(1, 'General Request', 'If you lost an item or found an item and want to
check it with the administrator.  Barcode or Serial Number of the item isn''t
required.', NULL),
(2, 'Report a problem', 'If you found there''s problem with an item and want
to report it to the administrator. You need to provide the Barcode or Serial
Number of the item.', 0),
(3, 'Return back', 'An item was returned. The barcode or serial number is
required.', 0),
(4, 'Moving', 'An item was moved from one location to another. Barcode or
serial number is required.', 0),
(5, 'Request for', 'Ask for an item. Barcode or serial number isn''t required.',
0),
(6, 'Discard', 'An item was discard or wtite off. Barcode or serial number is
required.', 0);

-- --------------------------------------------------------

--
-- Table structure for table `tablelist`
--

CREATE TABLE IF NOT EXISTS `tablelist` (
```

```
  `table_id` int(11) NOT NULL AUTO_INCREMENT COMMENT 'Table ID',
  `table_code` varchar(500) DEFAULT NULL COMMENT 'Table Code',
  `table_name` varchar(500) DEFAULT NULL COMMENT 'Table Name',
  `permissions` int(11) DEFAULT NULL COMMENT 'Permission',
  PRIMARY KEY (`table_id`)
) ENGINE=InnoDB DEFAULT CHARSET=latin1 AUTO_INCREMENT=1 ;

--
-- Dumping data for table `tablelist`
--


-- --------------------------------------------------------

--
-- Table structure for table `userinfo`
--

CREATE TABLE IF NOT EXISTS `userinfo` (
  `user_id` int(11) NOT NULL COMMENT 'User ID',
  `email` varchar(500) DEFAULT NULL COMMENT 'Email',
  `dob` date DEFAULT NULL COMMENT 'Birthday',
  `home_phone` varchar(50) DEFAULT NULL COMMENT 'Home Phone',
  `cell_phone` varchar(50) DEFAULT NULL COMMENT 'Cell Phone',
  `street_address` varchar(500) DEFAULT NULL COMMENT 'Address',
  PRIMARY KEY (`user_id`)
) ENGINE=InnoDB DEFAULT CHARSET=latin1;

--
-- Dumping data for table `userinfo`
--


-- --------------------------------------------------------

--
-- Table structure for table `userroles`
--

CREATE TABLE IF NOT EXISTS `userroles` (
  `user_role_id` int(11) NOT NULL AUTO_INCREMENT COMMENT 'User Role
ID',
  `user_id` int(11) DEFAULT NULL COMMENT 'User ID',
```

```
  `title_id` int(11) DEFAULT NULL COMMENT 'Title ID',
  `affln_id` int(11) DEFAULT NULL COMMENT 'Affiliation ID',
  `status` varchar(10) DEFAULT NULL COMMENT 'User Role Status',
  PRIMARY KEY (`user_role_id`)
) ENGINE=InnoDB  DEFAULT CHARSET=latin1 AUTO_INCREMENT=2 ;

--
-- Dumping data for table `userroles`
--

INSERT INTO `userroles` (`user_role_id`, `user_id`, `title_id`, `affln_id`,
`status`) VALUES
(1, 1, 1, 0, NULL);
-- --------------------------------------------------------

--
-- Table structure for table `users`
--

CREATE TABLE IF NOT EXISTS `users` (
  `user_id` int(11) NOT NULL AUTO_INCREMENT COMMENT 'User ID',
  `user_code` varchar(500) DEFAULT NULL COMMENT 'User Code',
  `last_name` varchar(500) DEFAULT NULL COMMENT 'Last Name',
  `first_name` varchar(500) DEFAULT NULL COMMENT 'First Name',
  `password` varchar(50) DEFAULT NULL COMMENT 'Password',
  `date_modified` timestamp NULL DEFAULT CURRENT_TIMESTAMP ON
UPDATE CURRENT_TIMESTAMP COMMENT 'Modified Date',
  `login_attempts` int(11) DEFAULT NULL COMMENT 'Login Attempts',
  `loc_id` int(11) DEFAULT NULL,
  PRIMARY KEY (`user_id`)
) ENGINE=InnoDB  DEFAULT CHARSET=latin1 AUTO_INCREMENT=2 ;

--
-- Dumping data for table `users`
--

INSERT INTO `users` (`user_id`, `user_code`, `last_name`, `first_name`,
`password`, `date_modified`, `login_attempts`, `loc_id`) VALUES
(1, 'admin', 'System', 'Administrator', 'teamtwo', '2010-04-22 02:14:42',
NULL, NULL);
```

**Appendix III.2. httpd.conf [5,6]**

```
#
# This is the main Apache server configuration file.  It contains the
# configuration directives that give the server its instructions.
# See <URL:http://httpd.apache.org/docs/2.2/> for detailed information.
# In particular, see
# <URL:http://httpd.apache.org/docs/2.2/mod/directives.html>
# for a discussion of each configuration directive.
#
#
# Do NOT simply read the instructions in here without understanding
# what they do.  They're here only as hints or reminders.  If you are unsure
# consult the online docs. You have been warned.
#
# The configuration directives are grouped into three basic sections:
#  1. Directives that control the operation of the Apache server process as a
#     whole (the 'global environment').
#  2. Directives that define the parameters of the 'main' or 'default' server,
#     which responds to requests that aren't handled by a virtual host.
#     These directives also provide default values for the settings
#     of all virtual hosts.
#  3. Settings for virtual hosts, which allow Web requests to be sent to
#     different IP addresses or hostnames and have them handled by the
#     same Apache server process.
#
# Configuration and logfile names: If the filenames you specify for many
# of the server's control files begin with "/" (or "drive:/" for Win32), the
# server will use that explicit path.  If the filenames do *not* begin
# with "/", the value of ServerRoot is prepended -- so "logs/foo.log"
# with ServerRoot set to "/etc/httpd" will be interpreted by the
# server as "/etc/httpd/logs/foo.log".
#

### Section 1: Global Environment
#
# The directives in this section affect the overall operation of Apache,
# such as the number of concurrent requests it can handle or where it
# can find its configuration files.
#

#
# Don't give away too much information about all the subcomponents
# we are running.  Comment out this line if you don't mind remote sites
# finding out what major optional modules you are running
```

```
ServerTokens OS

#
# ServerRoot: The top of the directory tree under which the server's
# configuration, error, and log files are kept.
#
# NOTE!  If you intend to place this on an NFS (or otherwise network)
# mounted filesystem then please read the LockFile documentation
# (available at
<URL:http://httpd.apache.org/docs/2.2/mod/mpm_common.html#lockfile>)
;
# you will save yourself a lot of trouble.
#
# Do NOT add a slash at the end of the directory path.
#
ServerRoot "/etc/httpd"

#
# PidFile: The file in which the server should record its process
# identification number when it starts.
#
PidFile run/httpd.pid

#
# Timeout: The number of seconds before receives and sends time out.
#
Timeout 120

#
# KeepAlive: Whether or not to allow persistent connections (more than
# one request per connection). Set to "Off" to deactivate.
#
KeepAlive Off

#
# MaxKeepAliveRequests: The maximum number of requests to allow
# during a persistent connection. Set to 0 to allow an unlimited amount.
# We recommend you leave this number high, for maximum performance.
#
MaxKeepAliveRequests 100

#
```

```
# KeepAliveTimeout: Number of seconds to wait for the next request from
the
# same client on the same connection.
#
KeepAliveTimeout 15

##
## Server-Pool Size Regulation (MPM specific)
##

# prefork MPM
# StartServers: number of server processes to start
# MinSpareServers: minimum number of server processes which are kept
spare
# MaxSpareServers: maximum number of server processes which are kept
spare
# ServerLimit: maximum value for MaxClients for the lifetime of the server
# MaxClients: maximum number of server processes allowed to start
# MaxRequestsPerChild: maximum number of requests a server process
serves
<IfModule prefork.c>
StartServers       8
MinSpareServers    5
MaxSpareServers   20
ServerLimit      256
MaxClients       256
MaxRequestsPerChild  4000
</IfModule>

# worker MPM
# StartServers: initial number of server processes to start
# MaxClients: maximum number of simultaneous client connections
# MinSpareThreads: minimum number of worker threads which are kept
spare
# MaxSpareThreads: maximum number of worker threads which are kept
spare
# ThreadsPerChild: constant number of worker threads in each server
process
# MaxRequestsPerChild: maximum number of requests a server process
serves
<IfModule worker.c>
StartServers         2
MaxClients         150
```

```
MinSpareThreads     25
MaxSpareThreads     75
ThreadsPerChild     25
MaxRequestsPerChild  0
</IfModule>

#
# Listen: Allows you to bind Apache to specific IP addresses and/or
# ports, in addition to the default. See also the <VirtualHost>
# directive.
#
# Change this to Listen on specific IP addresses as shown below to
# prevent Apache from glomming onto all bound IP addresses (0.0.0.0)
#
#Listen 12.34.56.78:80
Listen 80

#
# Dynamic Shared Object (DSO) Support
#
# To be able to use the functionality of a module which was built as a DSO
you
# have to place corresponding `LoadModule' lines at this location so the
# directives contained in it are actually available _before_ they are used.
# Statically compiled modules (those listed by `httpd -l') do not need
# to be loaded here.
#
# Example:
# LoadModule foo_module modules/mod_foo.so
#
LoadModule auth_basic_module modules/mod_auth_basic.so
LoadModule auth_digest_module modules/mod_auth_digest.so
LoadModule authn_file_module modules/mod_authn_file.so
LoadModule authn_alias_module modules/mod_authn_alias.so
LoadModule authn_anon_module modules/mod_authn_anon.so
LoadModule authn_dbm_module modules/mod_authn_dbm.so
LoadModule authn_default_module modules/mod_authn_default.so
LoadModule authz_host_module modules/mod_authz_host.so
LoadModule authz_user_module modules/mod_authz_user.so
LoadModule authz_owner_module modules/mod_authz_owner.so
LoadModule authz_groupfile_module modules/mod_authz_groupfile.so
LoadModule authz_dbm_module modules/mod_authz_dbm.so
LoadModule authz_default_module modules/mod_authz_default.so
```

```
LoadModule ldap_module modules/mod_ldap.so
LoadModule authnz_ldap_module modules/mod_authnz_ldap.so
LoadModule include_module modules/mod_include.so
LoadModule log_config_module modules/mod_log_config.so
LoadModule logio_module modules/mod_logio.so
LoadModule env_module modules/mod_env.so
LoadModule ext_filter_module modules/mod_ext_filter.so
LoadModule mime_magic_module modules/mod_mime_magic.so
LoadModule expires_module modules/mod_expires.so
LoadModule deflate_module modules/mod_deflate.so
LoadModule headers_module modules/mod_headers.so
LoadModule usertrack_module modules/mod_usertrack.so
LoadModule setenvif_module modules/mod_setenvif.so
LoadModule mime_module modules/mod_mime.so
LoadModule dav_module modules/mod_dav.so
LoadModule status_module modules/mod_status.so
LoadModule autoindex_module modules/mod_autoindex.so
LoadModule info_module modules/mod_info.so
LoadModule dav_fs_module modules/mod_dav_fs.so
LoadModule vhost_alias_module modules/mod_vhost_alias.so
LoadModule negotiation_module modules/mod_negotiation.so
LoadModule dir_module modules/mod_dir.so
LoadModule actions_module modules/mod_actions.so
LoadModule speling_module modules/mod_speling.so
LoadModule userdir_module modules/mod_userdir.so
LoadModule alias_module modules/mod_alias.so
LoadModule rewrite_module modules/mod_rewrite.so
LoadModule proxy_module modules/mod_proxy.so
LoadModule proxy_balancer_module modules/mod_proxy_balancer.so
LoadModule proxy_ftp_module modules/mod_proxy_ftp.so
LoadModule proxy_http_module modules/mod_proxy_http.so
LoadModule proxy_connect_module modules/mod_proxy_connect.so
LoadModule cache_module modules/mod_cache.so
LoadModule suexec_module modules/mod_suexec.so
LoadModule disk_cache_module modules/mod_disk_cache.so
LoadModule file_cache_module modules/mod_file_cache.so
LoadModule mem_cache_module modules/mod_mem_cache.so
LoadModule cgi_module modules/mod_cgi.so
LoadModule version_module modules/mod_version.so

#
# The following modules are not loaded by default:
#
```

```
#LoadModule cern_meta_module modules/mod_cern_meta.so
#LoadModule asis_module modules/mod_asis.so

#
# Load config files from the config directory "/etc/httpd/conf.d".
#
Include conf.d/*.conf

#
# ExtendedStatus controls whether Apache will generate "full" status
# information (ExtendedStatus On) or just basic information
(ExtendedStatus
# Off) when the "server-status" handler is called. The default is Off.
#
#ExtendedStatus On

#
# If you wish httpd to run as a different user or group, you must run
# httpd as root initially and it will switch.
#
# User/Group: The name (or #number) of the user/group to run httpd as.
#  . On SCO (ODT 3) use "User nouser" and "Group nogroup".
#  . On HPUX you may not be able to use shared memory as nobody, and
the
#    suggested workaround is to create a user www and use that user.
#  NOTE that some kernels refuse to setgid(Group) or semctl(IPC_SET)
#  when the value of (unsigned)Group is above 60000;
#  don't use Group #-1 on these systems!
#
User apache
Group apache

### Section 2: 'Main' server configuration
#
# The directives in this section set up the values used by the 'main'
# server, which responds to any requests that aren't handled by a
# <VirtualHost> definition.  These values also provide defaults for
# any <VirtualHost> containers you may define later in the file.
#
# All of these directives may appear inside <VirtualHost> containers,
# in which case these default settings will be overridden for the
# virtual host being defined.
#
```

```
#
# ServerAdmin: Your address, where problems with the server should be
# e-mailed.  This address appears on some server-generated pages, such
# as error documents.  e.g. admin@your-domain.com
#
ServerAdmin root@localhost

#
# ServerName gives the name and port that the server uses to identify
itself.
# This can often be determined automatically, but we recommend you
specify
# it explicitly to prevent problems during startup.
#
# If this is not set to valid DNS name for your host, server-generated
# redirections will not work.  See also the UseCanonicalName directive.
#
# If your host doesn't have a registered DNS name, enter its IP address
here.
# You will have to access it by its address anyway, and this will make
# redirections work in a sensible way.
#
#ServerName www.example.com:80

#
# UseCanonicalName: Determines how Apache constructs self-referencing
# URLs and the SERVER_NAME and SERVER_PORT variables.
# When set "Off", Apache will use the Hostname and Port supplied
# by the client.  When set "On", Apache will use the value of the
# ServerName directive.
#
UseCanonicalName Off

#
# DocumentRoot: The directory out of which you will serve your
# documents. By default, all requests are taken from this directory, but
# symbolic links and aliases may be used to point to other locations.
#
DocumentRoot "/var/www/html"

#
# Each directory to which Apache has access can be configured with respect
```

```
# to which services and features are allowed and/or disabled in that
# directory (and its subdirectories).
#
# First, we configure the "default" to be a very restrictive set of
# features.
#
<Directory />
   Options FollowSymLinks
   AllowOverride None
</Directory>

#
# Note that from this point forward you must specifically allow
# particular features to be enabled - so if something's not working as
# you might expect, make sure that you have specifically enabled it
# below.
#

#
# This should be changed to whatever you set DocumentRoot to.
#
<Directory "/var/www/html">

#
# Possible values for the Options directive are "None", "All",
# or any combination of:
#   Indexes Includes FollowSymLinks SymLinksifOwnerMatch ExecCGI
MultiViews
#
# Note that "MultiViews" must be named *explicitly* --- "Options All"
# doesn't give it to you.
#
# The Options directive is both complicated and important.  Please see
# http://httpd.apache.org/docs/2.2/mod/core.html#options
# for more information.
#
   Options Indexes FollowSymLinks

#
# AllowOverride controls what directives may be placed in .htaccess files.
# It can be "All", "None", or any combination of the keywords:
#   Options FileInfo AuthConfig Limit
#
```

```
    AllowOverride None

#
# Controls who can get stuff from this server.
#
    Order allow,deny
    Allow from all

</Directory>

#
# UserDir: The name of the directory that is appended onto a user's home
# directory if a ~user request is received.
#
# The path to the end user account 'public_html' directory must be
# accessible to the webserver userid.  This usually means that ~userid
# must have permissions of 711, ~userid/public_html must have
permissions
# of 755, and documents contained therein must be world-readable.
# Otherwise, the client will only receive a "403 Forbidden" message.
#
# See also: http://httpd.apache.org/docs/misc/FAQ.html#forbidden
#
<IfModule mod_userdir.c>
    #
    # UserDir is disabled by default since it can confirm the presence
    # of a username on the system (depending on home directory
    # permissions).
    #
    UserDir disable

    #
    # To enable requests to /~user/ to serve the user's public_html
    # directory, remove the "UserDir disable" line above, and uncomment
    # the following line instead:
    #
    #UserDir public_html

</IfModule>

#
# Control access to UserDir directories.  The following is an example
# for a site where these directories are restricted to read-only.
```

```
#
#<Directory /home/*/public_html>
#    AllowOverride FileInfo AuthConfig Limit
#    Options MultiViews Indexes SymLinksIfOwnerMatch IncludesNoExec
#    <Limit GET POST OPTIONS>
#        Order allow,deny
#        Allow from all
#    </Limit>
#    <LimitExcept GET POST OPTIONS>
#        Order deny,allow
#        Deny from all
#    </LimitExcept>
#</Directory>

#
# DirectoryIndex: sets the file that Apache will serve if a directory
# is requested.
#
# The index.html.var file (a type-map) is used to deliver content-
# negotiated documents.  The MultiViews Option can be used for the
# same purpose, but it is much slower.
#
DirectoryIndex index.html index.html.var

#
# AccessFileName: The name of the file to look for in each directory
# for additional configuration directives.  See also the AllowOverride
# directive.
#
AccessFileName .htaccess

#
# The following lines prevent .htaccess and .htpasswd files from being
# viewed by Web clients.
#
<Files ~ "^\.ht">
    Order allow,deny
    Deny from all
</Files>

#
# TypesConfig describes where the mime.types file (or equivalent) is
# to be found.
```

```
#
TypesConfig /etc/mime.types

#
# DefaultType is the default MIME type the server will use for a document
# if it cannot otherwise determine one, such as from filename extensions.
# If your server contains mostly text or HTML documents, "text/plain" is
# a good value.  If most of your content is binary, such as applications
# or images, you may want to use "application/octet-stream" instead to
# keep browsers from trying to display binary files as though they are
# text.
#
DefaultType text/plain

#
# The mod_mime_magic module allows the server to use various hints from
the
# contents of the file itself to determine its type.  The MIMEMagicFile
# directive tells the module where the hint definitions are located.
#
<IfModule mod_mime_magic.c>
#   MIMEMagicFile /usr/share/magic.mime
    MIMEMagicFile conf/magic
</IfModule>

#
# HostnameLookups: Log the names of clients or just their IP addresses
# e.g., www.apache.org (on) or 204.62.129.132 (off).
# The default is off because it'd be overall better for the net if people
# had to knowingly turn this feature on, since enabling it means that
# each client request will result in AT LEAST one lookup request to the
# nameserver.
#
HostnameLookups Off

#
# EnableMMAP: Control whether memory-mapping is used to deliver
# files (assuming that the underlying OS supports it).
# The default is on; turn this off if you serve from NFS-mounted
# filesystems.  On some systems, turning it off (regardless of
# filesystem) can improve performance; for details, please see
# http://httpd.apache.org/docs/2.2/mod/core.html#enablemmap
#
```

```
#EnableMMAP off

#
# EnableSendfile: Control whether the sendfile kernel support is
# used to deliver files (assuming that the OS supports it).
# The default is on; turn this off if you serve from NFS-mounted
# filesystems.  Please see
# http://httpd.apache.org/docs/2.2/mod/core.html#enablesendfile
#
#EnableSendfile off

#
# ErrorLog: The location of the error log file.
# If you do not specify an ErrorLog directive within a <VirtualHost>
# container, error messages relating to that virtual host will be
# logged here.  If you *do* define an error logfile for a <VirtualHost>
# container, that host's errors will be logged there and not here.
#
ErrorLog logs/error_log

#
# LogLevel: Control the number of messages logged to the error_log.
# Possible values include: debug, info, notice, warn, error, crit,
# alert, emerg.
#
LogLevel warn

#
# The following directives define some format nicknames for use with
# a CustomLog directive (see below).
#
LogFormat "%h %l %u %t \"%r\" %>s %b \"%{Referer}i\" \"%{User-
Agent}i\"" combined
LogFormat "%h %l %u %t \"%r\" %>s %b" common
LogFormat "%{Referer}i -> %U" referer
LogFormat "%{User-agent}i" agent

# "combinedio" includes actual counts of actual bytes received (%I) and
sent (%O); this
# requires the mod_logio module to be loaded.
#LogFormat "%h %l %u %t \"%r\" %>s %b \"%{Referer}i\" \"%{User-
Agent}i\" %I %O" combinedio
```

```
#
# The location and format of the access logfile (Common Logfile Format).
# If you do not define any access logfiles within a <VirtualHost>
# container, they will be logged here.  Contrariwise, if you *do*
# define per-<VirtualHost> access logfiles, transactions will be
# logged therein and *not* in this file.
#
#CustomLog logs/access_log common

#
# If you would like to have separate agent and referer logfiles, uncomment
# the following directives.
#
#CustomLog logs/referer_log referer
#CustomLog logs/agent_log agent

#
# For a single logfile with access, agent, and referer information
# (Combined Logfile Format), use the following directive:
#
CustomLog logs/access_log combined

#
# Optionally add a line containing the server version and virtual host
# name to server-generated pages (internal error documents, FTP directory
# listings, mod_status and mod_info output etc., but not CGI generated
# documents or custom error documents).
# Set to "EMail" to also include a mailto: link to the ServerAdmin.
# Set to one of:  On | Off | EMail
#
ServerSignature On

#
# Aliases: Add here as many aliases as you need (with no limit). The format
is
# Alias fakename realname
#
# Note that if you include a trailing / on fakename then the server will
# require it to be present in the URL.  So "/icons" isn't aliased in this
# example, only "/icons/".  If the fakename is slash-terminated, then the
# realname must also be slash terminated, and if the fakename omits the
# trailing slash, the realname must also omit it.
#
```

```
# We include the /icons/ alias for FancyIndexed directory listings.  If you
# do not use FancyIndexing, you may comment this out.
#
Alias /icons/ "/var/www/icons/"

<Directory "/var/www/icons">
    Options Indexes MultiViews
    AllowOverride None
    Order allow,deny
    Allow from all
</Directory>

#
# WebDAV module configuration section.
#
<IfModule mod_dav_fs.c>
    # Location of the WebDAV lock database.
    DAVLockDB /var/lib/dav/lockdb
</IfModule>

#
# ScriptAlias: This controls which directories contain server scripts.
# ScriptAliases are essentially the same as Aliases, except that
# documents in the realname directory are treated as applications and
# run by the server when requested rather than as documents sent to the
client.
# The same rules about trailing "/" apply to ScriptAlias directives as to
# Alias.
#
ScriptAlias /cgi-bin/ "/var/www/cgi-bin/"

#
# "/var/www/cgi-bin" should be changed to whatever your ScriptAliased
# CGI directory exists, if you have that configured.
#
<Directory "/var/www/cgi-bin">
    AllowOverride None
    Options None
    Order allow,deny
    Allow from all
</Directory>

#
```

```
# Redirect allows you to tell clients about documents which used to exist in
# your server's namespace, but do not anymore. This allows you to tell the
# clients where to look for the relocated document.
# Example:
# Redirect permanent /foo http://www.example.com/bar

#
# Directives controlling the display of server-generated directory listings.
#

#
# IndexOptions: Controls the appearance of server-generated directory
# listings.
#
IndexOptions FancyIndexing VersionSort NameWidth=* HTMLTable

#
# AddIcon* directives tell the server which icon to show for different
# files or filename extensions.  These are only displayed for
# FancyIndexed directories.
#
AddIconByEncoding (CMP,/icons/compressed.gif) x-compress x-gzip

AddIconByType (TXT,/icons/text.gif) text/*
AddIconByType (IMG,/icons/image2.gif) image/*
AddIconByType (SND,/icons/sound2.gif) audio/*
AddIconByType (VID,/icons/movie.gif) video/*

AddIcon /icons/binary.gif .bin .exe
AddIcon /icons/binhex.gif .hqx
AddIcon /icons/tar.gif .tar
AddIcon /icons/world2.gif .wrl .wrl.gz .vrml .vrm .iv
AddIcon /icons/compressed.gif .Z .z .tgz .gz .zip
AddIcon /icons/a.gif .ps .ai .eps
AddIcon /icons/layout.gif .html .shtml .htm .pdf
AddIcon /icons/text.gif .txt
AddIcon /icons/c.gif .c
AddIcon /icons/p.gif .pl .py
AddIcon /icons/f.gif .for
AddIcon /icons/dvi.gif .dvi
AddIcon /icons/uuencoded.gif .uu
AddIcon /icons/script.gif .conf .sh .shar .csh .ksh .tcl
AddIcon /icons/tex.gif .tex
```

```
AddIcon /icons/bomb.gif core

AddIcon /icons/back.gif ..
AddIcon /icons/hand.right.gif README
AddIcon /icons/folder.gif ^^DIRECTORY^^
AddIcon /icons/blank.gif ^^BLANKICON^^

#
# DefaultIcon is which icon to show for files which do not have an icon
# explicitly set.
#
DefaultIcon /icons/unknown.gif

#
# AddDescription allows you to place a short description after a file in
# server-generated indexes.  These are only displayed for FancyIndexed
# directories.
# Format: AddDescription "description" filename
#
#AddDescription "GZIP compressed document" .gz
#AddDescription "tar archive" .tar
#AddDescription "GZIP compressed tar archive" .tgz

#
# ReadmeName is the name of the README file the server will look for by
# default, and append to directory listings.
#
# HeaderName is the name of a file which should be prepended to
# directory indexes.
ReadmeName README.html
HeaderName HEADER.html

#
# IndexIgnore is a set of filenames which directory indexing should ignore
# and not include in the listing.  Shell-style wildcarding is permitted.
#
IndexIgnore .??* *~ *# HEADER* README* RCS CVS *,v *,t

#
# DefaultLanguage and AddLanguage allows you to specify the language of
# a document. You can then use content negotiation to give a browser a
# file in a language the user can understand.
#
```

```
# Specify a default language. This means that all data
# going out without a specific language tag (see below) will
# be marked with this one. You probably do NOT want to set
# this unless you are sure it is correct for all cases.
#
# * It is generally better to not mark a page as
# * being a certain language than marking it with the wrong
# * language!
#
# DefaultLanguage nl
#
# Note 1: The suffix does not have to be the same as the language
# keyword --- those with documents in Polish (whose net-standard
# language code is pl) may wish to use "AddLanguage pl .po" to
# avoid the ambiguity with the common suffix for perl scripts.
#
# Note 2: The example entries below illustrate that in some cases
# the two character 'Language' abbreviation is not identical to
# the two character 'Country' code for its country,
# E.g. 'Danmark/dk' versus 'Danish/da'.
#
# Note 3: In the case of 'ltz' we violate the RFC by using a three char
# specifier. There is 'work in progress' to fix this and get
# the reference data for rfc1766 cleaned up.
#
# Catalan (ca) - Croatian (hr) - Czech (cs) - Danish (da) - Dutch (nl)
# English (en) - Esperanto (eo) - Estonian (et) - French (fr) - German (de)
# Greek-Modern (el) - Hebrew (he) - Italian (it) - Japanese (ja)
# Korean (ko) - Luxembourgeois* (ltz) - Norwegian Nynorsk (nn)
# Norwegian (no) - Polish (pl) - Portugese (pt)
# Brazilian Portuguese (pt-BR) - Russian (ru) - Swedish (sv)
# Simplified Chinese (zh-CN) - Spanish (es) - Traditional Chinese (zh-TW)
#
AddLanguage ca .ca
AddLanguage cs .cz .cs
AddLanguage da .dk
AddLanguage de .de
AddLanguage el .el
AddLanguage en .en
AddLanguage eo .eo
AddLanguage es .es
AddLanguage et .et
AddLanguage fr .fr
```

```
AddLanguage he .he
AddLanguage hr .hr
AddLanguage it .it
AddLanguage ja .ja
AddLanguage ko .ko
AddLanguage ltz .ltz
AddLanguage nl .nl
AddLanguage nn .nn
AddLanguage no .no
AddLanguage pl .po
AddLanguage pt .pt
AddLanguage pt-BR .pt-br
AddLanguage ru .ru
AddLanguage sv .sv
AddLanguage zh-CN .zh-cn
AddLanguage zh-TW .zh-tw

#
# LanguagePriority allows you to give precedence to some languages
# in case of a tie during content negotiation.
#
# Just list the languages in decreasing order of preference. We have
# more or less alphabetized them here. You probably want to change this.
#
LanguagePriority en ca cs da de el eo es et fr he hr it ja ko ltz nl nn no pl pt
pt-BR ru sv zh-CN zh-TW

#
# ForceLanguagePriority allows you to serve a result page rather than
# MULTIPLE CHOICES (Prefer) [in case of a tie] or NOT ACCEPTABLE
(Fallback)
# [in case no accepted languages matched the available variants]
#
ForceLanguagePriority Prefer Fallback

#
# Specify a default charset for all content served; this enables
# interpretation of all content as UTF-8 by default.  To use the
# default browser choice (ISO-8859-1), or to allow the META tags
# in HTML content to override this choice, comment out this
# directive:
#
AddDefaultCharset UTF-8
```

```
#
# AddType allows you to add to or override the MIME configuration
# file mime.types for specific file types.
#
#AddType application/x-tar .tgz

#
# AddEncoding allows you to have certain browsers uncompress
# information on the fly. Note: Not all browsers support this.
# Despite the name similarity, the following Add* directives have nothing
# to do with the FancyIndexing customization directives above.
#
#AddEncoding x-compress .Z
#AddEncoding x-gzip .gz .tgz

# If the AddEncoding directives above are commented-out, then you
# probably should define those extensions to indicate media types:
#
AddType application/x-compress .Z
AddType application/x-gzip .gz .tgz

#
# AddHandler allows you to map certain file extensions to "handlers":
# actions unrelated to filetype. These can be either built into the server
# or added with the Action directive (see below)
#
# To use CGI scripts outside of ScriptAliased directories:
# (You will also need to add "ExecCGI" to the "Options" directive.)
#
#AddHandler cgi-script .cgi

#
# For files that include their own HTTP headers:
#
#AddHandler send-as-is asis

#
# For type maps (negotiated resources):
# (This is enabled by default to allow the Apache "It Worked" page
#  to be distributed in multiple languages.)
#
AddHandler type-map var
```

```
#
# Filters allow you to process content before it is sent to the client.
#
# To parse .shtml files for server-side includes (SSI):
# (You will also need to add "Includes" to the "Options" directive.)
#
AddType text/html .shtml
AddOutputFilter INCLUDES .shtml

#
# Action lets you define media types that will execute a script whenever
# a matching file is called. This eliminates the need for repeated URL
# pathnames for oft-used CGI file processors.
# Format: Action media/type /cgi-script/location
# Format: Action handler-name /cgi-script/location
#

#
# Customizable error responses come in three flavors:
# 1) plain text 2) local redirects 3) external redirects
#
# Some examples:
#ErrorDocument 500 "The server made a boo boo."
#ErrorDocument 404 /missing.html
#ErrorDocument 404 "/cgi-bin/missing_handler.pl"
#ErrorDocument 402 http://www.example.com/subscription_info.html
#

#
# Putting this all together, we can internationalize error responses.
#
# We use Alias to redirect any /error/HTTP_<error>.html.var response to
# our collection of by-error message multi-language collections.  We use
# includes to substitute the appropriate text.
#
# You can modify the messages' appearance without changing any of the
# default HTTP_<error>.html.var files by adding the line:
#
#   Alias /error/include/ "/your/include/path/"
#
# which allows you to create your own set of files by starting with the
# /var/www/error/include/ files and
```

```
# copying them to /your/include/path/, even on a per-VirtualHost basis.
#

Alias /error/ "/var/www/error/"

<IfModule mod_negotiation.c>
<IfModule mod_include.c>
   <Directory "/var/www/error">
      AllowOverride None
      Options IncludesNoExec
      AddOutputFilter Includes html
      AddHandler type-map var
      Order allow,deny
      Allow from all
      LanguagePriority en es de fr
      ForceLanguagePriority Prefer Fallback
   </Directory>

#    ErrorDocument 400 /error/HTTP_BAD_REQUEST.html.var
#    ErrorDocument 401 /error/HTTP_UNAUTHORIZED.html.var
#    ErrorDocument 403 /error/HTTP_FORBIDDEN.html.var
#    ErrorDocument 404 /error/HTTP_NOT_FOUND.html.var
#    ErrorDocument 405 /error/HTTP_METHOD_NOT_ALLOWED.html.var
#    ErrorDocument 408 /error/HTTP_REQUEST_TIME_OUT.html.var
#    ErrorDocument 410 /error/HTTP_GONE.html.var
#    ErrorDocument 411 /error/HTTP_LENGTH_REQUIRED.html.var
#    ErrorDocument 412 /error/HTTP_PRECONDITION_FAILED.html.var
#    ErrorDocument 413
/error/HTTP_REQUEST_ENTITY_TOO_LARGE.html.var
#    ErrorDocument 414 /error/HTTP_REQUEST_URI_TOO_LARGE.html.var
#    ErrorDocument 415 /error/HTTP_UNSUPPORTED_MEDIA_TYPE.html.var
#    ErrorDocument 500 /error/HTTP_INTERNAL_SERVER_ERROR.html.var
#    ErrorDocument 501 /error/HTTP_NOT_IMPLEMENTED.html.var
#    ErrorDocument 502 /error/HTTP_BAD_GATEWAY.html.var
#    ErrorDocument 503 /error/HTTP_SERVICE_UNAVAILABLE.html.var
#    ErrorDocument 506 /error/HTTP_VARIANT_ALSO_VARIES.html.var

</IfModule>
</IfModule>

#
# The following directives modify normal HTTP response behavior to
# handle known problems with browser implementations.
```

```
#
BrowserMatch "Mozilla/2" nokeepalive
BrowserMatch "MSIE 4\.0b2;" nokeepalive downgrade-1.0 force-response-
1.0
BrowserMatch "RealPlayer 4\.0" force-response-1.0
BrowserMatch "Java/1\.0" force-response-1.0
BrowserMatch "JDK/1\.0" force-response-1.0

#
# The following directive disables redirects on non-GET requests for
# a directory that does not include the trailing slash.  This fixes a
# problem with Microsoft WebFolders which does not appropriately handle
# redirects for folders with DAV methods.
# Same deal with Apple's DAV filesystem and Gnome VFS support for DAV.
#
BrowserMatch "Microsoft Data Access Internet Publishing Provider" redirect-
carefully
BrowserMatch "MS FrontPage" redirect-carefully
BrowserMatch "^WebDrive" redirect-carefully
BrowserMatch "^WebDAVFS/1.[0123]" redirect-carefully
BrowserMatch "^gnome-vfs/1.0" redirect-carefully
BrowserMatch "^XML Spy" redirect-carefully
BrowserMatch "^Dreamweaver-WebDAV-SCM1" redirect-carefully

#
# Allow server status reports generated by mod_status,
# with the URL of http://servername/server-status
# Change the ".example.com" to match your domain to enable.
#
#<Location /server-status>
#    SetHandler server-status
#    Order deny,allow
#    Deny from all
#    Allow from .example.com
#</Location>

#
# Allow remote server configuration reports, with the URL of
#  http://servername/server-info (requires that mod_info.c be loaded).
# Change the ".example.com" to match your domain to enable.
#
#<Location /server-info>
#    SetHandler server-info
```

```
#    Order deny,allow
#    Deny from all
#    Allow from .example.com
#</Location>

#
# Proxy Server directives. Uncomment the following lines to
# enable the proxy server:
#
#<IfModule mod_proxy.c>
#ProxyRequests On
#
#<Proxy *>
#    Order deny,allow
#    Deny from all
#    Allow from .example.com
#</Proxy>

#
# Enable/disable the handling of HTTP/1.1 "Via:" headers.
# ("Full" adds the server version; "Block" removes all outgoing Via: headers)
# Set to one of: Off | On | Full | Block
#
#ProxyVia On

#
# To enable a cache of proxied content, uncomment the following lines.
# See http://httpd.apache.org/docs/2.2/mod/mod_cache.html for more
details.
#
#<IfModule mod_disk_cache.c>
#   CacheEnable disk /
#   CacheRoot "/var/cache/mod_proxy"
#</IfModule>
#

#</IfModule>
# End of proxy directives.

### Section 3: Virtual Hosts
#
# VirtualHost: If you want to maintain multiple domains/hostnames on your
```

```
# machine you can setup VirtualHost containers for them. Most
configurations
# use only name-based virtual hosts so the server doesn't need to worry
about
# IP addresses. This is indicated by the asterisks in the directives below.
#
# Please see the documentation at
# <URL:http://httpd.apache.org/docs/2.2/vhosts/>
# for further details before you try to setup virtual hosts.
#
# You may use the command line option '-S' to verify your virtual host
# configuration.

#
# Use name-based virtual hosting.
#
#NameVirtualHost *:80
#
# NOTE: NameVirtualHost cannot be used without a port specifier
# (e.g. :80) if mod_ssl is being used, due to the nature of the
# SSL protocol.
#

#
# VirtualHost example:
# Almost any Apache directive may go into a VirtualHost container.
# The first VirtualHost section is used for requests without a known
# server name.
#
#<VirtualHost *:80>
#    ServerAdmin webmaster@dummy-host.example.com
#    DocumentRoot /www/docs/dummy-host.example.com
#    ServerName dummy-host.example.com
#    ErrorLog logs/dummy-host.example.com-error_log
#    CustomLog logs/dummy-host.example.com-access_log common
#</VirtualHost>
```

**Appendix III.3. php.ini**

```
[PHP]

;;;;;;;;;;;;;;;;;;;
; About php.ini   ;
```

```
;;;;;;;;;;;;;;;;;;;
; This file controls many aspects of PHP's behavior.  In order for PHP to
; read it, it must be named 'php.ini'.  PHP looks for it in the current
; working directory, in the path designated by the environment variable
; PHPRC, and in the path that was defined in compile time (in that order).
; Under Windows, the compile-time path is the Windows directory.  The
; path in which the php.ini file is looked for can be overridden using
; the -c argument in command line mode.
;
; The syntax of the file is extremely simple.  Whitespace and Lines
; beginning with a semicolon are silently ignored (as you probably guessed).
; Section headers (e.g. [Foo]) are also silently ignored, even though
; they might mean something in the future.
;
; Directives are specified using the following syntax:
; directive = value
; Directive names are *case sensitive* - foo=bar is different from FOO=bar.
;
; The value can be a string, a number, a PHP constant (e.g. E_ALL or M_PI),
one
; of the INI constants (On, Off, True, False, Yes, No and None) or an
expression
; (e.g. E_ALL & ~E_NOTICE), or a quoted string ("foo").
;
; Expressions in the INI file are limited to bitwise operators and
parentheses:
; |        bitwise OR
; &        bitwise AND
; ~        bitwise NOT
; !        boolean NOT
;
; Boolean flags can be turned on using the values 1, On, True or Yes.
; They can be turned off using the values 0, Off, False or No.
;
; An empty string can be denoted by simply not writing anything after the
equal
; sign, or by using the None keyword:
;
;  foo =         ; sets foo to an empty string
;  foo = none    ; sets foo to an empty string
;  foo = "none"  ; sets foo to the string 'none'
;
; If you use constants in your value, and these constants belong to a
```

```
; dynamically loaded extension (either a PHP extension or a Zend
extension),
; you may only use these constants *after* the line that loads the extension.
;
;
;;;;;;;;;;;;;;;;;;;
; About this file ;
;;;;;;;;;;;;;;;;;;;
; This is the recommended, PHP 5-style version of the php.ini-dist file.  It
; sets some non standard settings, that make PHP more efficient, more
secure,
; and encourage cleaner coding.
;
; The price is that with these settings, PHP may be incompatible with some
; applications, and sometimes, more difficult to develop with.  Using this
; file is warmly recommended for production sites.  As all of the changes
from
; the standard settings are thoroughly documented, you can go over each
one,
; and decide whether you want to use it or not.
;
; For general information about the php.ini file, please consult the php.ini-
dist
; file, included in your PHP distribution.
;
; This file is different from the php.ini-dist file in the fact that it features
; different values for several directives, in order to improve performance,
while
; possibly breaking compatibility with the standard out-of-the-box behavior
of
; PHP.  Please make sure you read what's different, and modify your scripts
; accordingly, if you decide to use this file instead.
;
; - register_globals = Off         [Security, Performance]
;     Global variables are no longer registered for input data (POST, GET,
cookies,
;     environment and other server variables).  Instead of using $foo, you
must use
;     you can use $_REQUEST["foo"] (includes any variable that arrives
through the
;     request, namely, POST, GET and cookie variables), or use one of the
specific
```

```
;    $_GET["foo"], $_POST["foo"], $_COOKIE["foo"] or $_FILES["foo"],
depending
;    on where the input originates.  Also, you can look at the
;    import_request_variables() function.
;    Note that register_globals is going to be depracated (i.e., turned off by
;    default) in the next version of PHP, because it often leads to security
bugs.
;    Read http://php.net/manual/en/security.registerglobals.php for further
;    information.
; - register_long_arrays = Off     [Performance]
;    Disables registration of the older (and deprecated) long predefined array
;    variables ($HTTP_*_VARS).  Instead, use the superglobals that were
;    introduced in PHP 4.1.0
; - display_errors = Off           [Security]
;    With this directive set to off, errors that occur during the execution of
;    scripts will no longer be displayed as a part of the script output, and
thus,
;    will no longer be exposed to remote users.  With some errors, the error
message
;    content may expose information about your script, web server, or
database
;    server that may be exploitable for hacking.  Production sites should
have this
;    directive set to off.
; - log_errors = On                [Security]
;    This directive complements the above one.  Any errors that occur during
the
;    execution of your script will be logged (typically, to your server's error
log,
;    but can be configured in several ways).  Along with setting
display_errors to off,
;    this setup gives you the ability to fully understand what may have gone
wrong,
;    without exposing any sensitive information to remote users.
; - output_buffering = 4096        [Performance]
;    Set a 4KB output buffer.  Enabling output buffering typically results in
less
;    writes, and sometimes less packets sent on the wire, which can often
lead to
;    better performance.  The gain this directive actually yields greatly
depends
;    on which Web server you're working with, and what kind of scripts
you're using.
```

```
; - register_argc_argv = Off       [Performance]
;     Disables registration of the somewhat redundant $argv and $argc global
;     variables.
; - magic_quotes_gpc = Off         [Performance]
;     Input data is no longer escaped with slashes so that it can be sent into
;     SQL databases without further manipulation.  Instead, you should use
the
;     function addslashes() on each input element you wish to send to a
database.
; - variables_order = "GPCS"       [Performance]
;     The environment variables are not hashed into the $_ENV.  To access
;     environment variables, you can use getenv() instead.
; - error_reporting = E_ALL        [Code Cleanliness, Security(?)]
;     By default, PHP surpresses errors of type E_NOTICE.  These error
messages
;     are emitted for non-critical errors, but that could be a symptom of a
bigger
;     problem.  Most notably, this will cause error messages about the use
;     of uninitialized variables to be displayed.
; - allow_call_time_pass_reference = Off     [Code cleanliness]
;     It's not possible to decide to force a variable to be passed by reference
;     when calling a function.  The PHP 4 style to do this is by making the
;     function require the relevant argument by reference.


;;;;;;;;;;;;;;;;;;;;
; Language Options ;
;;;;;;;;;;;;;;;;;;;;

; Enable the PHP scripting language engine under Apache.
engine = On

; Enable compatibility mode with Zend Engine 1 (PHP 4.x)
zend.ze1_compatibility_mode = Off

; Allow the <? tag.  Otherwise, only <?php and <script> tags are
recognized.
; NOTE: Using short tags should be avoided when developing applications or
; libraries that are meant for redistribution, or deployment on PHP
; servers which are not under your control, because short tags may not
; be supported on the target server. For portable, redistributable code,
; be sure not to use short tags.
short_open_tag = On
```

```
; Allow ASP-style <% %> tags.
asp_tags = Off

; The number of significant digits displayed in floating point numbers.
precision    =  14

; Enforce year 2000 compliance (will cause problems with non-compliant
browsers)
y2k_compliance = On

; Output buffering allows you to send header lines (including cookies) even
; after you send body content, at the price of slowing PHP's output layer a
; bit.  You can enable output buffering during runtime by calling the output
; buffering functions.  You can also enable output buffering for all files by
; setting this directive to On.  If you wish to limit the size of the buffer
; to a certain size - you can use a maximum number of bytes instead of 'On',
as
; a value for this directive (e.g., output_buffering=4096).
output_buffering = 4096

; You can redirect all of the output of your scripts to a function.  For
; example, if you set output_handler to "mb_output_handler", character
; encoding will be transparently converted to the specified encoding.
; Setting any output handler automatically turns on output buffering.
; Note: People who wrote portable scripts should not depend on this ini
;       directive. Instead, explicitly set the output handler using ob_start().
;       Using this ini directive may cause problems unless you know what
script
;       is doing.
; Note: You cannot use both "mb_output_handler" with "ob_iconv_handler"
;       and you cannot use both "ob_gzhandler" and
"zlib.output_compression".
; Note: output_handler must be empty if this is set 'On' !!!!
;       Instead you must use zlib.output_handler.
;output_handler =

; Transparent output compression using the zlib library
; Valid values for this option are 'off', 'on', or a specific buffer size
; to be used for compression (default is 4KB)
; Note: Resulting chunk size may vary due to nature of compression. PHP
;       outputs chunks that are few hundreds bytes each as a result of
;       compression. If you prefer a larger chunk size for better
```

```
;       performance, enable output_buffering in addition.
; Note: You need to use zlib.output_handler instead of the standard
;       output_handler, or otherwise the output will be corrupted.
zlib.output_compression = Off

; You cannot specify additional output handlers if zlib.output_compression
; is activated here. This setting does the same as output_handler but in
; a different order.
;zlib.output_handler =

; Implicit flush tells PHP to tell the output layer to flush itself
; automatically after every output block.  This is equivalent to calling the
; PHP function flush() after each and every call to print() or echo() and each
; and every HTML block.  Turning this option on has serious performance
; implications and is generally recommended for debugging purposes only.
implicit_flush = Off

; The unserialize callback function will be called (with the undefined class'
; name as parameter), if the unserializer finds an undefined class
; which should be instantiated.
; A warning appears if the specified function is not defined, or if the
; function doesn't include/implement the missing class.
; So only set this entry, if you really want to implement such a
; callback-function.
unserialize_callback_func=

; When floats & doubles are serialized store serialize_precision significant
; digits after the floating point. The default value ensures that when floats
; are decoded with unserialize, the data will remain the same.
serialize_precision = 100

; Whether to enable the ability to force arguments to be passed by reference
; at function call time.  This method is deprecated and is likely to be
; unsupported in future versions of PHP/Zend.  The encouraged method of
; specifying which arguments should be passed by reference is in the
function
; declaration.  You're encouraged to try and turn this option Off and make
; sure your scripts work properly with it in order to ensure they will work
; with future versions of the language (you will receive a warning each time
; you use this feature, and the argument will be passed by value instead of
by
; reference).
allow_call_time_pass_reference = Off
```

```
;
; Safe Mode
;
safe_mode = Off

; By default, Safe Mode does a UID compare check when
; opening files. If you want to relax this to a GID compare,
; then turn on safe_mode_gid.
safe_mode_gid = Off

; When safe_mode is on, UID/GID checks are bypassed when
; including files from this directory and its subdirectories.
; (directory must also be in include_path or full path must
; be used when including)
safe_mode_include_dir =

; When safe_mode is on, only executables located in the
safe_mode_exec_dir
; will be allowed to be executed via the exec family of functions.
safe_mode_exec_dir =

; Setting certain environment variables may be a potential security breach.
; This directive contains a comma-delimited list of prefixes.  In Safe Mode,
; the user may only alter environment variables whose names begin with the
; prefixes supplied here.  By default, users will only be able to set
; environment variables that begin with PHP_ (e.g. PHP_FOO=BAR).
;
; Note:  If this directive is empty, PHP will let the user modify ANY
; environment variable!
safe_mode_allowed_env_vars = PHP_

; This directive contains a comma-delimited list of environment variables
that
; the end user won't be able to change using putenv().  These variables will
be
; protected even if safe_mode_allowed_env_vars is set to allow to change
them.
safe_mode_protected_env_vars = LD_LIBRARY_PATH

; open_basedir, if set, limits all file operations to the defined directory
; and below.  This directive makes most sense if used in a per-directory
; or per-virtualhost web server configuration file. This directive is
```

```
; *NOT* affected by whether Safe Mode is turned On or Off.
;open_basedir =

; This directive allows you to disable certain functions for security reasons.
; It receives a comma-delimited list of function names. This directive is
; *NOT* affected by whether Safe Mode is turned On or Off.
disable_functions =

; This directive allows you to disable certain classes for security reasons.
; It receives a comma-delimited list of class names. This directive is
; *NOT* affected by whether Safe Mode is turned On or Off.
disable_classes =

; Colors for Syntax Highlighting mode.  Anything that's acceptable in
; <span style="color: ???????"> would work.
;highlight.string  = #DD0000
;highlight.comment = #FF9900
;highlight.keyword = #007700
;highlight.bg      = #FFFFFF
;highlight.default = #0000BB
;highlight.html    = #000000

; If enabled, the request will be allowed to complete even if the user aborts
; the request. Consider enabling it if executing long request, which may end
up
; being interrupted by the user or a browser timing out.
; ignore_user_abort = On

; Determines the size of the realpath cache to be used by PHP. This value
should
; be increased on systems where PHP opens many files to reflect the
quantity of
; the file operations performed.
; realpath_cache_size=16k

; Duration of time, in seconds for which to cache realpath information for a
given
; file or directory. For systems with rarely changing files, consider increasing
this
; value.
; realpath_cache_ttl=120

;
```

```
; Misc
;
; Decides whether PHP may expose the fact that it is installed on the server
; (e.g. by adding its signature to the Web server header).  It is no security
; threat in any way, but it makes it possible to determine whether you use
PHP
; on your server or not.
expose_php = On


;;;;;;;;;;;;;;;;;;;
; Resource Limits ;
;;;;;;;;;;;;;;;;;;;

max_execution_time = 30     ; Maximum execution time of each script, in
seconds
max_input_time = 60   ; Maximum amount of time each script may spend
parsing request data
memory_limit = 16M      ; Maximum amount of memory a script may
consume


;;;;;;;;;;;;;;;;;;;;;;;;;;;;;;
; Error handling and logging ;
;;;;;;;;;;;;;;;;;;;;;;;;;;;;;;

; error_reporting is a bit-field.  Or each number up to get desired error
; reporting level
; E_ALL             - All errors and warnings (doesn't include E_STRICT)
; E_ERROR           - fatal run-time errors
; E_WARNING         - run-time warnings (non-fatal errors)
; E_PARSE           - compile-time parse errors
; E_NOTICE          - run-time notices (these are warnings which often result
;                   from a bug in your code, but it's possible that it was
;                   intentional (e.g., using an uninitialized variable and
;                   relying on the fact it's automatically initialized to an
;                   empty string)
; E_STRICT          - run-time notices, enable to have PHP suggest changes
;                   to your code which will ensure the best interoperability
;                   and forward compatibility of your code
; E_CORE_ERROR      - fatal errors that occur during PHP's initial startup
; E_CORE_WARNING    - warnings (non-fatal errors) that occur during PHP's
;                   initial startup
```

```
; E_COMPILE_ERROR   - fatal compile-time errors
; E_COMPILE_WARNING - compile-time warnings (non-fatal errors)
; E_USER_ERROR      - user-generated error message
; E_USER_WARNING    - user-generated warning message
; E_USER_NOTICE     - user-generated notice message
;
; Examples:
;
;   - Show all errors, except for notices and coding standards warnings
;
;error_reporting = E_ALL & ~E_NOTICE
;
;   - Show all errors, except for notices
;
;error_reporting = E_ALL & ~E_NOTICE | E_STRICT
;
;   - Show only errors
;
;error_reporting = E_COMPILE_ERROR|E_ERROR|E_CORE_ERROR
;
;   - Show all errors, except coding standards warnings
;
error_reporting  =  E_ALL

; Print out errors (as a part of the output).  For production web sites,
; you're strongly encouraged to turn this feature off, and use error logging
; instead (see below).  Keeping display_errors enabled on a production web
site
; may reveal security information to end users, such as file paths on your
Web
; server, your database schema or other information.
;display_errors = Off
display_errors = On

; Even when display_errors is on, errors that occur during PHP's startup
; sequence are not displayed.  It's strongly recommended to keep
; display_startup_errors off, except for when debugging.
display_startup_errors = Off

; Log errors into a log file (server-specific log, stderr, or error_log (below))
; As stated above, you're strongly advised to use error logging in place of
; error displaying on production web sites.
log_errors = On
```

; Set maximum length of log_errors. In error_log information about the source is
; added. The default is 1024 and 0 allows to not apply any maximum length at all.
log_errors_max_len = 1024

; Do not log repeated messages. Repeated errors must occur in same file on same
; line until ignore_repeated_source is set true.
ignore_repeated_errors = Off

; Ignore source of message when ignoring repeated messages. When this setting
; is On you will not log errors with repeated messages from different files or
; sourcelines.
ignore_repeated_source = Off

; If this parameter is set to Off, then memory leaks will not be shown (on
; stdout or in the log). This has only effect in a debug compile, and if
; error reporting includes E_WARNING in the allowed list
report_memleaks = On

; Store the last error/warning message in $php_errormsg (boolean).
track_errors = Off

; Disable the inclusion of HTML tags in error messages.
; Note: Never use this feature for production boxes.
;html_errors = Off

; If html_errors is set On PHP produces clickable error messages that direct
; to a page describing the error or function causing the error in detail.
; You can download a copy of the PHP manual from http://www.php.net/docs.php
; and change docref_root to the base URL of your local copy including the
; leading '/'. You must also specify the file extension being used including
; the dot.
; Note: Never use this feature for production boxes.
;docref_root = "/phpmanual/"
;docref_ext = .html

; String to output before an error message.
;error_prepend_string = "<font color=ff0000>"

```
; String to output after an error message.
;error_append_string = "</font>"

; Log errors to specified file.
;error_log = filename

; Log errors to syslog (Event Log on NT, not valid in Windows 95).
;error_log = syslog


;;;;;;;;;;;;;;;;;
; Data Handling ;
;;;;;;;;;;;;;;;;;
;
; Note - track_vars is ALWAYS enabled as of PHP 4.0.3

; The separator used in PHP generated URLs to separate arguments.
; Default is "&".
;arg_separator.output = "&amp;"

; List of separator(s) used by PHP to parse input URLs into variables.
; Default is "&".
; NOTE: Every character in this directive is considered as separator!
;arg_separator.input = ";&"

; This directive describes the order in which PHP registers GET, POST,
Cookie,
; Environment and Built-in variables (G, P, C, E & S respectively, often
; referred to as EGPCS or GPC).  Registration is done from left to right,
newer
; values override older values.
variables_order = "EGPCS"

; Whether or not to register the EGPCS variables as global variables.  You
may
; want to turn this off if you don't want to clutter your scripts' global scope
; with user data.  This makes most sense when coupled with track_vars - in
which
; case you can access all of the GPC variables through the $HTTP_*_VARS[],
; variables.
;
; You should do your best to write your scripts so that they do not require
```

```
; register_globals to be on;  Using form variables as globals can easily lead
; to possible security problems, if the code is not very well thought of.
register_globals = Off

; Whether or not to register the old-style input arrays, HTTP_GET_VARS
; and friends.  If you're not using them, it's recommended to turn them off,
; for performance reasons.
register_long_arrays = Off

; This directive tells PHP whether to declare the argv&argc variables (that
; would contain the GET information).  If you don't use these variables, you
; should turn it off for increased performance.
register_argc_argv = Off

; When enabled, the SERVER and ENV variables are created when they're
first
; used (Just In Time) instead of when the script starts. If these variables
; are not used within a script, having this directive on will result in a
; performance gain. The PHP directives register_globals,
register_long_arrays,
; and register_argc_argv must be disabled for this directive to have any
affect.
auto_globals_jit = On

; Maximum size of POST data that PHP will accept.
post_max_size = 8M

; Magic quotes
;

; Magic quotes for incoming GET/POST/Cookie data.
magic_quotes_gpc = Off

; Magic quotes for runtime-generated data, e.g. data from SQL, from exec(),
etc.
magic_quotes_runtime = Off

; Use Sybase-style magic quotes (escape ' with '' instead of \').
magic_quotes_sybase = Off

; Automatically add files before or after any PHP document.
auto_prepend_file =
auto_append_file =
```

```
; As of 4.0b4, PHP always outputs a character encoding by default in
; the Content-type: header.  To disable sending of the charset, simply
; set it to be empty.
;
; PHP's built-in default is text/html
default_mimetype = "text/html"
;default_charset = "iso-8859-1"

; Always populate the $HTTP_RAW_POST_DATA variable.
;always_populate_raw_post_data = On


;;;;;;;;;;;;;;;;;;;;;;;;;
; Paths and Directories ;
;;;;;;;;;;;;;;;;;;;;;;;;;

; UNIX: "/path1:/path2"
;include_path = ".:/php/includes"
;
; Windows: "\path1;\path2"
;include_path = ".;c:\php\includes"

; The root of the PHP pages, used only if nonempty.
; if PHP was not compiled with FORCE_REDIRECT, you SHOULD set doc_root
; if you are running php as a CGI under any web server (other than IIS)
; see documentation for security issues.  The alternate is to use the
; cgi.force_redirect configuration below
doc_root =

; The directory under which PHP opens the script using /~username used
only
; if nonempty.
user_dir =

; Directory in which the loadable extensions (modules) reside.
extension_dir = "/usr/lib/php/modules"

; Whether or not to enable the dl() function.  The dl() function does NOT
work
; properly in multithreaded servers, such as IIS or Zeus, and is
automatically
; disabled on them.
```

```
enable_dl = On

; cgi.force_redirect is necessary to provide security running PHP as a CGI
under
; most web servers.  Left undefined, PHP turns this on by default.  You can
; turn it off here AT YOUR OWN RISK
; **You CAN safely turn this off for IIS, in fact, you MUST.**
; cgi.force_redirect = 1

; if cgi.nph is enabled it will force cgi to always sent Status: 200 with
; every request.
; cgi.nph = 1

; if cgi.force_redirect is turned on, and you are not running under Apache or
Netscape
; (iPlanet) web servers, you MAY need to set an environment variable name
that PHP
; will look for to know it is OK to continue execution.  Setting this variable
MAY
; cause security issues, KNOW WHAT YOU ARE DOING FIRST.
; cgi.redirect_status_env = ;

; FastCGI under IIS (on WINNT based OS) supports the ability to
impersonate
; security tokens of the calling client.  This allows IIS to define the
; security context that the request runs under.  mod_fastcgi under Apache
; does not currently support this feature (03/17/2002)
; Set to 1 if running under IIS.  Default is zero.
; fastcgi.impersonate = 1;

; Disable logging through FastCGI connection
; fastcgi.log = 0

; cgi.rfc2616_headers configuration option tells PHP what type of headers to
; use when sending HTTP response code. If it's set 0 PHP sends Status:
header that
; is supported by Apache. When this option is set to 1 PHP will send
; RFC2616 compliant header.
; Default is zero.
;cgi.rfc2616_headers = 0


;;;;;;;;;;;;;;;;
```

```
; File Uploads ;
;;;;;;;;;;;;;;;;

; Whether to allow HTTP file uploads.
file_uploads = On

; Temporary directory for HTTP uploaded files (will use system default if not
; specified).
;upload_tmp_dir =

; Maximum allowed size for uploaded files.
upload_max_filesize = 2M


;;;;;;;;;;;;;;;;;;
; Fopen wrappers ;
;;;;;;;;;;;;;;;;;;

; Whether to allow the treatment of URLs (like http:// or ftp://) as files.
allow_url_fopen = On

; Define the anonymous ftp password (your email address)
;from="john@doe.com"

; Define the User-Agent string
; user_agent="PHP"

; Default timeout for socket based streams (seconds)
default_socket_timeout = 60

; If your scripts have to deal with files from Macintosh systems,
; or you are running on a Mac and need to deal with files from
; unix or win32 systems, setting this flag will cause PHP to
; automatically detect the EOL character in those files so that
; fgets() and file() will work regardless of the source of the file.
; auto_detect_line_endings = Off


;;;;;;;;;;;;;;;;;;;;;;
; Dynamic Extensions ;
;;;;;;;;;;;;;;;;;;;;;;
;
; If you wish to have an extension loaded automatically, use the following
```

```
; syntax:
;
;   extension=modulename.extension
;
; For example:
;
;   extension=msql.so
;
; Note that it should be the name of the module only; no directory
information
; needs to go here.  Specify the location of the extension with the
; extension_dir directive above.


;;;;
; Note: packaged extension modules are now loaded via the .ini files
; found in the directory /etc/php.d; these are loaded by default.
;;;;


;;;;;;;;;;;;;;;;;;;
; Module Settings ;
;;;;;;;;;;;;;;;;;;;

[Date]
; Defines the default timezone used by the date functions
;date.timezone =

[Syslog]
; Whether or not to define the various syslog variables (e.g. $LOG_PID,
; $LOG_CRON, etc.).  Turning it off is a good idea performance-wise.  In
; runtime, you can define these variables by calling
define_syslog_variables().
define_syslog_variables  = Off

[mail function]
; For Win32 only.
SMTP = localhost
smtp_port = 25

; For Win32 only.
;sendmail_from = me@example.com
```

; For Unix only. You may supply arguments as well (default: "sendmail -t -i").
sendmail_path = /usr/sbin/sendmail -t -i

; Force the addition of the specified parameters to be passed as extra parameters
; to the sendmail binary. These parameters will always replace the value of
; the 5th parameter to mail(), even in safe mode.
;mail.force_extra_parameters =

[SQL]
sql.safe_mode = Off

[ODBC]
;odbc.default_db = Not yet implemented
;odbc.default_user = Not yet implemented
;odbc.default_pw = Not yet implemented

; Allow or prevent persistent links.
odbc.allow_persistent = On

; Check that a connection is still valid before reuse.
odbc.check_persistent = On

; Maximum number of persistent links. -1 means no limit.
odbc.max_persistent = -1

; Maximum number of links (persistent + non-persistent). -1 means no limit.
odbc.max_links = -1

; Handling of LONG fields. Returns number of bytes to variables. 0 means
; passthru.
odbc.defaultlrl = 4096

; Handling of binary data. 0 means passthru, 1 return as is, 2 convert to char.
; See the documentation on odbc_binmode and odbc_longreadlen for an explanation
; of uodbc.defaultlrl and uodbc.defaultbinmode
odbc.defaultbinmode = 1

[MySQL]

; Allow or prevent persistent links.
mysql.allow_persistent = On

; Maximum number of persistent links. -1 means no limit.
mysql.max_persistent = -1

; Maximum number of links (persistent + non-persistent). -1 means no limit.
mysql.max_links = -1

; Default port number for mysql_connect(). If unset, mysql_connect() will use
; the $MYSQL_TCP_PORT or the mysql-tcp entry in /etc/services or the
; compile-time value defined MYSQL_PORT (in that order). Win32 will only look
; at MYSQL_PORT.
mysql.default_port =

; Default socket name for local MySQL connects. If empty, uses the built-in
; MySQL defaults.
mysql.default_socket =

; Default host for mysql_connect() (doesn't apply in safe mode).
mysql.default_host =

; Default user for mysql_connect() (doesn't apply in safe mode).
mysql.default_user =

; Default password for mysql_connect() (doesn't apply in safe mode).
; Note that this is generally a *bad* idea to store passwords in this file.
; *Any* user with PHP access can run 'echo get_cfg_var("mysql.default_password")
; and reveal this password! And of course, any users with read access to this
; file will be able to reveal the password as well.
mysql.default_password =

; Maximum time (in secondes) for connect timeout. -1 means no limit
mysql.connect_timeout = 60

; Trace mode. When trace_mode is active (=On), warnings for table/index scans and
; SQL-Errors will be displayed.

```
mysql.trace_mode = Off

[MySQLi]

; Maximum number of links.  -1 means no limit.
mysqli.max_links = -1

; Default port number for mysqli_connect().  If unset, mysqli_connect() will
use
; the $MYSQL_TCP_PORT or the mysql-tcp entry in /etc/services or the
; compile-time value defined MYSQL_PORT (in that order).  Win32 will only
look
; at MYSQL_PORT.
mysqli.default_port = 3306

; Default socket name for local MySQL connects.  If empty, uses the built-in
; MySQL defaults.
mysqli.default_socket =

; Default host for mysql_connect() (doesn't apply in safe mode).
mysqli.default_host =

; Default user for mysql_connect() (doesn't apply in safe mode).
mysqli.default_user =

; Default password for mysqli_connect() (doesn't apply in safe mode).
; Note that this is generally a *bad* idea to store passwords in this file.
; *Any* user with PHP access can run 'echo
get_cfg_var("mysqli.default_pw")
; and reveal this password!  And of course, any users with read access to
this
; file will be able to reveal the password as well.
mysqli.default_pw =

; Allow or prevent reconnect
mysqli.reconnect = Off

[mSQL]
; Allow or prevent persistent links.
msql.allow_persistent = On

; Maximum number of persistent links.  -1 means no limit.
msql.max_persistent = -1
```

; Maximum number of links (persistent+non persistent).  -1 means no limit.
msql.max_links = -1

[PostgresSQL]
; Allow or prevent persistent links.
pgsql.allow_persistent = On

; Detect broken persistent links always with pg_pconnect().
; Auto reset feature requires a little overheads.
pgsql.auto_reset_persistent = Off

; Maximum number of persistent links.  -1 means no limit.
pgsql.max_persistent = -1

; Maximum number of links (persistent+non persistent).  -1 means no limit.
pgsql.max_links = -1

; Ignore PostgreSQL backends Notice message or not.
; Notice message logging require a little overheads.
pgsql.ignore_notice = 0

; Log PostgreSQL backends Noitce message or not.
; Unless pgsql.ignore_notice=0, module cannot log notice message.
pgsql.log_notice = 0

[Sybase]
; Allow or prevent persistent links.
sybase.allow_persistent = On

; Maximum number of persistent links.  -1 means no limit.
sybase.max_persistent = -1

; Maximum number of links (persistent + non-persistent).  -1 means no
limit.
sybase.max_links = -1

;sybase.interface_file = "/usr/sybase/interfaces"

; Minimum error severity to display.
sybase.min_error_severity = 10

; Minimum message severity to display.

```
sybase.min_message_severity = 10

; Compatability mode with old versions of PHP 3.0.
; If on, this will cause PHP to automatically assign types to results according
; to their Sybase type, instead of treating them all as strings.  This
; compatability mode will probably not stay around forever, so try applying
; whatever necessary changes to your code, and turn it off.
sybase.compatability_mode = Off

[Sybase-CT]
; Allow or prevent persistent links.
sybct.allow_persistent = On

; Maximum number of persistent links.  -1 means no limit.
sybct.max_persistent = -1

; Maximum number of links (persistent + non-persistent).  -1 means no
limit.
sybct.max_links = -1

; Minimum server message severity to display.
sybct.min_server_severity = 10

; Minimum client message severity to display.
sybct.min_client_severity = 10

[bcmath]
; Number of decimal digits for all bcmath functions.
bcmath.scale = 0

[browscap]
;browscap = extra/browscap.ini

[Informix]
; Default host for ifx_connect() (doesn't apply in safe mode).
ifx.default_host =

; Default user for ifx_connect() (doesn't apply in safe mode).
ifx.default_user =

; Default password for ifx_connect() (doesn't apply in safe mode).
ifx.default_password =
```

```
; Allow or prevent persistent links.
ifx.allow_persistent = On

; Maximum number of persistent links.  -1 means no limit.
ifx.max_persistent = -1

; Maximum number of links (persistent + non-persistent).  -1 means no
limit.
ifx.max_links = -1

; If on, select statements return the contents of a text blob instead of its id.
ifx.textasvarchar = 0

; If on, select statements return the contents of a byte blob instead of its id.
ifx.byteasvarchar = 0

; Trailing blanks are stripped from fixed-length char columns.  May help the
; life of Informix SE users.
ifx.charasvarchar = 0

; If on, the contents of text and byte blobs are dumped to a file instead of
; keeping them in memory.
ifx.blobinfile = 0

; NULL's are returned as empty strings, unless this is set to 1.  In that case,
; NULL's are returned as string 'NULL'.
ifx.nullformat = 0

[Session]
; Handler used to store/retrieve data.
session.save_handler = files

; Argument passed to save_handler.  In the case of files, this is the path
; where data files are stored. Note: Windows users have to change this
; variable in order to use PHP's session functions.
;
; As of PHP 4.0.1, you can define the path as:
;
;     session.save_path = "N;/path"
;
; where N is an integer.  Instead of storing all the session files in
; /path, what this will do is use subdirectories N-levels deep, and
; store the session data in those directories.  This is useful if you
```

```
; or your OS have problems with lots of files in one directory, and is
; a more efficient layout for servers that handle lots of sessions.
;
; NOTE 1: PHP will not create this directory structure automatically.
;         You can use the script in the ext/session dir for that purpose.
; NOTE 2: See the section on garbage collection below if you choose to
;         use subdirectories for session storage
;
; The file storage module creates files using mode 600 by default.
; You can change that by using
;
;     session.save_path = "N;MODE;/path"
;
; where MODE is the octal representation of the mode. Note that this
; does not overwrite the process's umask.
session.save_path = "/var/lib/php/session"

; Whether to use cookies.
session.use_cookies = 1

; This option enables administrators to make their users invulnerable to
; attacks which involve passing session ids in URLs; defaults to 0.
; session.use_only_cookies = 1

; Name of the session (used as cookie name).
session.name = PHPSESSID

; Initialize session on request startup.
session.auto_start = 0

; Lifetime in seconds of cookie or, if 0, until browser is restarted.
session.cookie_lifetime = 0

; The path for which the cookie is valid.
session.cookie_path = /

; The domain for which the cookie is valid.
session.cookie_domain =

; Handler used to serialize data.  php is the standard serializer of PHP.
session.serialize_handler = php

; Define the probability that the 'garbage collection' process is started
```

```
; on every session initialization.
; The probability is calculated by using gc_probability/gc_divisor,
; e.g. 1/100 means there is a 1% chance that the GC process starts
; on each request.

session.gc_probability = 1
session.gc_divisor     = 1000

; After this number of seconds, stored data will be seen as 'garbage' and
; cleaned up by the garbage collection process.
session.gc_maxlifetime = 1440

; NOTE: If you are using the subdirectory option for storing session files
;       (see session.save_path above), then garbage collection does *not*
;       happen automatically.  You will need to do your own garbage
;       collection through a shell script, cron entry, or some other method.
;       For example, the following script would is the equivalent of
;       setting session.gc_maxlifetime to 1440 (1440 seconds = 24 minutes):
;          cd /path/to/sessions; find -cmin +24 | xargs rm

; PHP 4.2 and less have an undocumented feature/bug that allows you to
; to initialize a session variable in the global scope, albeit register_globals
; is disabled.  PHP 4.3 and later will warn you, if this feature is used.
; You can disable the feature and the warning separately. At this time,
; the warning is only displayed, if bug_compat_42 is enabled.

session.bug_compat_42 = 0
session.bug_compat_warn = 1

; Check HTTP Referer to invalidate externally stored URLs containing ids.
; HTTP_REFERER has to contain this substring for the session to be
; considered as valid.
session.referer_check =

; How many bytes to read from the file.
session.entropy_length = 0

; Specified here to create the session id.
session.entropy_file =

;session.entropy_length = 16

;session.entropy_file = /dev/urandom
```

```
; Set to {nocache,private,public,} to determine HTTP caching aspects
; or leave this empty to avoid sending anti-caching headers.
session.cache_limiter = nocache

; Document expires after n minutes.
session.cache_expire = 180

; trans sid support is disabled by default.
; Use of trans sid may risk your users security.
; Use this option with caution.
; - User may send URL contains active session ID
;   to other person via. email/irc/etc.
; - URL that contains active session ID may be stored
;   in publically accessible computer.
; - User may access your site with the same session ID
;   always using URL stored in browser's history or bookmarks.
session.use_trans_sid = 0

; Select a hash function
; 0: MD5   (128 bits)
; 1: SHA-1 (160 bits)
session.hash_function = 0

; Define how many bits are stored in each character when converting
; the binary hash data to something readable.
;
; 4 bits: 0-9, a-f
; 5 bits: 0-9, a-v
; 6 bits: 0-9, a-z, A-Z, "-", ","
session.hash_bits_per_character = 5

; The URL rewriter will look for URLs in a defined set of HTML tags.
; form/fieldset are special; if you include them here, the rewriter will
; add a hidden <input> field with the info which is otherwise appended
; to URLs.  If you want XHTML conformity, remove the form entry.
; Note that all valid entries require a "=", even if no value follows.
url_rewriter.tags = "a=href,area=href,frame=src,input=src,form=fakeentry"

[MSSQL]
; Allow or prevent persistent links.
mssql.allow_persistent = On
```

```
; Maximum number of persistent links.  -1 means no limit.
mssql.max_persistent = -1

; Maximum number of links (persistent+non persistent).  -1 means no limit.
mssql.max_links = -1

; Minimum error severity to display.
mssql.min_error_severity = 10

; Minimum message severity to display.
mssql.min_message_severity = 10

; Compatability mode with old versions of PHP 3.0.
mssql.compatability_mode = Off

; Connect timeout
;mssql.connect_timeout = 5

; Query timeout
;mssql.timeout = 60

; Valid range 0 - 2147483647.  Default = 4096.
;mssql.textlimit = 4096

; Valid range 0 - 2147483647.  Default = 4096.
;mssql.textsize = 4096

; Limits the number of records in each batch.  0 = all records in one batch.
;mssql.batchsize = 0

; Specify how datetime and datetim4 columns are returned
; On => Returns data converted to SQL server settings
; Off => Returns values as YYYY-MM-DD hh:mm:ss
;mssql.datetimeconvert = On

; Use NT authentication when connecting to the server
mssql.secure_connection = Off

; Specify max number of processes. -1 = library default
; msdlib defaults to 25
; FreeTDS defaults to 4096
;mssql.max_procs = -1
```

```
; Specify client character set.
; If empty or not set the client charset from freetds.comf is used
; This is only used when compiled with FreeTDS
;mssql.charset = "ISO-8859-1"

[Assertion]
; Assert(expr); active by default.
;assert.active = On

; Issue a PHP warning for each failed assertion.
;assert.warning = On

; Don't bail out by default.
;assert.bail = Off

; User-function to be called if an assertion fails.
;assert.callback = 0

; Eval the expression with current error_reporting().  Set to true if you want
; error_reporting(0) around the eval().
;assert.quiet_eval = 0

[Verisign Payflow Pro]
; Default Payflow Pro server.
pfpro.defaulthost = "test-payflow.verisign.com"

; Default port to connect to.
pfpro.defaultport = 443

; Default timeout in seconds.
pfpro.defaulttimeout = 30

; Default proxy IP address (if required).
;pfpro.proxyaddress =

; Default proxy port.
;pfpro.proxyport =

; Default proxy logon.
;pfpro.proxylogon =

; Default proxy password.
;pfpro.proxypassword =
```

```
[COM]
; path to a file containing GUIDs, IIDs or filenames of files with TypeLibs
;com.typelib_file =
; allow Distributed-COM calls
;com.allow_dcom = true
; autoregister constants of a components typlib on com_load()
;com.autoregister_typelib = true
; register constants casesensitive
;com.autoregister_casesensitive = false
; show warnings on duplicate constat registrations
;com.autoregister_verbose = true

[mbstring]
; language for internal character representation.
;mbstring.language = Japanese

; internal/script encoding.
; Some encoding cannot work as internal encoding.
; (e.g. SJIS, BIG5, ISO-2022-*)
;mbstring.internal_encoding = EUC-JP

; http input encoding.
;mbstring.http_input = auto

; http output encoding. mb_output_handler must be
; registered as output buffer to function
;mbstring.http_output = SJIS

; enable automatic encoding translation according to
; mbstring.internal_encoding setting. Input chars are
; converted to internal encoding by setting this to On.
; Note: Do _not_ use automatic encoding translation for
;       portable libs/applications.
;mbstring.encoding_translation = Off

; automatic encoding detection order.
; auto means
;mbstring.detect_order = auto

; substitute_character used when character cannot be converted
; one from another
;mbstring.substitute_character = none;
```

```
; overload(replace) single byte functions by mbstring functions.
; mail(), ereg(), etc are overloaded by mb_send_mail(), mb_ereg(),
; etc. Possible values are 0,1,2,4 or combination of them.
; For example, 7 for overload everything.
; 0: No overload
; 1: Overload mail() function
; 2: Overload str*() functions
; 4: Overload ereg*() functions
;mbstring.func_overload = 0

; enable strict encoding detection.
;mbstring.strict_encoding = Off

[FrontBase]
;fbsql.allow_persistent = On
;fbsql.autocommit = On
;fbsql.default_database =
;fbsql.default_database_password =
;fbsql.default_host =
;fbsql.default_password =
;fbsql.default_user = "_SYSTEM"
;fbsql.generate_warnings = Off
;fbsql.max_connections = 128
;fbsql.max_links = 128
;fbsql.max_persistent = -1
;fbsql.max_results = 128
;fbsql.batchSize = 1000

[gd]
; Tell the jpeg decode to libjpeg warnings and try to create
; a gd image. The warning will then be displayed as notices
; disabled by default
;gd.jpeg_ignore_warning = 0

[exif]
; Exif UNICODE user comments are handled as UCS-2BE/UCS-2LE and JIS as
JIS.
; With mbstring support this will automatically be converted into the
encoding
; given by corresponding encode setting. When empty
mbstring.internal_encoding
; is used. For the decode settings you can distinguish between motorola and
```

```
; intel byte order. A decode setting cannot be empty.
;exif.encode_unicode = ISO-8859-15
;exif.decode_unicode_motorola = UCS-2BE
;exif.decode_unicode_intel    = UCS-2LE
;exif.encode_jis =
;exif.decode_jis_motorola = JIS
;exif.decode_jis_intel    = JIS

[Tidy]
; The path to a default tidy configuration file to use when using tidy
;tidy.default_config = /usr/local/lib/php/default.tcfg

; Should tidy clean and repair output automatically?
; WARNING: Do not use this option if you are generating non-html content
; such as dynamic images
tidy.clean_output = Off

[soap]
; Enables or disables WSDL caching feature.
soap.wsdl_cache_enabled=1
; Sets the directory name where SOAP extension will put cache files.
soap.wsdl_cache_dir="/tmp"
; (time to live) Sets the number of second while cached file will be used
; instead of original one.
soap.wsdl_cache_ttl=86400

; Local Variables:
; tab-width: 4
; End:


;***** Added by go-pear
include_path=".:/usr/share/pear/PEAR"
;*****
```

**Appendix III.4. config.php.**

```
<?php
$host="spec106.encs.concordia.ca";
$user="dbuser";
$pass="teamtwo";
$db01="uuisdb";
?>
```